\documentclass[aps]{revtex4}
\usepackage{eurosym}
\usepackage{amsfonts}
\usepackage{amsmath}
\usepackage{amssymb,epsf}
\usepackage{color}
\usepackage{graphicx}
\usepackage{float}
\usepackage{caption}
\usepackage{subfig}
\usepackage{epstopdf}

\begin{document}

\title{Gauss-Bonnet Dyonic Black Holes:\\
geometry, thermodynamics and test particles' trajectories}
\author{S. Panahiyan$^{1,2}$\footnote{
email address: shahram.panahiyan@uni-jena.de}, S. H. Hendi$^{3,4}$\footnote{
email address: hendi@shirazu.ac.ir} and N. Riazi$^{1}$\footnote{
email address: n$_{-}$riazi@sbu.ac.ir}}
\affiliation{$^1$ Physics Department, Shahid Beheshti University, Tehran 19839, Iran\\
$^2$ Helmholtz-Institut Jena, Fr\"{o}belstieg 3, Jena D-07743 Germany \\
$^3$ Physics Department and Biruni Observatory, College of Sciences, Shiraz
University, Shiraz 71454, Iran\\
$^4$ Research Institute for Astronomy and Astrophysics of Maragha (RIAAM),
P.O. Box 55134-441, Maragha, Iran}

\begin{abstract}
In this paper, we investigate a class of $5$-dimensional black holes in the
presence of Gauss-Bonnet gravity with dyonic charges. At first step,
thermodynamical quantities of the black holes and their behaviors are
explored for different limits. Thermal stability and the possibility of the
van der Waals like phase transition are addressed and the effects
of different parameters on them are
investigated. The second part is devoted to simulation of the trajectory of
particles around these black holes and investigation of the angular frequency
of particles' motion. The main goal is
understanding the effects of higher curvature gravity (Gauss-Bonnet gravity)
and magnetic charge on the structure of black holes and the geodesic paths
of particles moving around these black holes.
\end{abstract}

\maketitle

\section{Introduction}

Lovelock gravity is one of the well established/known
generalizations of the Einstein gravity
\cite{Lovelock1,Lovelock2}. The first three terms of this
generalization include three degrees of curvature term including:
I) a constant realized as the cosmological constant, II) a first
order curvature term known as the Einstein Lagrangian, III) and
finally a curvature-squared term called Gauss-Bonnet (GB)
Lagrangian. The Gauss-Bonnet term is a topological term in $4$ and
lower dimensions. Its structure results into the appearance of up
to the second order derivations of metric functions in the field
equations
\cite{secondorder1,secondorder2,secondorder3,secondorder4}. It is
a higher derivative gravity enjoying the absence of ghost
instability \cite{ghostfree1,ghostfree2}. Its Lagrangian could be
obtained in the low-energy limit of heterotic string theory
\cite{string1,string2,string3,string4,string5}. Black holes and
their properties in the presence of GB gravity have been
intensively investigated in literature
\cite{blackholesGB1,blackholesGB01,blackholesGB011,blackholesGB2,blackholesGB3,blackholesGB4,blackholesGB5,blackholesGB6,blackholesGB7,blackholesGB8,blackholesGB9,blackholesGB10,blackholesGB11,blackholesGB12,blackholesGB121,blackholesGB13,blackholesGB14,blackholesGB16,blackholesGB17,blackholesGB18,blackholesGB19}.
In addition, the black string solutions with GB generalization and
their stabilities have been studied in Refs.
\cite{blackSt1,blackSt2,blackSt3,blackSt4,blackSt5}. In
holographical context, the effects of the GB gravity on finite
coupling \cite{H1}, second order transport \cite{H2}, entanglement
entropy \cite{H31,H32,H33,H34} and superconductivity
\cite{H4,H5,H6} have been addressed. It is worthwhile to mention
that inclusion of the GB term in $5$ dimensions could results into
causality violation unless you have a tower of massive higher-spin
particles with a delicately tuned interaction, which happens in
string theory \cite{EL}. Considering the diverse applications of
GB gravity in classical black hole thermodynamics, holography and
string theory, we investigate a specific type of GB black holes in
the presence of dyonic charge.

The \textit{dyonic charge} corresponds to the existence of
magnetic charge (in addition to electric charge) in the structure
of black holes. The dyonic black hole solutions have been vastly
used as popular models for investigations in the context of
AdS/CFT duality. One of the early applications of the dyonic black
holes was studying the Hall conductivity and zero momentum
hydrodynamic response functions in the context of AdS/CFT
\cite{Hartnoll}. In addition, it was confirmed that large dyonic
black holes in AdS spacetime correspond to stationary solutions of
the equations of relativistic magnetohydrodynamics on the
conformal boundary of AdS \cite{Caldarelli}. Of other
applications/investigations of/on the dyonic black holes in
AdS/CFT context, one can name the following ones: I) Inducing
external magnetic field on superconductors which results into
magnetic dependency similar to the Meissner effect \cite{Albash},
II) Their effects on the holographical properties of solutions
such as transport coefficients, Hall conductance and DC
longitudinal conductivity \cite{Goldstein}. III) The
paramagnetism/ferromagnetism phase transitions in case of dyonic
Reissner-Nordstrom black holes \cite{HD1,HD2,HD3} and massive
dyonic black holes \cite{HD4}.

In the context of string theory, the exact dyonic black hole
solutions were constructed in Refs. \cite{DS1,DS2}. In addition,
the dyonic black hole solutions were also investigated in
effective \cite{DS3} and hetroctic \cite{DS4} string theories. The
extreme solutions \cite{DS5} and asymptotic degeneracy of dyonic
$N=4$ string states \cite{DS6} were studied as well. From
supergravity point of view, dyonic black hole solutions were
obtained in Refs. \cite{SGD1,SGD2} and their thermodynamics were
investigated in Ref. \cite{SGD3}. Furthermore, the existence of
dyonic black holes in the Einstein-Yang-Mills theory was
investigated in Refs. \cite{YD1,YD2,YD3} and their gravitating
properties \cite{GD1,GD2} were studied.

Recently, the dyonic black holes with the Born-Infeld nonlinear
electromagnetic field in diverse dimensions were constructed
\cite{BHD1}. Thermodynamic of the dyonic black holes was
thoroughly investigated in Ref. \cite{BHD2}. Ref. \cite{BHD3}
presents a beautiful systematic study regarding computation of
conserved quantities of the dyonic black holes in $AdS_{4}$. The
discreteness of dyonic dilaton black holes \cite{BHD4},
thermodynamics of dyonic black holes with Thurston horizon
geometries \cite{BHD5} and dyonic black holes at arbitrary
locations \cite{BHD6} were also studied. A general scheme for
obtaining dyonic solutions in nonlinear electrodynamics was
proposed in Ref. \cite{BHD7}.

Considering the vast applications of both GB gravity and dyonic
configuration in the context of black holes, this paper is
dedicated to the investigation of dyonic black holes in the
presence of GB gravity. To our knowledge, so far the consideration
of dyonic configuration with GB gravity was done only for the
cases where dyonic contribution is confined to $4$-dimensions
while GB gravity was contributing from higher dimensions
\cite{GDG1,GDG2}. In this paper, we intend to construct a
$5$-dimensional black hole solution in the presence of dyonic
matter-field and GB gravity. We first introduce action and
corresponding field equations. Then, we extract black hole
solutions and investigate the effects of dyonic and GB gravity on
geometrical properties of the black holes. We continue our study
by obtaining thermodynamical properties of the solutions,
investigating the stability and thermal phase transition of the
solutions. Later, we focus on trajectories of a test particle
around these black holes. We finalize our paper with some closing
remarks.

\section{Lagrangian Ansatz}

Our line of work in this paper includes investigating dyonic black holes
with Gauss-Bonnet gravity generalization. The main goal is to understand the
geometrical and thermodynamical properties of the black holes with magnetic
and/or electric fields and higher curvature terms generalization
(Gauss-Bonnet gravity).

In this section, we first introduce the Lagrangian governing
Gauss-Bonnet-dyonic black holes. Then, we obtain the metric function and
confirm the existence of black holes. Finally, we investigate the effects of
magnetic, electric and GB parameters on geometrical properties of the black
holes.

\subsection{Constructing the solutions}

The Lagrangian of Einstein gravity in the presence of cosmological constant
is given by $\mathcal{L}_{E\Lambda }=R-2\Lambda $, which contains the first
order curvature scalar. Using the variational principle, one finds that its
related field equation contains first order or at most second order
derivation of the metric functions.

On the other hand, GB Lagrangian includes second order scalar curvature
terms which is given by
\begin{equation}
\mathcal{L}_{GB} = R_{\mu \nu \gamma \delta }R^{\mu \nu \gamma \delta
}-4R_{\mu \nu }R^{\mu \nu }+R^{2},  \label{GB}
\end{equation}
where $R_{\mu \nu \gamma \delta }$ and $R_{\mu \nu}$ are respectively, the
Riemann and the Ricci tensors. GB Lagrangian has two important properties:
I) Due to the specific factors considered for different terms of Lagrangian,
only up to second order derivations of the metric functions could appear in
the field equations. II) The Lagrangian is effective for $dimensions>4$ and
it behaves as a topological term for $dimensions \leq 4$. Considering the
latter, we will conduct our study in $5-$dimension in this paper. Sure our
results can be generalized to higher dimensions.

The black holes that we are interested in, are topological black holes with
various horizon topologies. The static metric ansatz is given by
\begin{equation}
ds^{2}=-f(r)dt^{2}+\frac{dr^{2}}{f(r)}+r^{2}d\Omega _{k}^{2},  \label{Metric}
\end{equation}
in which $f(r)$ is the metric function and $d\Omega _{k}^{2}$ is the line
element of a $3$-dimensional hypersurface with the constant curvature $6k$
and volume $V_{3}$, in which its explicit form is
\begin{equation*}
d\Omega _{k}^{2}=\left\{
\begin{array}{cc}
d\theta^{2}+\sin^{2}\theta d\psi^{2}+\sin^{2}\theta \sin^{2}\psi d\phi^{2},
& k=1 \text{ (spherical)} \\
d\theta^{2}+\sinh ^{2}\theta d\psi^{2}+\sinh ^{2}\theta \sinh ^{2}\psi
d\phi^{2}, & k=-1 \text{ (hyperbolic)} \\
d\theta^{2}+ d\psi^{2}+ d\phi^{2}, & k=0 \text{ (flat)}%
\end{array}%
\right. .
\end{equation*}

The matter field of our interest is of dyonic nature. The word
"dyonic" corresponds to the presence of the magnetic charge
alongside of the electric charge in the matter field. There are
different methods for introducing magnetic charge into the
structure of the black holes. In this paper, we follow the method
introduced in Ref. \cite{HRPE}. The electromagnetic tensor is an
antisymmetric second rank tensor with the following components of
electric and magnetic field in $5-$dimension
\begin{equation}
F_{\mu \nu }=\left[
\begin{array}{ccccc}
0 & F_{tr} & F_{t\theta } & F_{t\phi } & F_{t\psi } \\
-F_{tr} & 0 & F_{r\theta } & F_{r\phi } & F_{r\psi } \\
-F_{t\theta } & -F_{r\theta } & 0 & F_{\theta \phi } & F_{\theta \psi } \\
-F_{t\phi } & -F_{r\phi } & -F_{\theta \phi } & 0 & F_{\phi \psi } \\
-F_{t\psi } & -F_{r\psi } & -F_{\theta \psi } & -F_{\phi \psi } & 0%
\end{array}%
\right] ,  \label{Fmn1}
\end{equation}%
where its nonzero components are functions of all independent
coordinates. In this paper, we assume that Eq. (\ref{Fmn1})
satisfies the Maxwell field equations, while the effective
components of electromagnetic tensor in the energy-momentum tensor
are $F_{tr}$ and $F_{\theta \phi }$ with the
following forms%
\begin{equation}
F_{tr}=-F_{rt}=\frac{q_{E}}{r^{3}}\text{ \ \ \ \ \ \ \& \ \ \ \ \ \ }%
F_{\theta \phi }=-F_{\phi \theta }=\frac{q_{M}}{r}\times \left\{
\begin{array}{cc}
\sin \theta  & k=1 \\
\theta  & k=0 \\
\sinh \theta  & k=-1%
\end{array}%
\right.   \label{Ftr}
\end{equation}%
in which $q_{E}$ and $q_{M}$ are, respectively, electric and
magnetic charges related to the total electric and magnetic
charges of the solutions. It is notable to mention that although
other off-diagonal components of Eq. (\ref{Fmn1}) do not vanish,
necessarily, their influences in the energy-momentum tensor are
negligible comparing with $F_{tr}$ and $F_{\theta \phi }$, $T_{\mu
\nu }$. In other words, Eq. (\ref{Ftr}) denotes the leading terms
of $F_{\mu \nu }$ affect the nonzero components of $T_{\mu \nu }$,
dominantly. Thus, we use the gravitational field equation of
$5$-dimensional GB-dyonic black holes which is given by
\begin{equation}
e_{\mu \nu }\equiv G_{\mu \nu }+\Lambda g_{\mu \nu }+\alpha H_{\mu
\nu }- \left[ 2F_{\mu \lambda }F_{\nu }^{\lambda
}-\frac{1}{2}g_{\mu \nu }F^{\sigma \rho }F_{\sigma \rho }\right]
=0,  \label{fieldeq}
\end{equation}%
in which $\alpha $ is GB parameter, $g_{\mu \nu }$ and $G_{\mu \nu
}$ are, respectively, the metric and Einstein tensors, and $H_{\mu
\nu }$ is the GB tensor given by
\begin{eqnarray*}
H_{\mu \nu } &=&-\frac{1}{2}\left( 8R^{\rho \sigma }R_{\mu \rho \nu \sigma
}-4R_{\mu }^{\rho \sigma \lambda }R_{\nu \rho \sigma \lambda }-4RR_{\mu \nu
}+8R_{\mu \lambda }R_{\nu }^{\lambda }+g_{\mu \nu }L_{GB}\right).
\end{eqnarray*}

Using the metric (\ref{Metric}) with the obtained field equation (\ref%
{fieldeq}), one can extract the nonzero components of field equation. The
full form of these components are given in the appendix (see Eqs. (\ref{e1})
and (\ref{e2})). It is a matter of calculation to obtain the metric function
by solving the components of field equation, simultaneously, as

\begin{equation}
f(r)=k+\frac{r^{2}}{4\alpha }\left\{ 1-\sqrt{1+\frac{4\alpha }{3}\left[
\Lambda +\frac{6m}{2r^{4}}-\frac{2(q_{E}^{2}+q_{M}^{2})}{r^{6}}\right] }%
\right\} ,  \label{metric function}
\end{equation}%
in which $m$ is an integration constant known as the geometrical mass.

\subsection{Properties of the solution}

Here, we investigate different properties of the obtained metric function (%
\ref{metric function}). Before we start, one point should be noted.
Existence of black holes has two conditions: one is the presence of
singularity for scalar curvatures and the other one is presence of horizon
covering the singularity.

I) Since the electric and magnetic charges are decoupled, in the limit of $%
q_{E} \rightarrow 0$, one can have purely magnetically charged
black holes in the presence of GB gravity. It is also worthwhile
to mention that solutions enjoy the electric-magnetic duality and
the metric function has the full symmetry of swapping the
subscripts "E" and "M".

II) Existence of real metric function is restricted to satisfaction of the
following condition
\begin{equation*}
\frac{3}{4 \alpha}- \Lambda \geq \frac{2 \left(q_{E}^{2} +q_{M}^{2}\right)}{%
r^6}-\frac{6 m}{r^4}.
\end{equation*}

III) For the case
\begin{equation*}
\frac{3}{4 \alpha}- \Lambda = \frac{2 \left(q_{E}^{2} +q_{M}^{2}\right)}{r^6}%
-\frac{6 m}{r^4},
\end{equation*}
the metric function reduces to
\begin{equation}
f\left( r\right) =k+\frac{r^{2}}{4\alpha },
\end{equation}
which yields $r=\sqrt{-k \alpha}$ as the root of the metric function. Roots
of the metric function are horizons of the black holes. Therefore, in this
case, only hyperbolic solutions have real positive non-zero root and black
holes exist only for hyperbolic horizon (other horizons admit naked
singularity not black hole solutions).

\begin{figure*}[!htbp]
    \centering
    \includegraphics[width=0.4\linewidth]{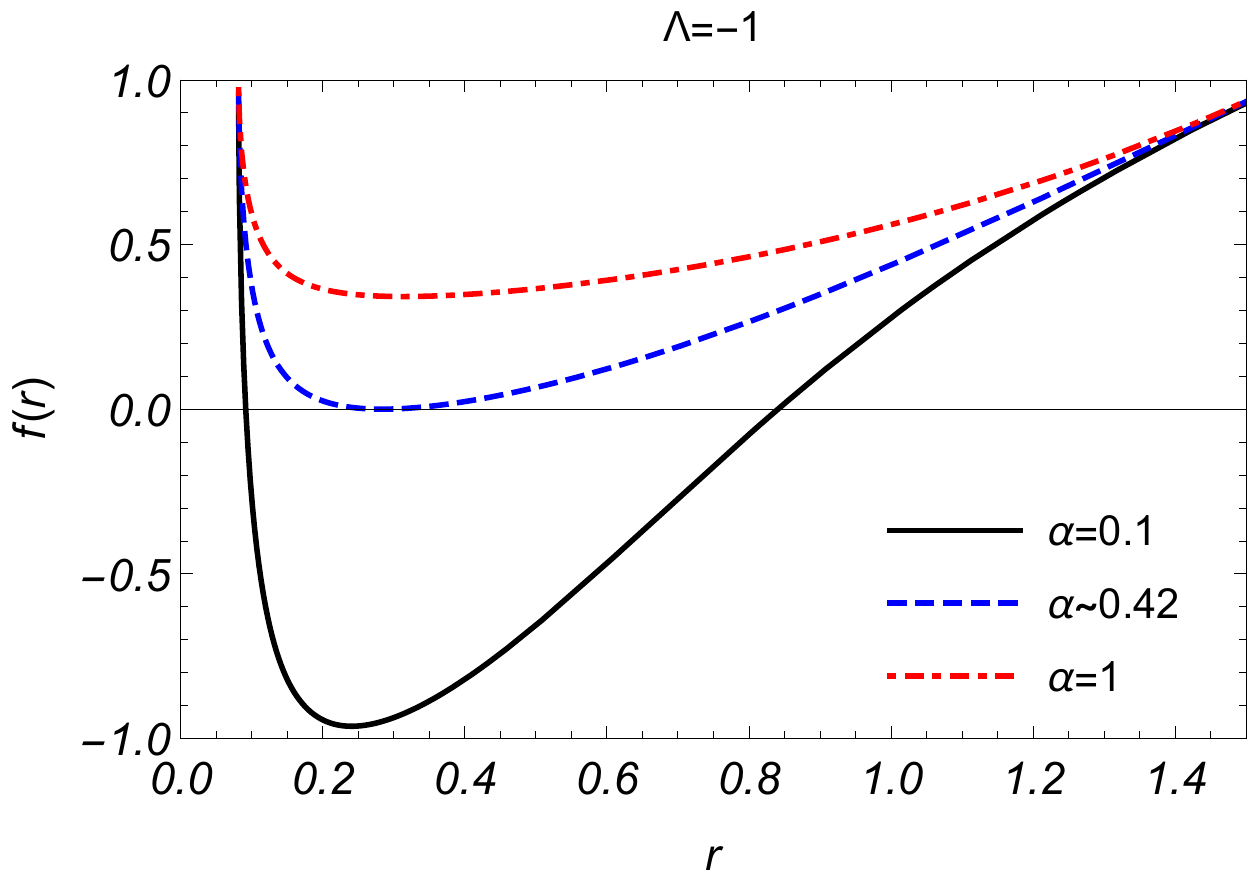}
    \includegraphics[width=0.4\linewidth]{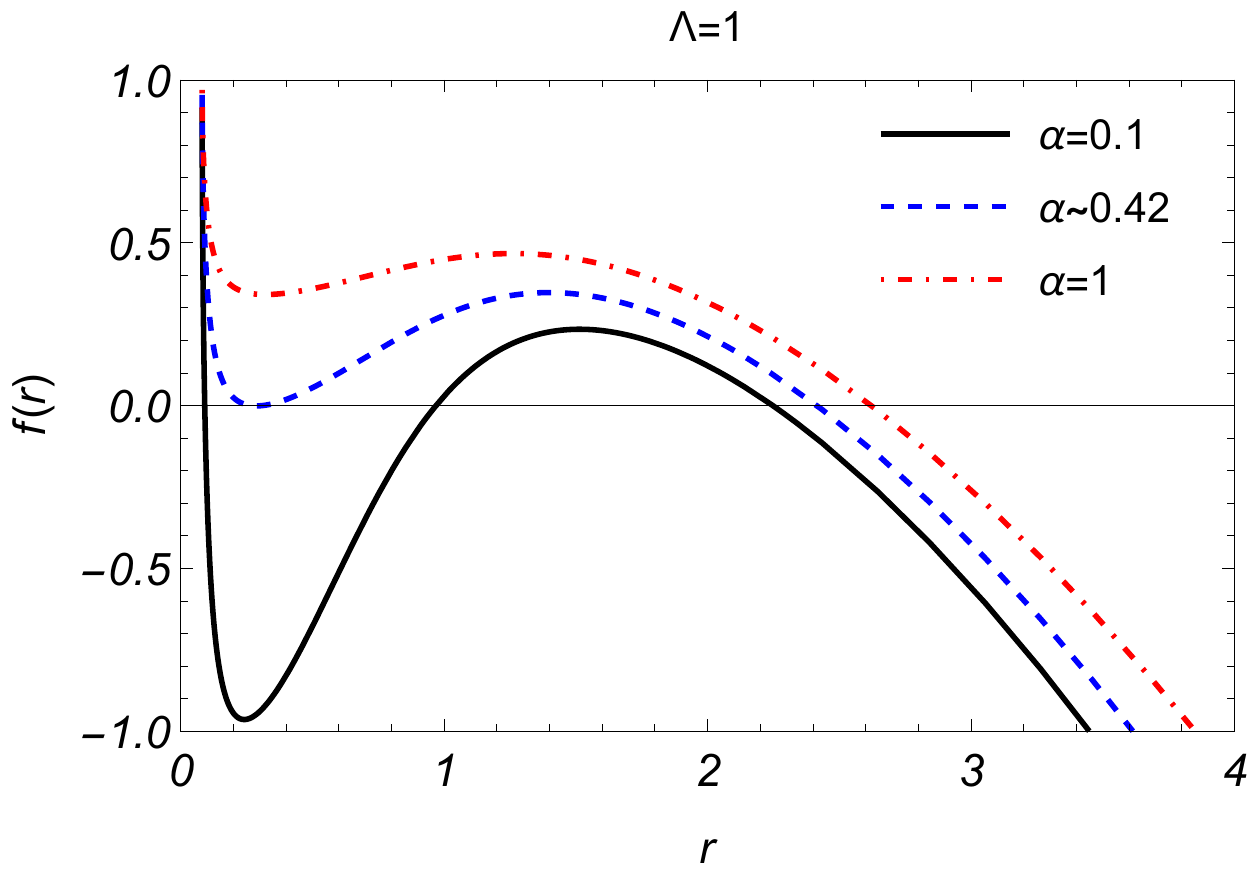}
    \caption{Metric function versus $r$ for $q_{M}=q_{E}=0.1$ and $k=m=1$. }
    \label{Fig0}
\end{figure*}

IV) In general, the metric function could admit up to three roots. The roots
are given in the appendix (see Eq. \ref{roots of metric}). These roots
contain square root functions. These square root functions put certain
conditions for having real positive non-zero roots for the metric function.
If these conditions are satisfied, the metric function could have one to
three roots. Otherwise, the solutions will admit naked singularity. In order
to elaborate these issues, we have plotted Fig. \ref{Fig0} which confirms
our discussion. Existence of roots (horizons) for metric function indicates
that the second condition for having black hole solutions is satisfied.

V) The presence of singularity could be detected by studying the
divergencies of curvature scalars. In this regard, we study a curvature
scalar known as Kretschmann scalar. In general, for $5$-dimensional
topological black holes, the Kretschmann scalar is given by
\begin{equation}
R_{\alpha \beta \gamma \delta }R^{\alpha \beta \gamma \delta } =\left( \frac{%
d^{2}f\left( r\right) }{dr^{2}}\right) +\frac{4}{r^{2}}\left( \frac{df\left(
r\right) }{dr}\right) ^{2}+\frac{12} {r^{4}}\left( f\left( r\right)
-k\right) ^{2}.  \label{Kretschmann}
\end{equation}

Inserting metric function (\ref{metric function}) in the Kretschmann scalar (%
\ref{Kretschmann}) leads to
\begin{equation}
R_{\alpha \beta \gamma \delta }R^{\alpha \beta \gamma \delta } \propto O(%
\frac{1}{r^{6}}).  \label{KretschmannA}
\end{equation}

The full form of the Kretschmann scalar is given in appendix (see Eq. \ref%
{Full Kretschmann}). Eq. (\ref{KretschmannA}) shows that
\begin{equation*}
\lim_{r\longrightarrow 0}R_{\alpha \beta \gamma \delta }R^{\alpha \beta
\gamma \delta }\longrightarrow \infty ,
\end{equation*}%
which confirms that there is a curvature singularity at $r=0$. By series
expanding the obtained Kretschmann scalar for small values of $r$, one can
find the following relation
\begin{equation*}
\lim_{r\longrightarrow 0}R_{\alpha \beta \gamma \delta }R^{\alpha \beta
\gamma \delta }\propto -\frac{3\left( q_{E}^{2}+q_{M}^{2}\right) }{\alpha
r^{6}}+\frac{3m}{\alpha r^{4}}+O(\frac{1}{\alpha r^{3}}).
\end{equation*}

This limit confirms that the singularity and its nature (being temporal or
spatial singularity) is affected by GB gravity, electric and magnetic fields.

VI) The asymptotic behavior of Kretschmann scalar is a measure that hints us
to find whether solutions have asymptotic AdS/dS behavior or not. Using the
obtained Kretschmann scalar (Eq. \ref{Full Kretschmann}) and its series
expansion for large $r$ result into
\begin{equation*}
\lim_{r\longrightarrow \infty }R_{\alpha \beta \gamma \delta }R^{\alpha
\beta \gamma \delta }\propto \frac{(-\alpha -9)\sqrt{1+\frac{4\alpha \Lambda
}{3}}}{2\alpha ^{2}}+\frac{(1+6\Lambda )\alpha +9}{2\alpha ^{2}}+O(\frac{1}{r%
}).
\end{equation*}

This indicates that the asymptotic behavior of the solutions is not anymore
the regular AdS/dS, but in fact an effective relation is obtained for it.

\section{Thermodynamics}

In this section, we focus on thermodynamical aspects of the GB-dyonic black
holes. Our goal is to understand the effects of magnetic charge alongside
the GB gravity generalization. We conduct our study in three subsections,
each focusing on a specific thermodynamical property of these black holes.
First, we extract thermodynamical quantities. Then, we investigate the
possible van der Waals like phase transition. Finally, we study thermal
stability of solutions through canonical ensemble.

\subsection{Thermodynamical quantities}

In this subsection, we obtain the thermodynamical quantities of the
solutions and check the validity of first law of black hole thermodynamics.

The first item of our interest is the entropy. In Einsteinian black holes,
without higher curvature terms, the entropy could be obtained by using the
area law \cite{Beckenstein,Hawking}. But, since our solutions are
GB-included, the area law is no longer valid for the calculation of entropy.
Instead, one can use the Wald formula given by \cite{Wald}
\begin{equation}
S=\frac{1}{4}\int d^3x\sqrt{\gamma}(1+2 \alpha \tilde{R}),
\end{equation}
in which $\tilde{R}$ is the Ricci scalar of the induced metric $\gamma_{ab}$
on the $3$-dimensional boundary. It is a matter of calculation to show that
in general, for topological GB black holes, the entropy per volume $V_{3}$
is obtained as
\begin{equation}
S=\frac{1}{4} r_{+}^3 \left(1+\frac{12 \alpha k}{r_{+}^2}\right).
\label{entropy}
\end{equation}

This expression for the entropy confirms the following important points:

I) The entropy is not directly affected by electric and magnetic charges.
This quantity is affected by topological structure of the black holes and GB
parameter (metric and left hand side of gravitational field equation).

II) The GB parameter and topological factor are coupled. Since GB parameter
is positive valued, the effects of GB parameter and topological factor on
entropy is determined by the topological factor, $k$. Therefore, entropy is:
1) a decreasing function of GB parameter for the hyperbolic black holes. 2)
not affected by GB gravity for the flat horizon. 3) and an increasing
function of GB parameter for the spherical horizon.

III) For hyperbolic black holes, the entropy has a root at $r_{+}=2\sqrt{3
\alpha}$. Whereas, for flat and spherical horizon black holes, entropy is a
smooth function of the horizon radius without any root. It is worthwhile to
mention that for flat horizon, the effects of the higher curvature terms on
entropy disappears and the area law is valid.

Next, we calculate the temperature. To do so, one should first obtain the
surface gravity through \cite{HawkingT}
\begin{equation*}
\kappa =\sqrt{-\frac{1}{2}\left( \nabla _{\mu }\chi _{\nu }\right) \left(
\nabla ^{\mu }\chi ^{\nu }\right) },
\end{equation*}
in which $\chi ^{\nu }$ is the Killing vector. Considering the metric
employed in this paper (\ref{Metric}), the Killing vector will be temporal
in the form of $\chi =\partial _{t}$. It is a matter of calculation to show
that surface gravity is obtained as
\begin{equation}
\kappa =\frac{1 }{2}\frac{df\left( r\right) }{dr}.  \label{sur}
\end{equation}

Using the surface gravity (\ref{sur}) with metric function (\ref{metric
function}), the temperature is calculated as
\begin{equation}
T=\frac{\kappa }{2\pi }=\frac{1}{4\pi } \frac{df\left(r_{+}\right) }{dr}=%
\frac{3 k r_{+}^4-8 \Lambda r_{+}^6 -q_{E}^2-q_{M}^2}{24 \pi \alpha k
r_{+}^3+6 \pi r_{+}^5}.  \label{temp}
\end{equation}

As one can see, obtained temperature gives us the following details:

I) For spherical and flat horizon black holes, the temperature is divergence
free. Whereas, for hyperbolic black holes, the temperature has a divergency
at $r_{+}=2\sqrt{ \alpha }$.

II) Temperature is a decreasing function of electric and magnetic charges.
The effects of cosmological constant and topological factor depend on the
AdS/dS nature and horizon of the solutions. The effect of GB parameter
depends on the sign of $k$ (due to coupling with topological factor).

III) Considering the structure of numerator of the temperature, it is
possible to obtain a root for the temperature. The full form of the root(s)
is given in the appendix (see Eq. \ref{roots of temp}).

IV) By series expanding the temperature for small horizon radius, one can
obtain the following relation
\begin{equation*}
\lim_{r_{+} \longrightarrow 0} T \propto -\frac{q_{E}^2+q_{M}^2}{24 \pi
\alpha k r_{+}^3}+O(\frac{1}{r_{+}}),
\end{equation*}
which confirms that: 1) Temperature diverges at $r_{+} \rightarrow 0$. 2)
The only quantity which does not have direct effect on the high energy limit
of the temperature is cosmological constant. 3) For small black holes, the
temperature is highly affected and governed by the topological factor, GB
parameter, electric and magnetic charges.

V) The asymptotic behavior of the temperature is given by
\begin{equation*}
\lim_{r_{+} \longrightarrow \infty}T \propto -\frac{\Lambda r_{+}}{6 \pi }+O(%
\frac{1}{r_{+}}),
\end{equation*}
which shows that for large black holes, the cosmological constant is the
governing factor in temperature. The other details about the temperature
will be discussed in the following subsections.

Now, we focus on the conserved quantities of the matter field. The first
items of interest are the total electric and magnetic charges. For
calculating these quantities, one can employ the Gauss law. Using this
method results into the following total electric and magnetic charges per
unit volume $V_{3}$
\begin{equation}
Q_{E}=\frac{q_{E}}{4\pi },  \label{electric charge}
\end{equation}
\begin{equation}
Q_{M}=\frac{q_{M}}{4\pi }.  \label{magnetic charge}
\end{equation}

For calculating the total mass, one can use the ADM (Arnowitt-Deser-Misner)
method \cite{Brewin}, finding the following total mass per unit volume $%
V_{3} $ for the solutions
\begin{equation}
M=\frac{3 }{16}m.
\end{equation}

In order to obtain the general form of the mass, one should obtain
geometrical mass, $m$. To do so, one can evaluate the metric function on its
outer horizon ($f(r=r_{+})=0$). It is a matter of calculation to show that
the total mass per unit volume $V_{3}$ for these black holes is
\begin{equation}
M=\frac{4 \alpha k^2+3 k r_{+}^2}{16}+\frac{q_{E}^2+q_{M}^2}{16 r_{+}^2}-%
\frac{\Lambda r_{+}^4}{32}.  \label{mass}
\end{equation}

The obtained total mass gives us the following information:

I) Electromagnetic part of the solutions and GB generalization have positive
contributions on the value of mass. As for the topological factor, it
depends on the horizon of black hole under consideration. The contribution
of cosmological constant depends on the AdS/dS nature of our solutions.

\begin{figure}[!tbp]
    \centering
    \subfloat[$q_{M}=0.1$ and $\Lambda=-1$]
    {\includegraphics[width=0.3\textwidth]{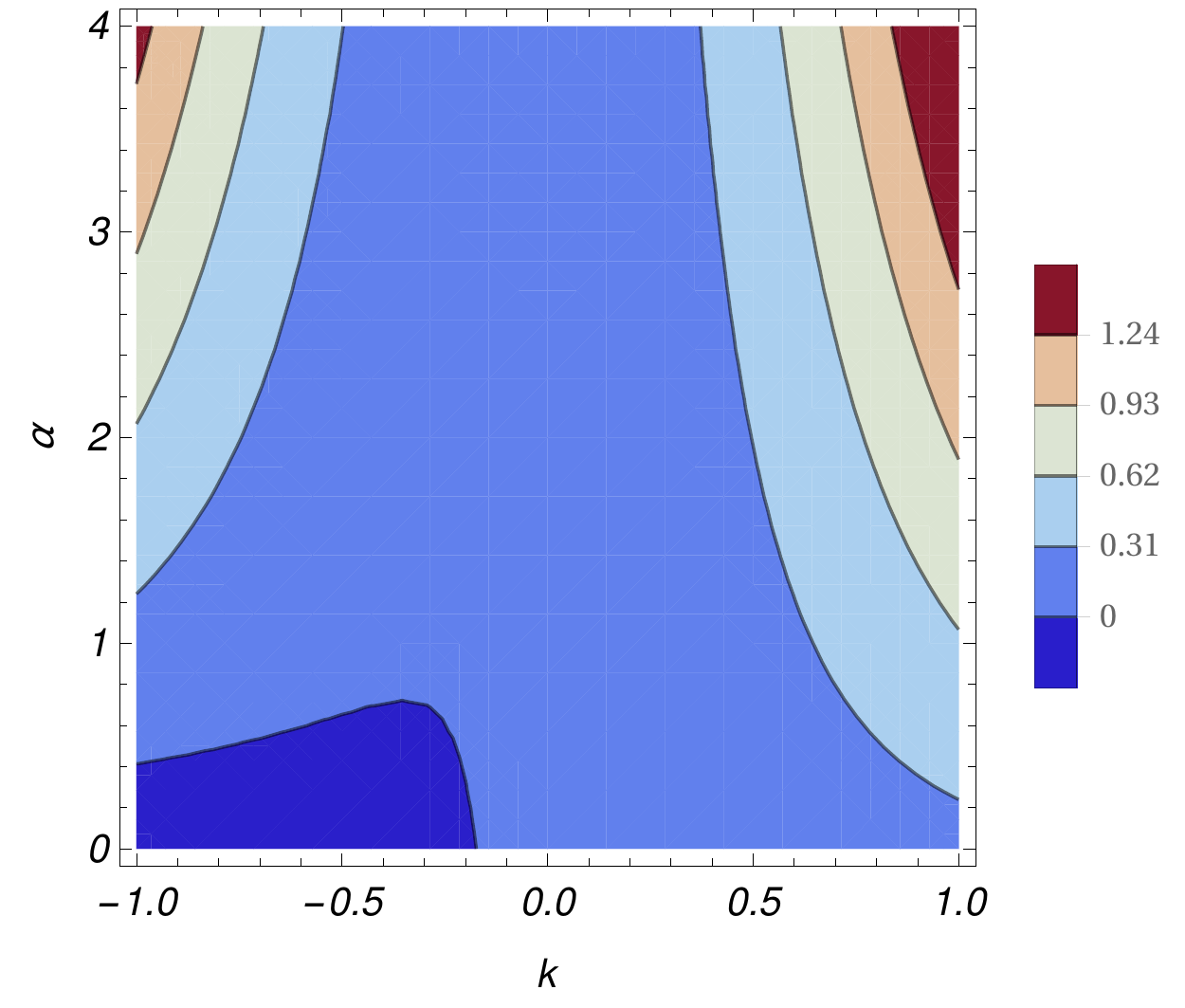}}
    \subfloat[$\alpha=0.1$ and $\Lambda=-1$]
    {\includegraphics[width=0.3\textwidth]{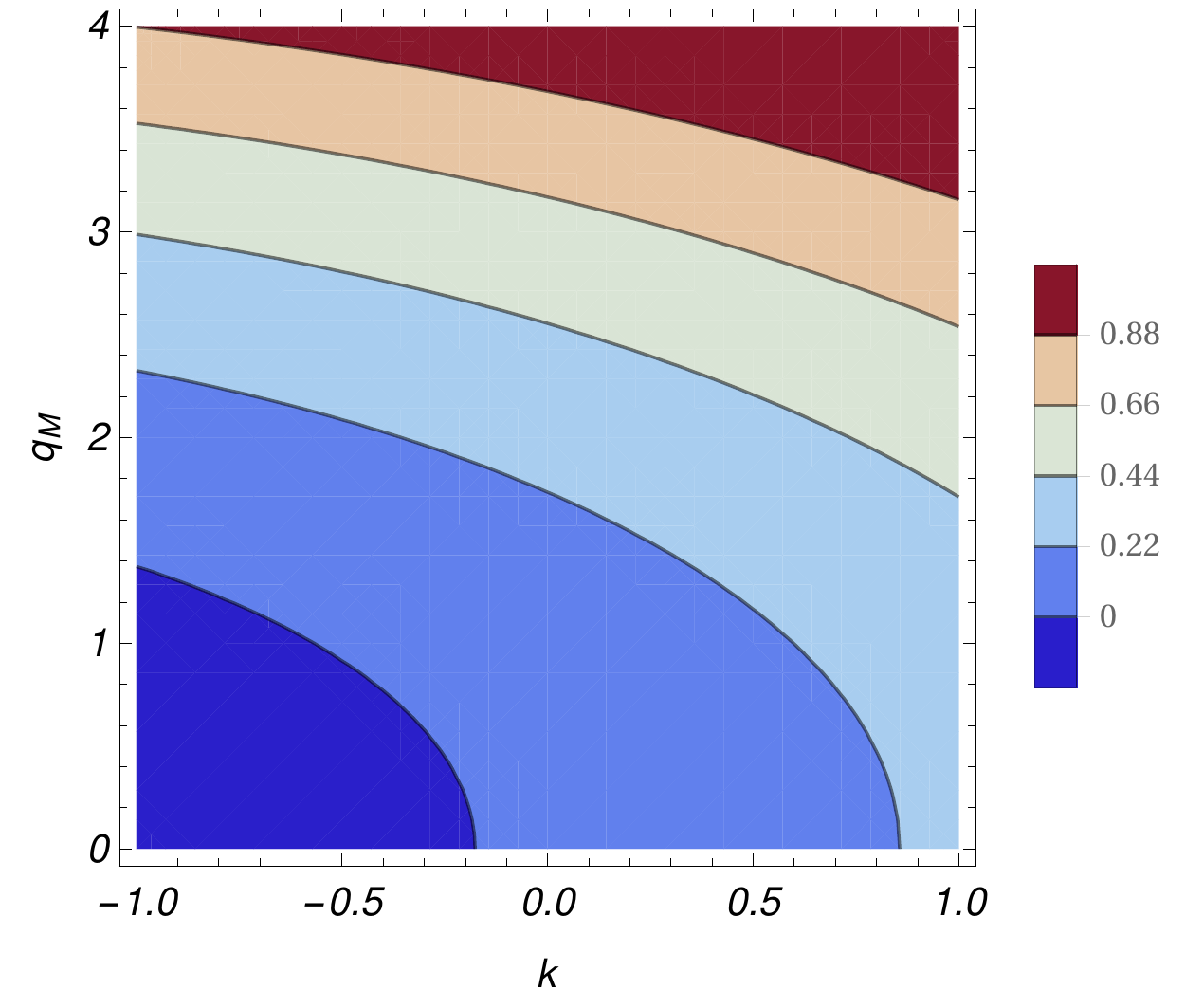}}
    \subfloat[$q_{M}=0.1$ and $\alpha=0.1$]
    {\includegraphics[width=0.311\textwidth]{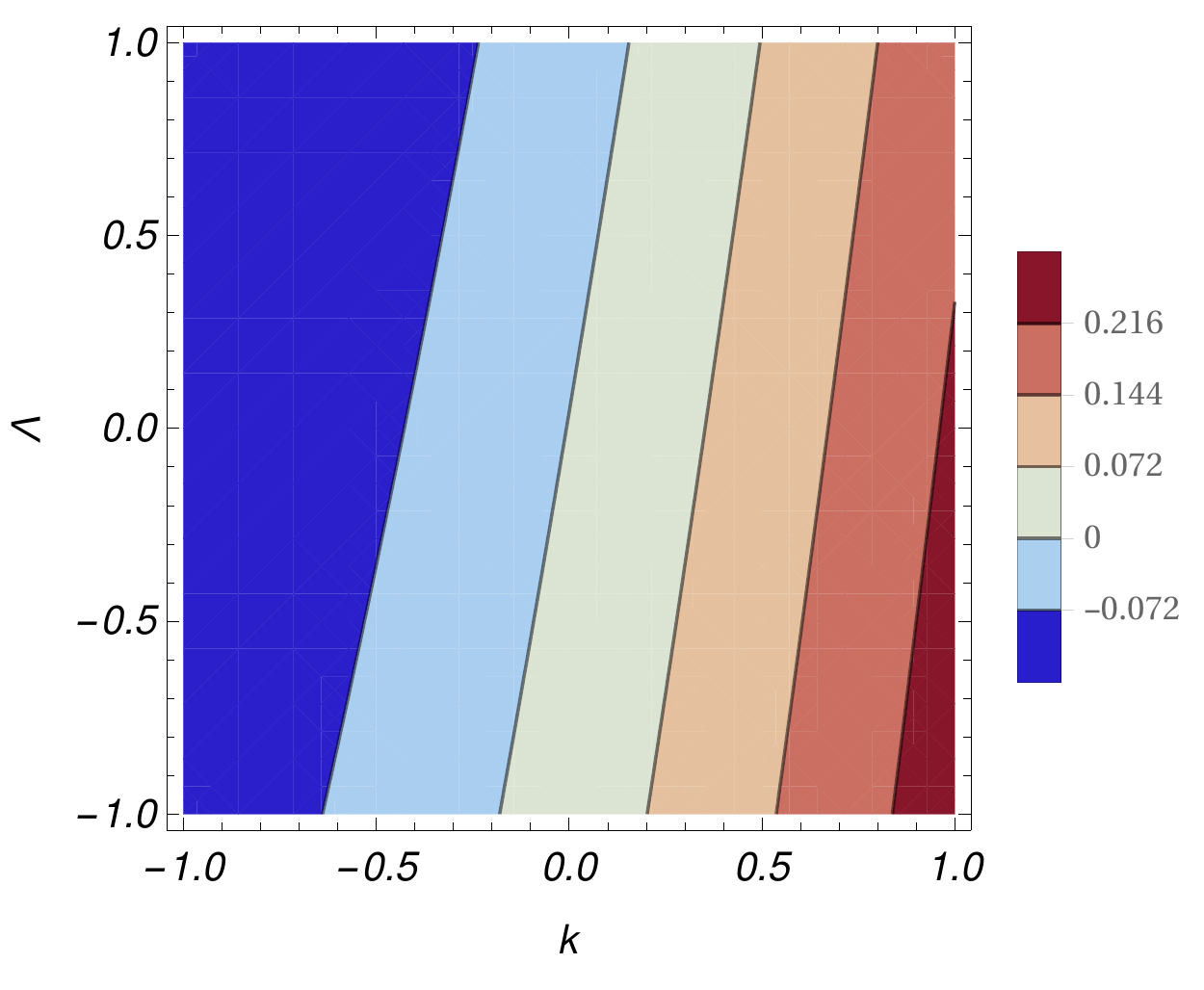}} \\
    \caption{Variation of the mass as a function of different parameters for $q_{E}=0.1$ and $r_{+}=1$.} \label{Fig00}
\end{figure}

II) Due to structure of the expression for mass, it is possible to obtain
root(s) and a region of negativity for the mass. The full form of the mass's
root is given in the appendix (see Eq. (\ref{roots of mass})). In order to
elaborate the possible cases, we have plotted the variation of mass for
different parameters in Fig. \ref{Fig00}. It is worthwhile to mention that
in the classical black hole thermodynamics, mass would be interpreted as the
internal energy and it should be positive in order for the solutions to be
physical. Plotted diagrams confirm that depending on different choices, one
can observe regions of negative mass, hence non-physical region for these
black holes.

III) Series expanding the mass for small values of the horizon radius
results into
\begin{equation*}
\lim_{r_{+} \longrightarrow 0}M \propto \frac{q_{E}^2+q_{M}^2}{16 r_{+}^2}+%
\frac{3 \alpha k^2}{8}+O\left(r_{+}^2\right),
\end{equation*}
which confirms three important results: 1) Due to presence of the electric
and magnetic charges, the total mass diverges for vanishing $r_{+}$. 2) In
the absence of electric and magnetic charges, the mass is non-zero and
finite at $r_{+}=0$. It is in fact a function of GB parameter coupled with
topological factor. This is a very important consequence of generalization
to GB gravity which is absent in the Einstein theory of the gravity. 3) In
the presence of electric and/or magnetic charges, the high energy limit of
the mass is governed by them. In their absence, the high energy limit of
mass is determined by GB gravity and horizon of the solutions.

IV) The asymptotic behavior of the mass, similar to the temperature, is
governed by cosmological constant and it is given by
\begin{equation*}
\lim_{r_{+} \longrightarrow \infty}M \propto -\frac{\Lambda r_{+}^4}{32}%
+O\left(r_{+}^2\right),
\end{equation*}

It is a well established fact that black holes admit the first law of black
hole thermodynamics. In other words, the following relation is valid for
black holes
\begin{equation}
dM=TdS+\Phi _{E}dQ_{E}+ \Phi _{M}dQ_{M},  \label{first law}
\end{equation}%
in which $\Phi _{E}$ and $\Phi _{M}$ are electric and magnetic potentials.
Using the obtained mass (\ref{mass}) with total electric (\ref{electric
charge}) and magnetic (\ref{magnetic charge}) charges, one may calculate the
electric and magnetic potentials as
\begin{eqnarray}
\Phi _{E} &=&\frac{\pi q_{E}}{ 2 r_{+}^{2}},  \label{electric potential} \\
&&  \notag \\
\Phi _{M} &=&\frac{\pi q_{M}}{ 2 r_{+}^{2}},  \label{magnetic potential}
\end{eqnarray}
which confirm the independency of all conserved quantities of the matter
from GB gravity generalization and horizon of the solutions.

\subsection{van der Waals like behavior}

In this section, we investigate the possibility of van der Waals like phase
transition for the solutions. To do so, we employ the proposal in which
cosmological constant is considered as a thermodynamical variable known as
pressure \cite{Kubiznak,Rapid}. The relation between cosmological constant
and pressure is given by
\begin{equation*}
P=-\frac{\Lambda }{8\pi }.
\end{equation*}

Replacing the cosmological constant with pressure in total mass results into
changing the role of the total mass from internal energy to enthalpy. Using
this enthalpy, it is a matter of calculation to obtain the volume of these
black holes (per $V_{3}$) as
\begin{equation}
V=\left( \frac{\partial H}{\partial P}\right) _{q_{M},q_{E}}=\frac{\pi
r_{+}^{4}}{4},  \label{volume}
\end{equation}

Evidently, there is a direct relation between total volume of the black
holes and horizon radius. This indicates that instead of using volume in
calculations, one can use the horizon radius (which is related to specific
volume, linearly \cite{Kubiznak}). For the sake of uniformity, we use the
horizon radius instead of total volume in our calculations.

By replacing the cosmological constant in the temperature and using horizon
radius instead of volume, one can obtain the equation of state as
\begin{equation}
P=T \frac{12 \alpha k+3 r_{+}^2}{4 r_{+}^3}-\frac{3 k}{8 \pi r_{+}^2}+\frac{%
q_{E}^2+q_{M}^2}{8 \pi r_{+}^6}.  \label{P}
\end{equation}

Obtained equation of state gives us the following information regarding the
pressure:

I) The pressure is an increasing function of the electric charge and
magnetic charges. The contribution of temperature term (which includes GB
gravity as well) depends on the horizon under consideration; 1) Flat
horizon: the GB gravity's effects on the pressure do not appear and pressure
is an increasing function of temperature. 2) Spherical horizon: both
temperature and GB gravity contribute positively to the pressure. 3)
Hyperbolic horizon: the GB gravity has negative affect on the pressure,
while for the temperature, if $r_{+}=2 \sqrt{\alpha}$, the effects of
temperature term is vanished. For $r_{+}<2 \sqrt{\alpha}$, the temperature
term has negative effect on the pressure. The opposite is true for the case $%
r_{+}>2 \sqrt{\alpha}$.

\begin{figure*}[!htbp]
    \centering
    \includegraphics[width=0.3\linewidth]{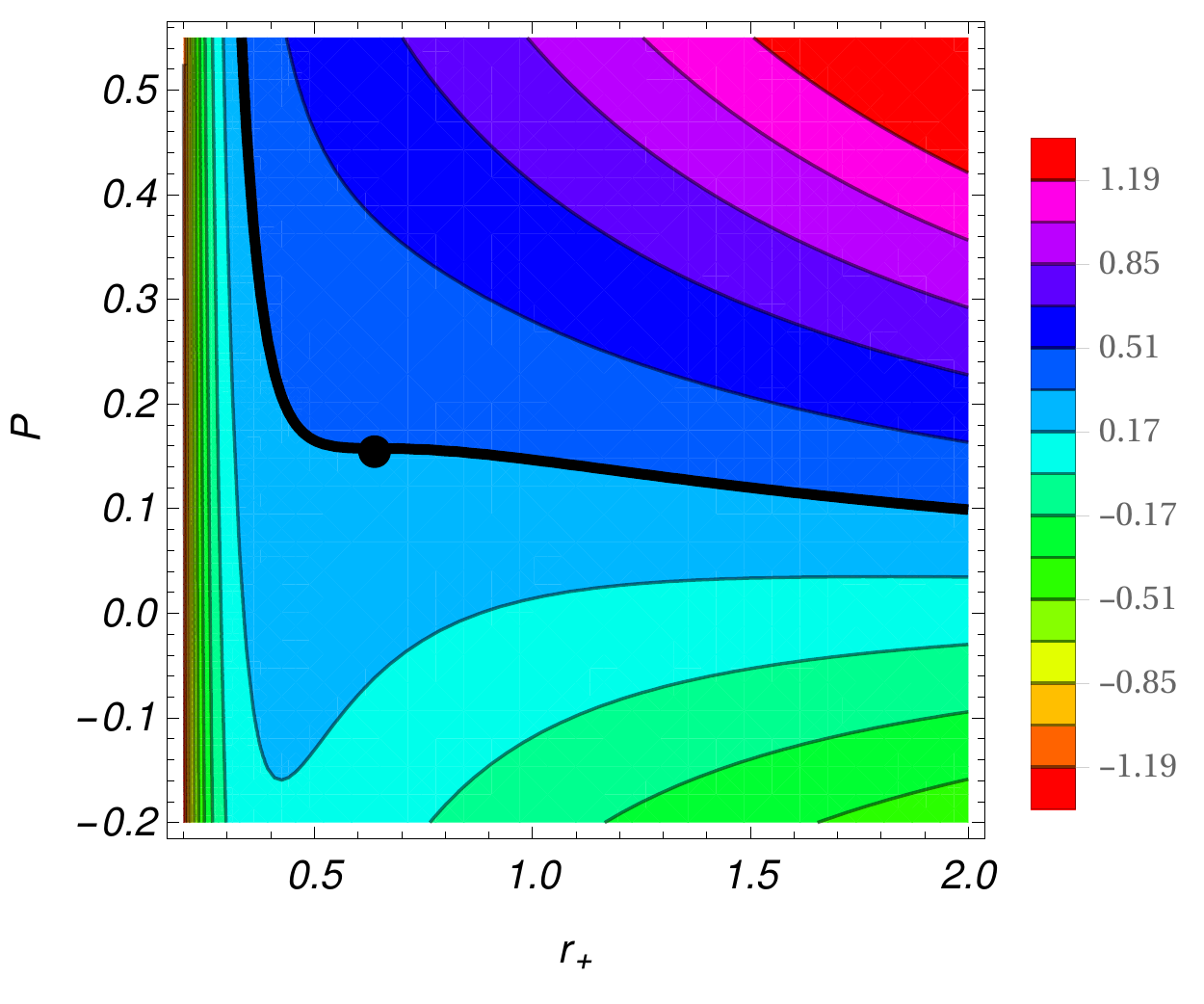}\hfil
    \includegraphics[width=0.3\linewidth]{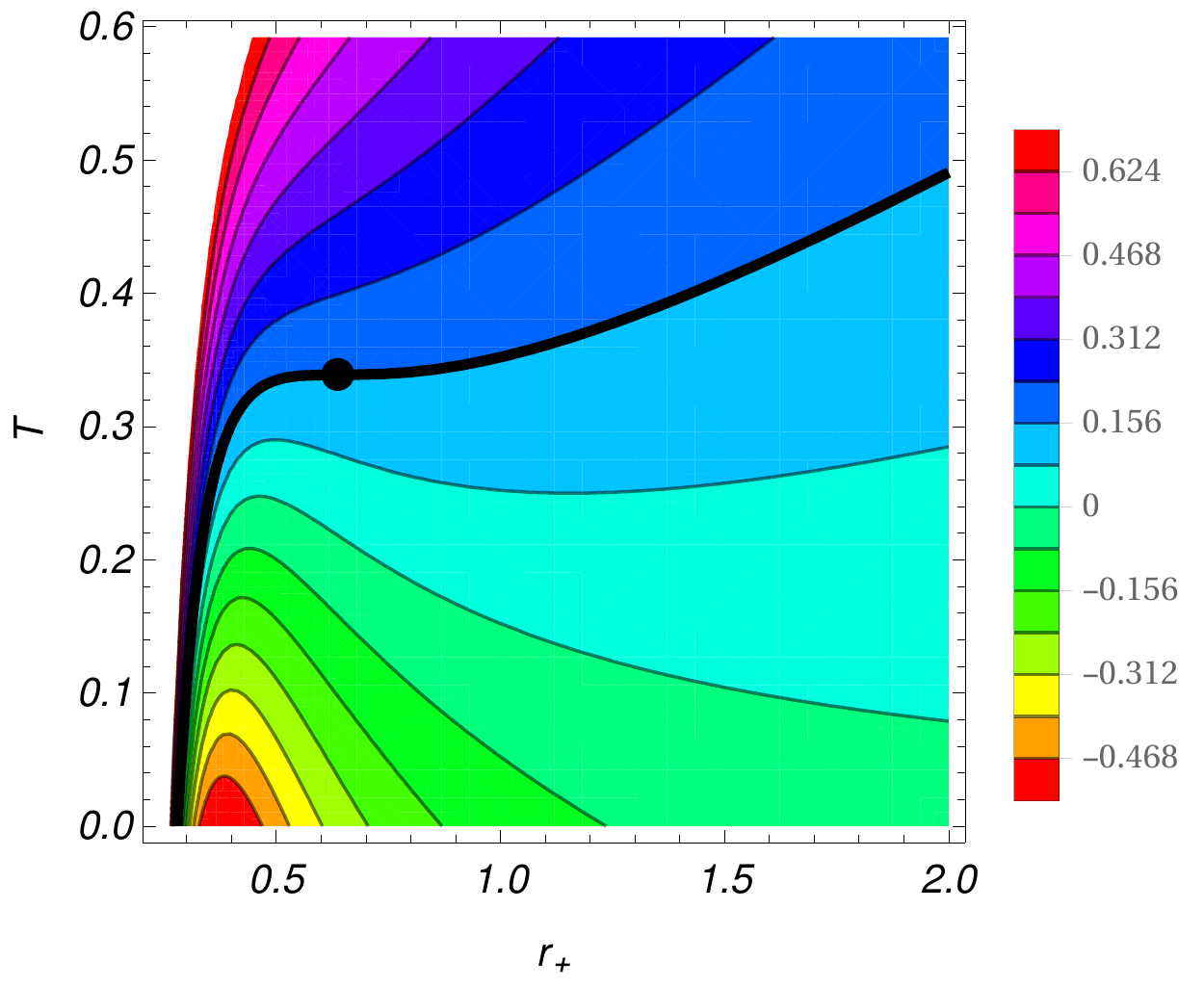}\hfil
    \includegraphics[width=0.35\linewidth]{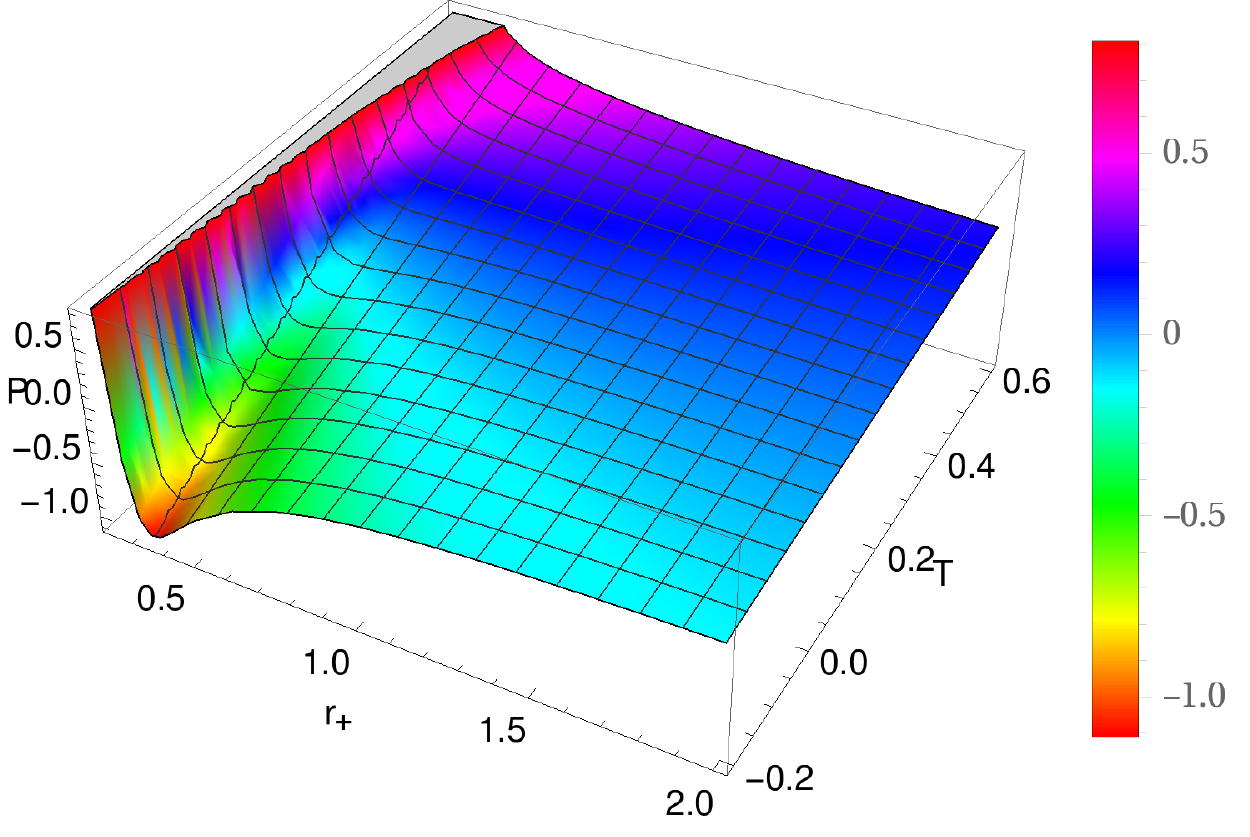}
    \caption{Van der Waals like phase diagrams for $q_{M}=q_{E}=0.1$, $\alpha=0.01$ and $k=1$.
    The dot corresponds to phase transition point between small and large black holes. \textbf{Left Panel:}
    Bold line corresponds to $T=T_{c}$. \textbf{Middle Panel:} Bold line corresponds to $P=P_{c}$. }
    \label{Fig1}
\end{figure*}

II) Regarding the expression of the pressure, one can obtain its real valued
root(s). Later, in phase transition diagrams, we confirm the presence of
roots for the pressure (see Fig. \ref{Fig1}).

III) Series expanding the pressure for small values of the horizon radius
results into
\begin{equation*}
\lim_{r_{+} \longrightarrow 0}P \propto \frac{q_{E}^2+q_{M}^2}{8 \pi r_{+}^6}%
+ \frac{3k \alpha T}{r_{+}^3}+O(\frac{1}{r_{+}^2}),
\end{equation*}
which confirms the following points: 1) The high energy limit of pressure
(also the behavior of pressure for large black holes) is governed only by
the electric and magnetic charges. 2) In the absence of the electric and
magnetic fields, the high energy limit of the pressure is governed by the
temperature, topological factor and GB parameter with the same order of
magnitude. 3) Evidently, for vanishing $r_{+}$, pressure diverges.

IV) The asymptotic behavior of the pressure (also the behavior of pressure
for small black holes) is only governed by the temperature in the following
form
\begin{equation*}
\lim_{r_{+} \longrightarrow \infty}P \propto \frac{3 T}{4 r_{+}}+O(\frac{1}{%
r_{+}^2}).
\end{equation*}

In order to investigate the existence of van der Waals like phase
transition, one can use the properties of inflection point of isotherm $P-V$
diagrams, as
\begin{equation}
\left( \frac{\partial P}{\partial r_{+}}\right)_{T,q_{M},q_{E}} =\left(
\frac{\partial ^{2}P}{\partial r_{+}^{2}}\right)_{T,q_{M},q_{E}} =0.
\label{infel}
\end{equation}
which leads to the following equation for obtaining the critical horizon
radius (specific volume)
\begin{equation}
r_{+}^2 (k r_{+}^4-5 q_{E}^2 -5 q_{M}^2)-12 \alpha k (r_{+}^4+3 q_{E}^2+3
q_{M}^2)=0.  \label{critical expression}
\end{equation}

Evidently, only for the flat horizon, the effects of the GB gravity
disappear in critical horizon radius. It is easy to obtain the critical
horizon radius. The full form is given in the appendix (see Eq. \ref%
{critical horizon}). The obtained form contains square root functions
indicating that certain condition must be satisfied in order to have real
valued positive critical horizon radius, hence critical behavior. Using
obtained critical horizon radius (Eq. \ref{critical horizon}), one can
obtain the critical temperature and pressure as well. (the full forms are
given in the appendix (see Eqs. \ref{critical temperature} and \ref{critical
pressure})). In order to elaborate the possibility of van der Waals like
phase transition, we have plotted Fig. \ref{Fig1}.

\begin{figure*}[!htbp]
    \centering
    \includegraphics[width=0.3\linewidth]{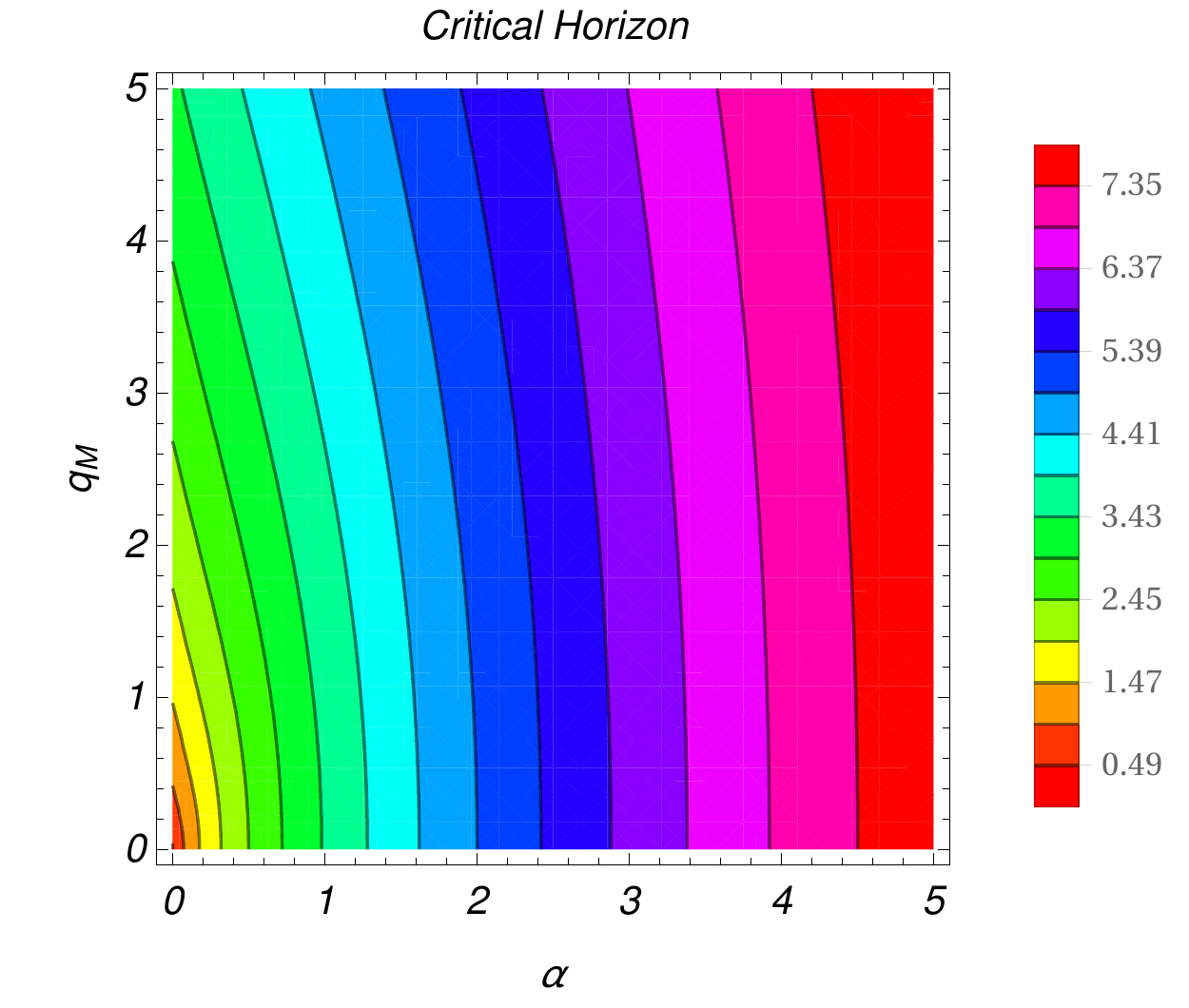}\hfil
    \includegraphics[width=0.3\linewidth]{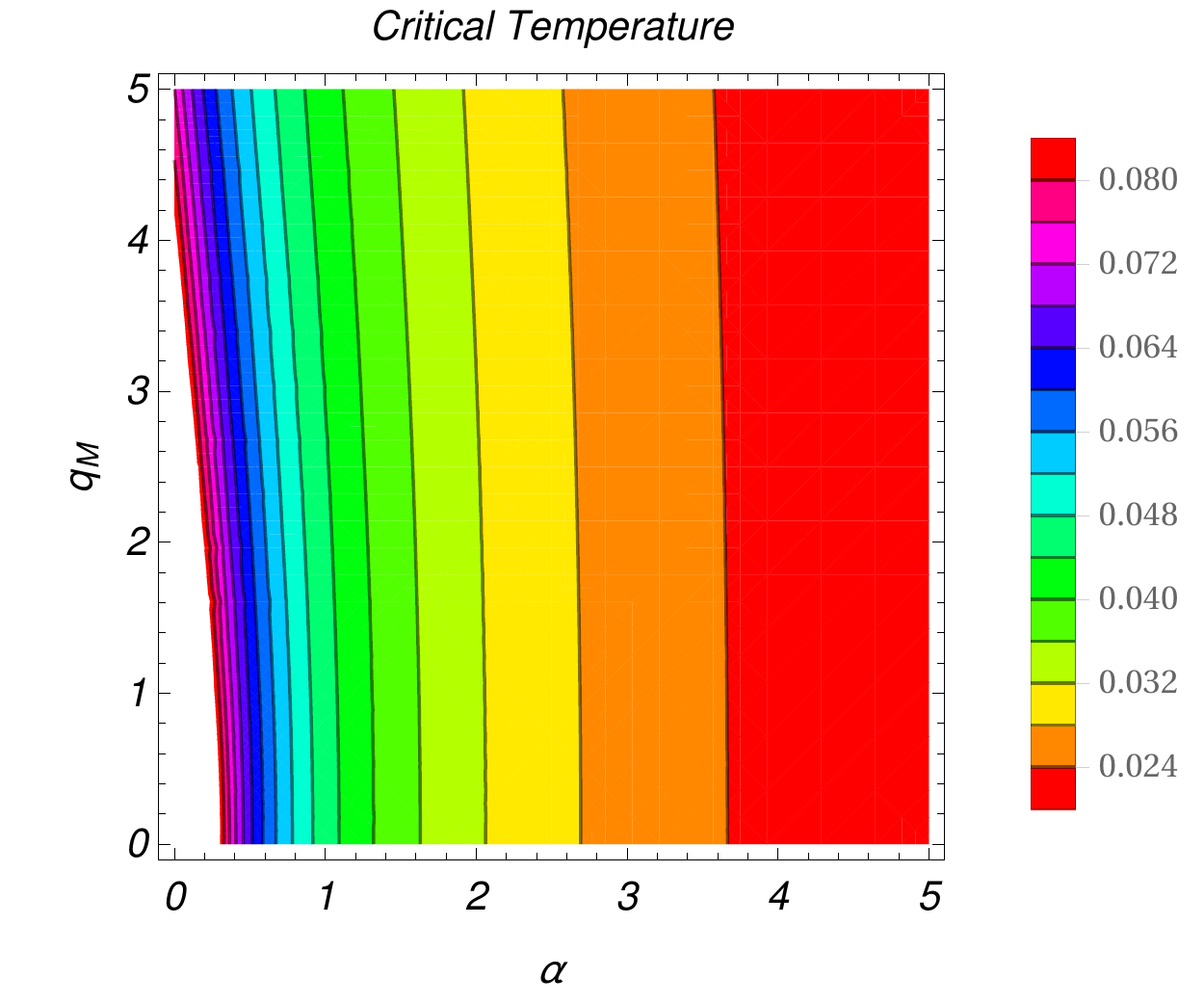}\hfil
    \includegraphics[width=0.31\linewidth]{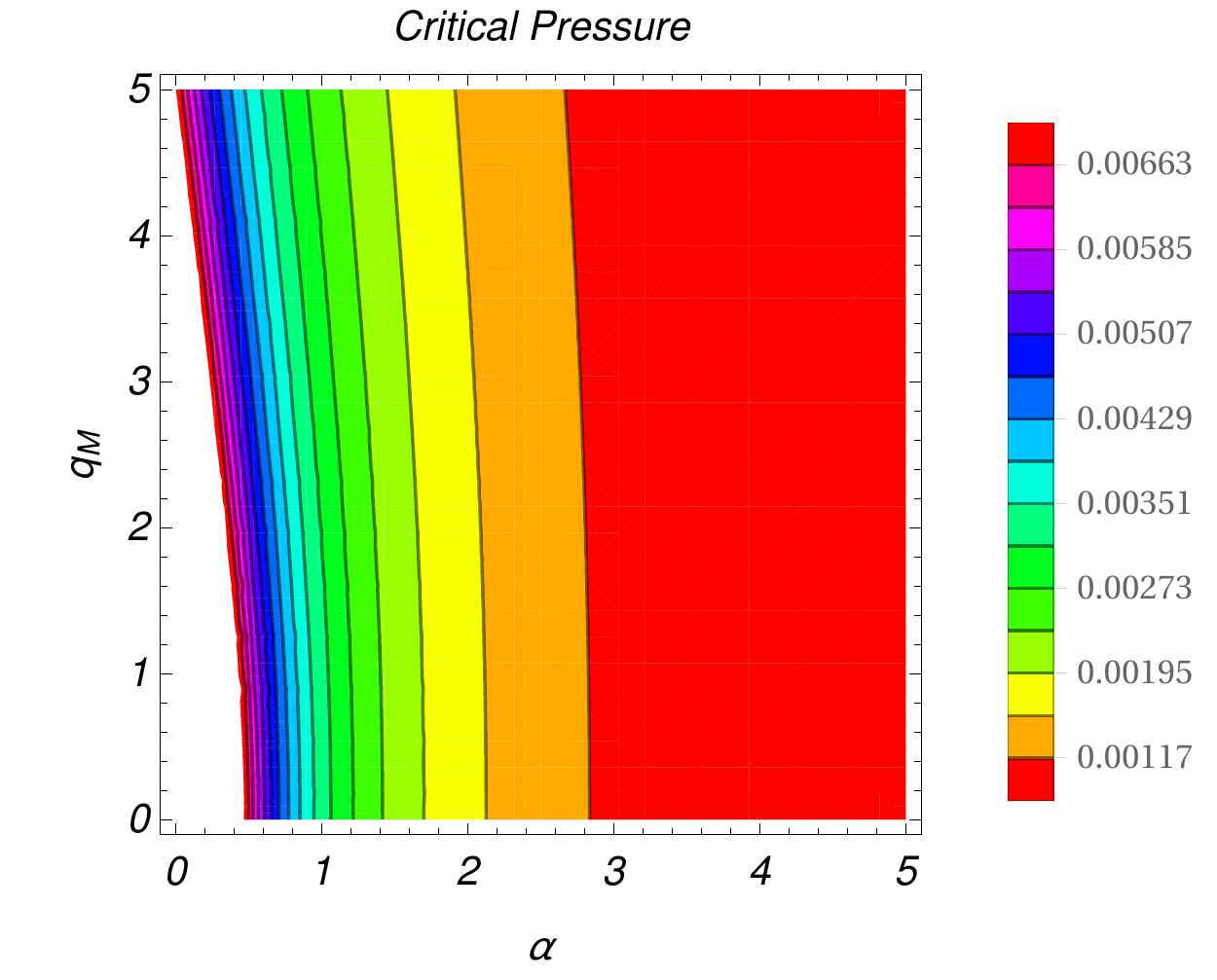}
    \caption{Variation of critical horizon radius, temperature and pressure as
     functions of Gauss-Bonnet parameter and magnetic charge for: $q_{E}=0.1$ and $k=1$.}
    \label{Fig2}
\end{figure*}

Evidently, by suitable choices of different parameters, one can obtain van
der Waals like phase transition for these black holes. The presence of
subcritical isobars in $T-r_{+}$ and isothermal diagrams in $P-r_{+}$
confirm the existence of van der Waals like phase transition. In order to
study the effects of magnetic charge and GB parameter on critical values, we
have plotted another set of diagrams (see Fig. \ref{Fig2}).

As one can see, the critical horizon radius is an increasing function of the
magnetic charge and GB parameter. Whereas, the critical temperature and
pressure are decreasing functions of the magnetic charge and GB parameter.
The effects of variation of the GB parameter on these critical values are
more significant compared to magnetic charge. The blank space in critical
temperature and pressure diagrams correspond to having non-real value for
the critical temperature and pressure (for more details see Eqs. (\ref%
{critical temperature}) and (\ref{critical pressure}) in the appendix). In
other words, these blank spaces correspond to the absence of van der Waals
like phase transition for these black holes.

\subsection{Stability of the solutions}

Now, we focus on thermal stability of the solutions. In the canonical
ensemble, we consider the heat capacity of the solutions. The heat capacity
could be used to extract two important properties of the solutions: I) Phase
transition points which could be detected as discontinuity in the heat
capacity. II) Thermal stability of the solutions. If the heat capacity is
positive valued, solutions are thermally stable, while the opposite stands
for the unstable black holes.

In general, the heat capacity of these black holes could be obtained as
\begin{equation}
C=T\left( \frac{\partial S}{\partial T}\right) _{q_{M},q_{E},P} \\
=\frac{3 r_{+} \left(4 \alpha k+r_{+}^2\right)^2 \left(3 k r_{+}^4+8 \pi P
r_{+}^6-q_{E}^2-q_{M}^2\right)}{48 \alpha k \left(k r_{+}^4+8 \pi P
r_{+}^6+q_{E}^2+q_{M}^2\right)+4 r_{+}^2 \left(-3 k r_{+}^4+8 \pi P
r_{+}^6+5 \left(q_{E}^2+q_{M}^2\right)\right)}.  \label{heat}
\end{equation}

The obtained heat capacity provides the following information:

I) The GB parameter is coupled with the cosmological factor. This indicates
that for the horizon flat black holes, heat capacity is free of the GB
gravity, so it is the same as that in Einstein gravity.

\begin{figure*}[!htbp]
    \centering
    \includegraphics[width=0.4\linewidth]{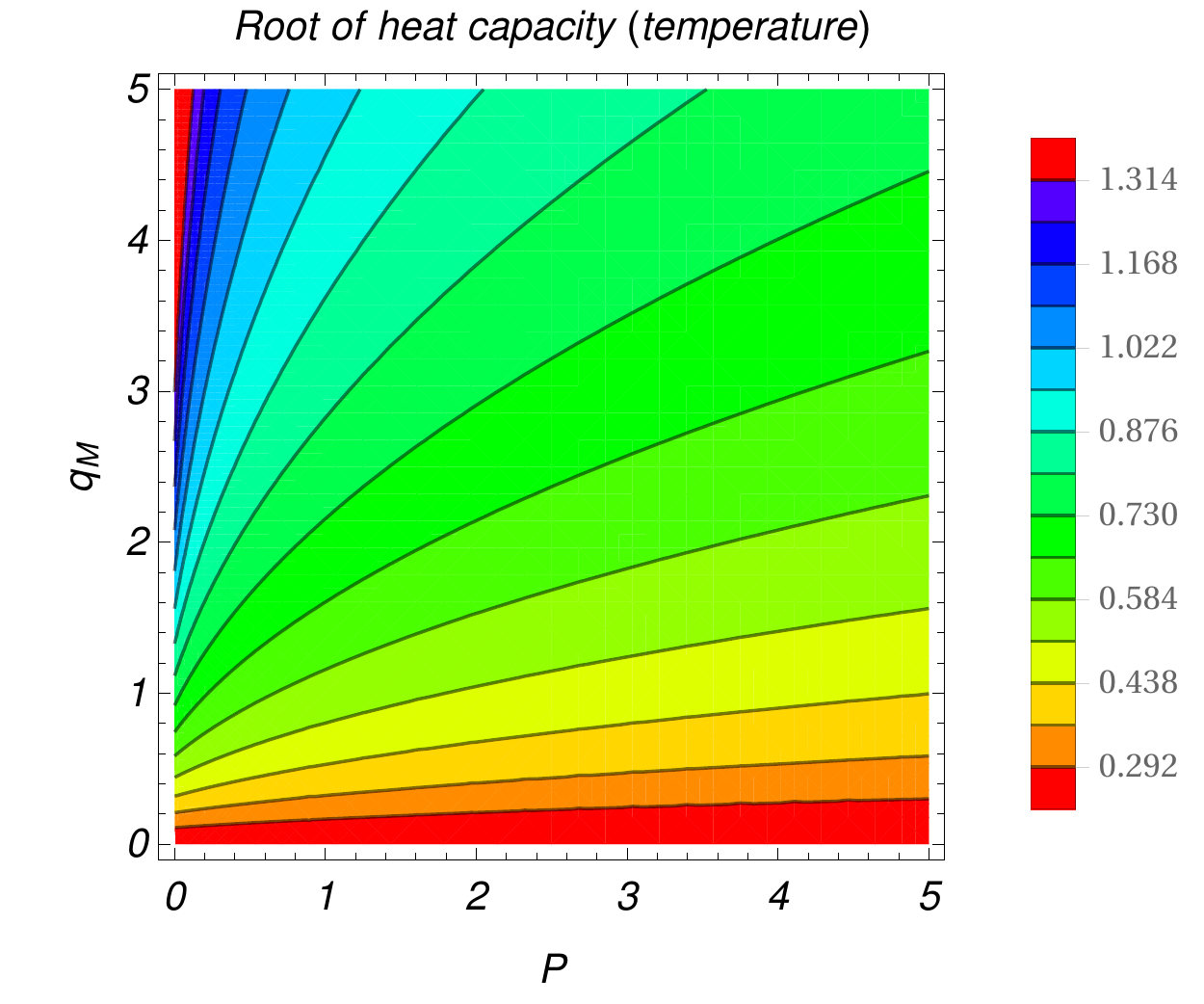}
    \caption{Variation of heat capacity's root as a function of pressure and magnetic charge for: $q_{E}=0.1$ and $k=1$.}
    \label{Fig3}
\end{figure*}

II) Regarding the roots of heat capacity, we have two distinctive terms from
which we can extract the roots. 1) One of these expressions is $4 \alpha
k+r_{+}^2$. Evidently, for this expression, the root of heat capacity could
be extracted in form of $r_{+}=2\sqrt{- \alpha k}$. This shows that only for
hyperbolic black holes, GB parameter could contribute to the root of heat
capacity. Otherwise, the root of heat capacity is independent of $\alpha$.
2) The other expression is $3 k r_{+}^4+8 \pi P r_{+}^6-q_{E}^2-q_{M}^2$. It
is possible to extract the root of heat capacity analytically from this
expression. The full form is given in appendix (see Eq. \ref{roots of heat
capacity}). The obtained root is independent of the GB parameter. In order
to understand the effects of magnetic charge and pressure on the root, we
have plotted Fig. \ref{Fig3}. Evidently, the root of heat capacity is an
increasing (a decreasing) function of the magnetic charge (pressure). It is
worthwhile to mention that both the heat capacity and temperature share the
same root.

III) In general, divergencies of the heat capacity are GB parameter
dependent. But, it is possible to omit such effect. The GB parameter is
present in denominator of the heat capacity as a factor of the expression $k
r_{+}^4+8 \pi P r_{+}^6+q_{E}^2+q_{M}^2$. By suitable tuning, this
expression could vanish. Such a case could only be achieved for hyperbolic
black holes. This is due to the fact that only the topological term with
hyperbolic horizon could be negative, otherwise the mentioned expression is
positive valued and non-zero.

IV) The high energy limit of heat capacity is governed by topological factor
and GB gravity in the following form
\begin{equation*}
\lim_{r_{+} \longrightarrow 0}C \propto -\alpha k r_{+}-\frac{r_{+}^3}{12}%
+O\left(r_{+}^5\right).
\end{equation*}

As one can see, the next leading term in the high energy limit is only a
function of the horizon radius. This comes from the contribution of the
Einstein gravity. This shows that the effects of GB gravity are more
dominant in the high energy limit of the heat capacity comparing to the
Einstein gravity. In the limit of $r_{+}=0$, the heat capacity completely
vanishes which is in contrast to the same cases of enthalpy, temperature and
pressure where they diverge.

V) The asymptotic behavior is given by
\begin{equation*}
\lim_{r_{+} \longrightarrow \infty}C \propto \frac{3 r_{+}^3}{4}+\left(\frac{%
9 k}{16 \pi P}-3 \alpha k\right)r_{+}+O\left(\frac{1}{r_{+}}\right).
\end{equation*}

Evidently, the dominant term in asymptotic behavior includes only horizon
radius. This comes from the contribution of Einstein gravity. Therefore, the
dominant factor in asymptotic behavior of the heat capacity is Einstein
gravity. The second dominant term includes pressure, GB parameter and
topological factor. As we see, the magnetic and electric charge effects are
not significant in both high energy limit and asymptotic behavior of the
heat capacity.

\begin{figure}[!tbp]
    \centering
    \subfloat[$q_{M}=0.1$ and $P=0.01$]{\includegraphics[width=0.3\textwidth]{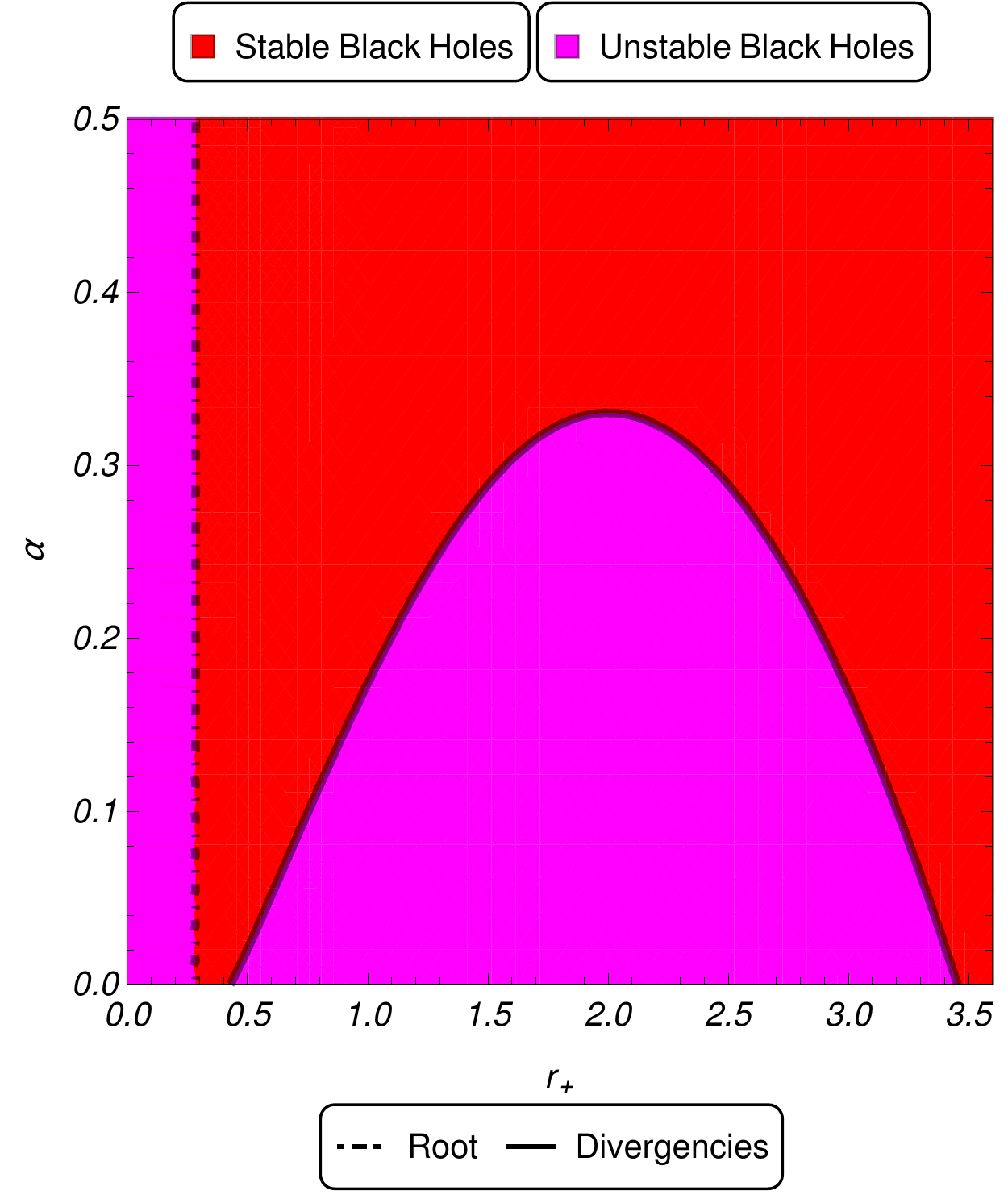}}\hfill
    \subfloat[$\alpha=0.1$ and $P=0.01$]{\includegraphics[width=0.3\textwidth]{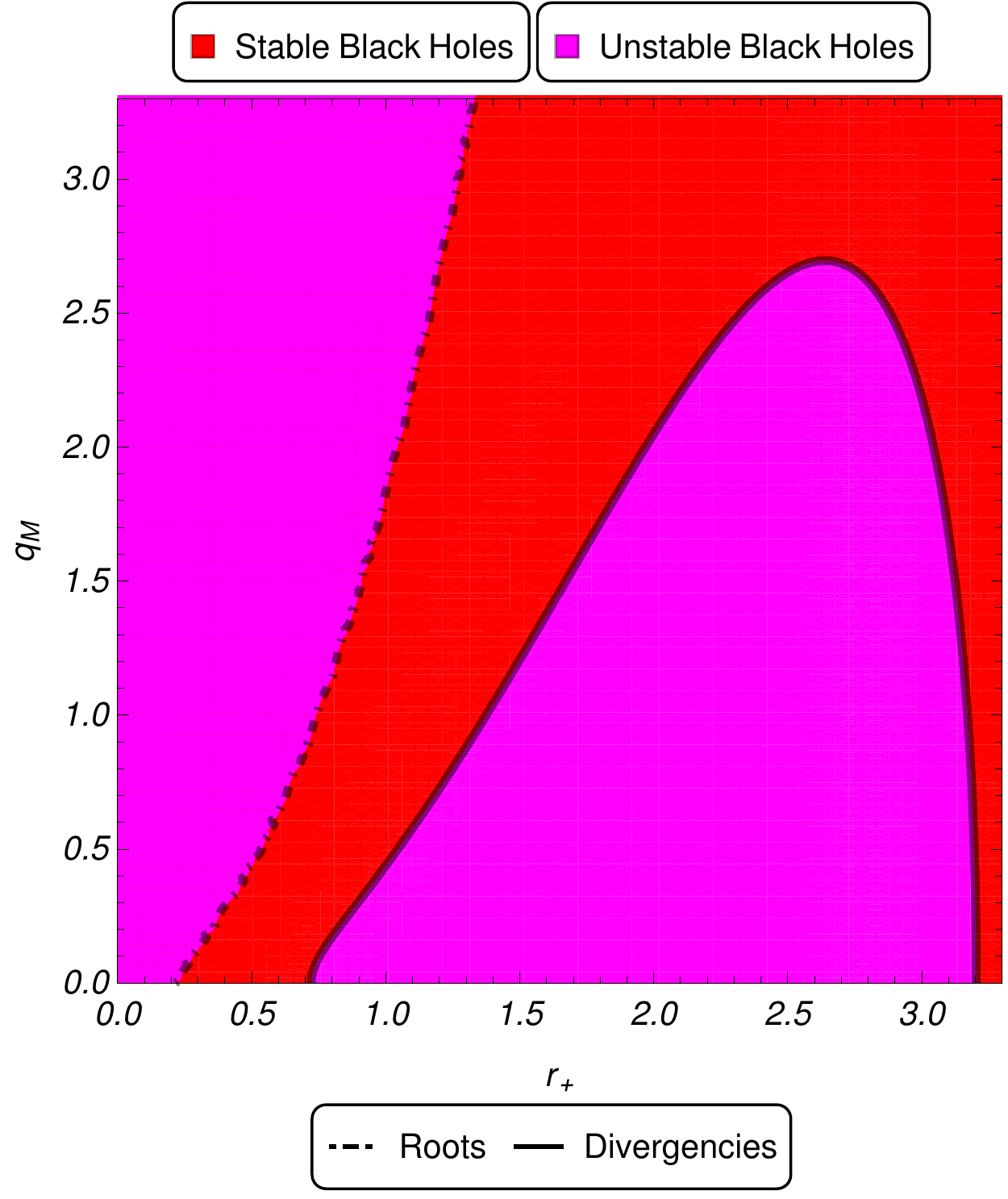}}\hfill
    \subfloat[$q_{M}=0.1$ and $\alpha=0.01$]{\includegraphics[width=0.3\textwidth]{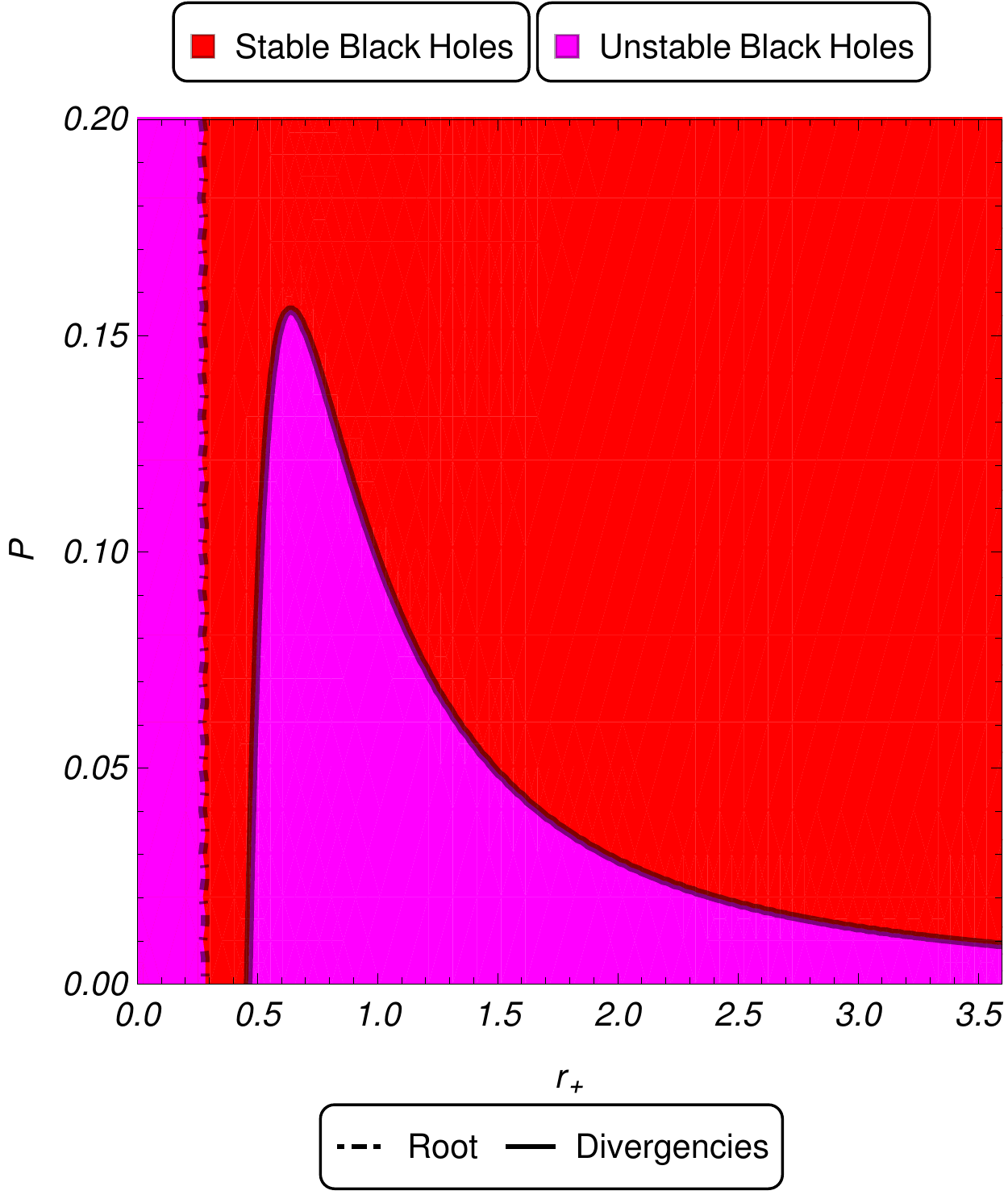}} \\
    \caption{Thermally stable and/or unstable regions of the black holes for $k=1$ and $q_{E}=0.1$.} \label{Fig4}
\end{figure}

Thermal stability in canonical ensemble is determined by the sign of heat
capacity. Positivity (negativity) shows that black holes are thermally
stable (unstable). Considering this issue, we have plotted Fig. \ref{Fig4}
to study the effects of different parameters and possible cases of
stability/instability for these black holes.

In general, there are three possibilities for the heat capacity:

I) Existence of two divergencies and one root: in this case, the stable
regions are between root and smaller divergency, and after larger divergency.

II) One root and one divergency: in which the regions of the stability are
between root and divergency, and after the divergency.

III) Only one root: where the stable black holes are after the root.

Plotted diagrams shows that there is a critical value for the GB parameter
where for values smaller than it, two divergencies exist for the heat
capacity. For the GB parameters larger than this critical value, the heat
capacity enjoys the absence of divergency in its structure. The same applies
to magnetic charge and pressure. Since we have plotted diagrams for
spherical black holes, the root of heat capacity is independent of the GB
parameter. Whereas, for the magnetic charge, the root is a sensitive
function of it.

\section{Trajectory of the particle}

Investigation of particle geodesics is one of the interesting subjects in
the context of black hole physics. This is due to the fact that analyzing
the accretion disk and out flows of black holes have a direct relation to
the trajectory of particles around black holes. Our final study in this
paper is devoted to the trajectory of a particle around GB-dyonic black
hole. Our main motivation is to investigate the effects of different
parameters on such a trajectory. In addition, we would like to address the
possible scenarios for the trajectory of the particle to some extent. In
order to obtain the trajectory of the particle, we use Euler-Lagrange
equation in the following form \cite{Wang,Berti}
\begin{equation}
\frac{d}{d\tau }\frac{\partial }{\partial \dot{x}^{\alpha }}\mathcal{L}-%
\frac{\partial }{\partial x^{\alpha }}\mathcal{L}=0,  \label{EL}
\end{equation}%
in which $\mathcal{L}$ is the particle Lagrangian, $x^{\alpha }$ is
coordinate, $\dot{x}^{\alpha }=dx^{\alpha }/d\tau $ and $\tau $ is the
affine parameter. The Lagrangian of the particle can be obtained as
\begin{equation}
2\mathcal{L}=g_{\mu \nu }\frac{dx^{\mu }}{d\tau }\frac{dx^{\nu }}{d\tau }%
=-f(r)(\frac{dt}{d\tau })^{2}+\frac{1}{f(r)}(\frac{dr}{d\tau })^{2}+r^{2}[(%
\frac{d\theta }{d\tau })^{2}+\sin ^{2}\theta (\frac{d\psi }{d\tau }%
)^{2}+\sin ^{2}\theta \sin ^{2}\psi (\frac{d\phi }{d\tau })^{2}].
\label{Lag}
\end{equation}

We choose $\theta=\psi=\pi /2$ to confine the motion of the particle to
equatorial plane $\theta=\psi=\pi /2$ at all times. Considering the
structure of the metric (\ref{Metric}), one can confirm that the Lagrangian
is independent of $t$ and $\phi$ coordinates. This indicates that $t$ and $%
\phi$ components yield conserved quantities as
\begin{eqnarray}
E &=& g_{tt}\frac{dt}{d\tau }=f(r) \frac{dt}{d\tau }\,,  \label{Energy} \\
L &=& g_{\phi \phi }\frac{d\phi } {d\tau }=r^{2}\frac{d\phi }{ d\tau }\,,
\label{Momenta}
\end{eqnarray}
in which $E$ is the energy (per unit mass for massive particles) of the
particle and $L$ is the orbital angular momentum. Using the obtained
relations and replacing them in Eq. (\ref{Lag}) result into
\begin{equation}
\mathcal{L} =-\frac{E^2}{2f(r)}+\frac{1}{2f(r) }( \frac{dr}{d\tau } ) ^{2}+%
\frac{L^2}{2 r^2}.  \label{Lag2}
\end{equation}

\begin{figure}[!tbp]
    \centering
    \subfloat[$q_{M}=1$, $E=3$ and $L=5$.]{
        \includegraphics[width=0.27\textwidth]{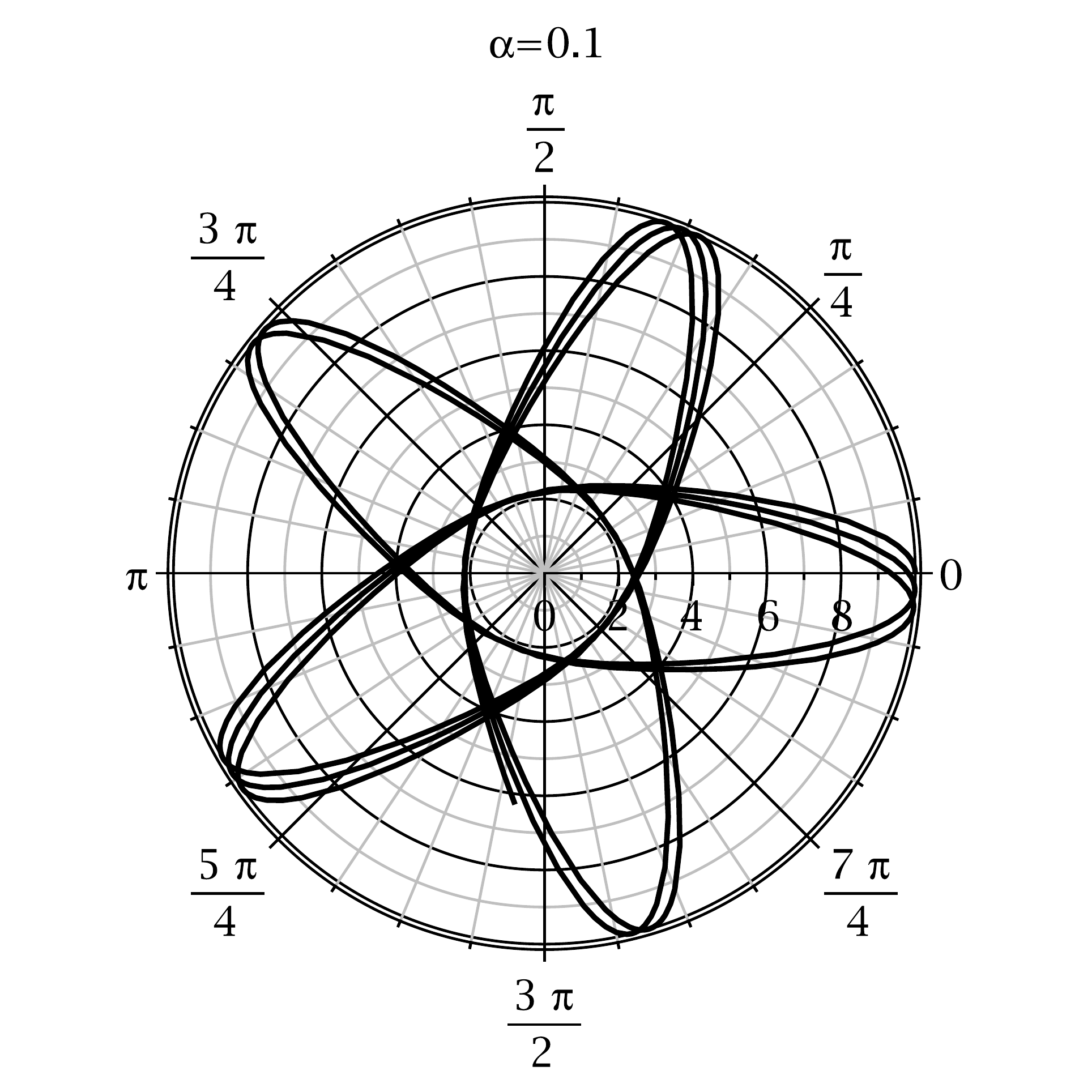}
        \includegraphics[width=0.27\textwidth]{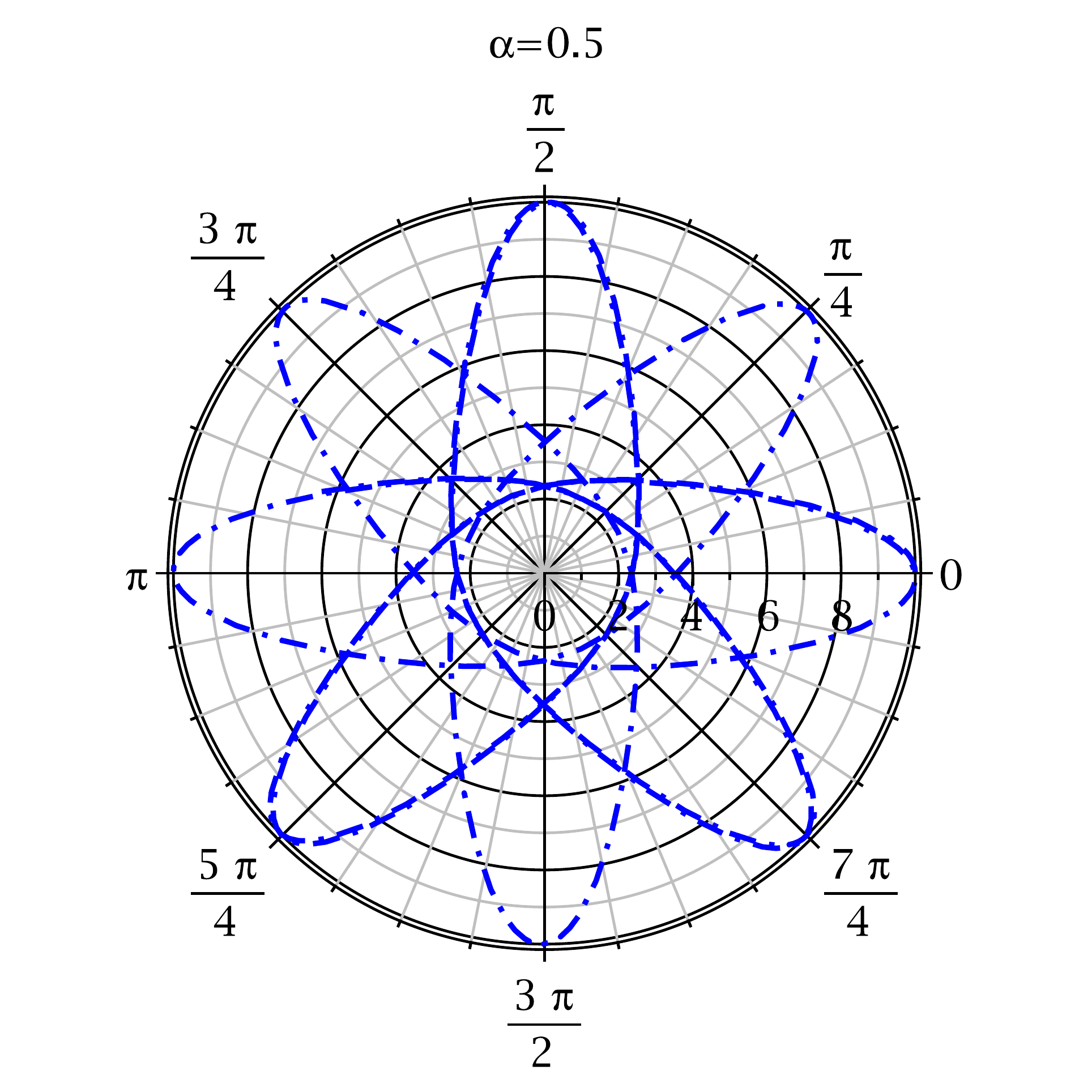}
        \includegraphics[width=0.27\textwidth]{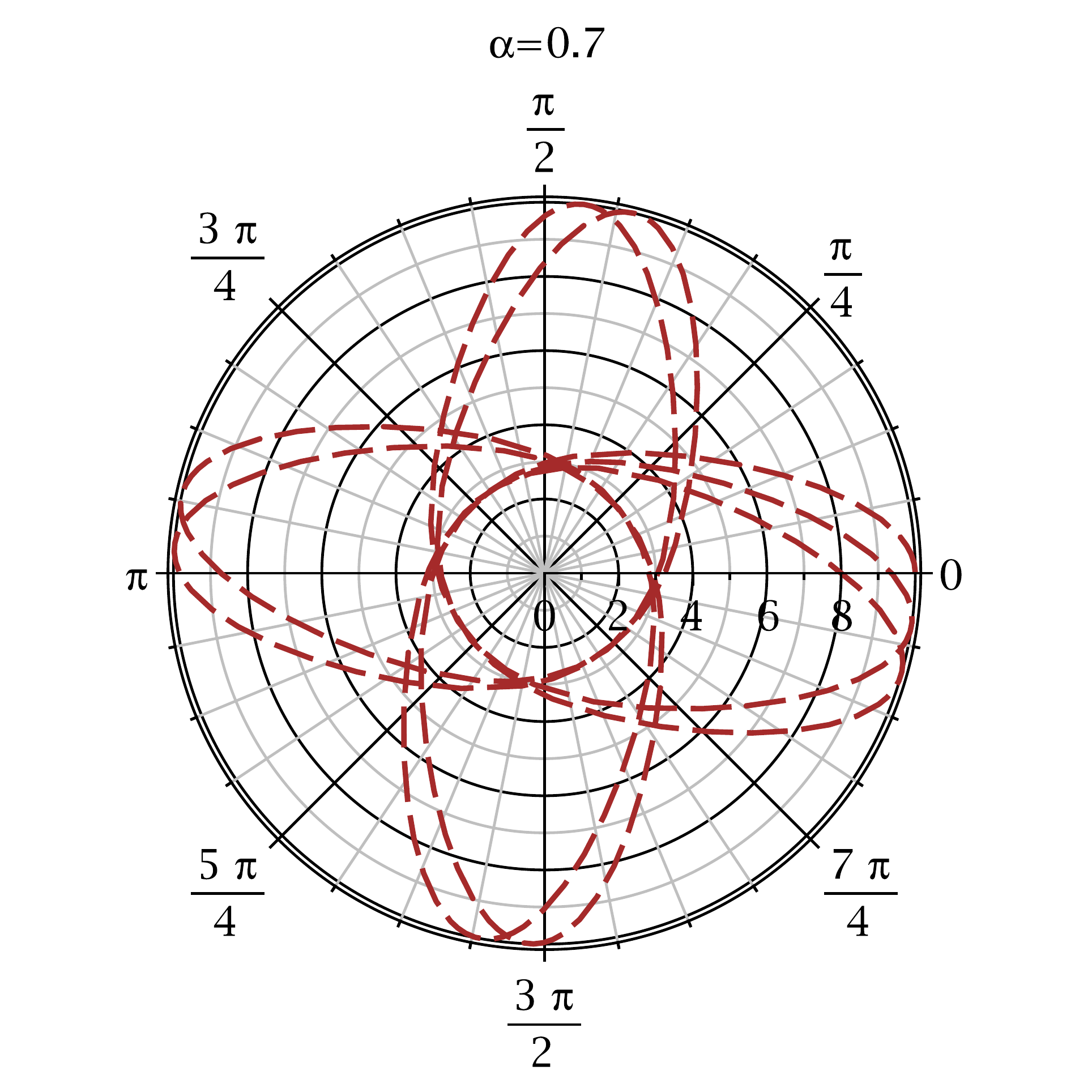}
        \includegraphics[width=0.27\textwidth]{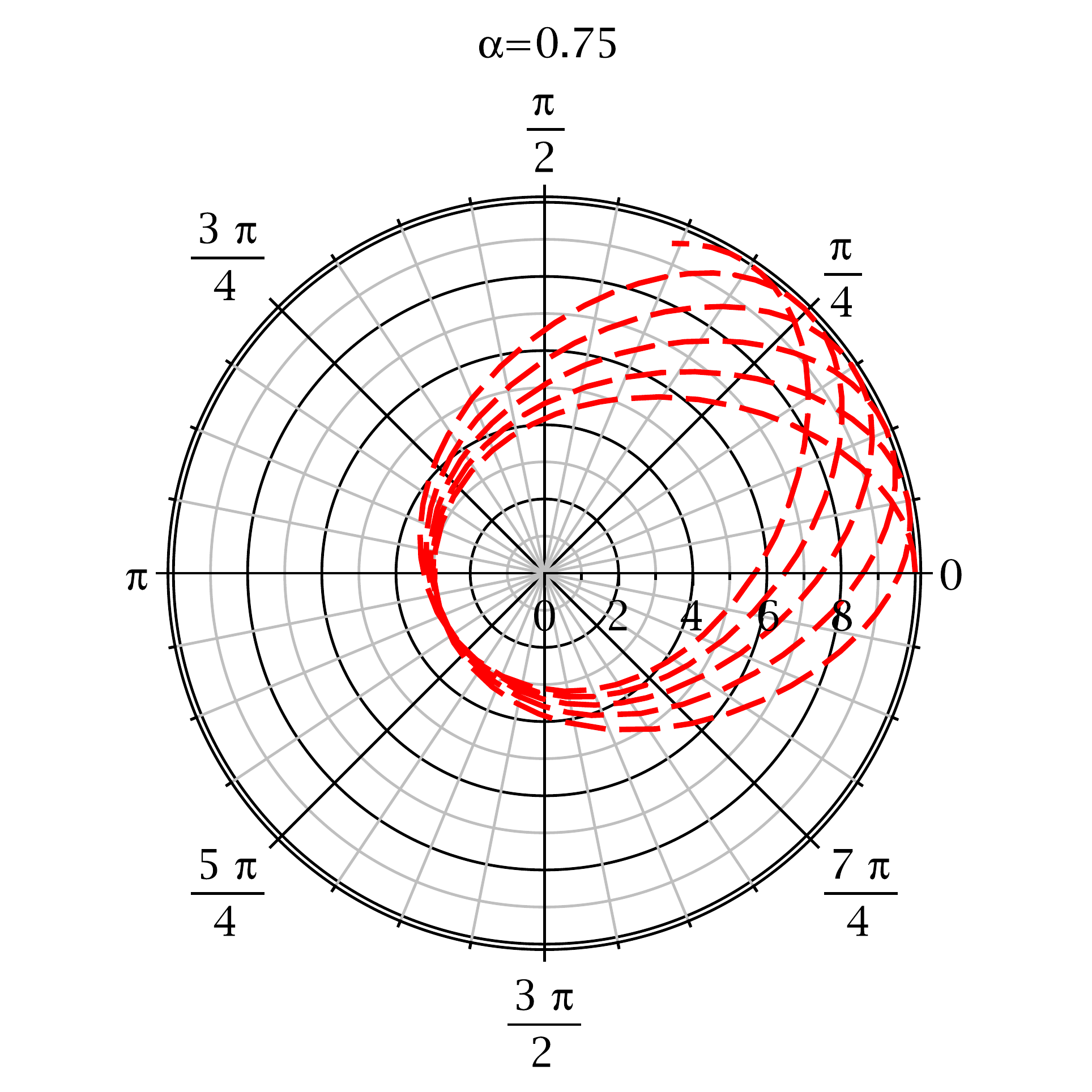}
        \label{1}} \\

    \subfloat[$\alpha=0.5$, $E=3$ and $L=5$.]{
        \includegraphics[width=0.27\textwidth]{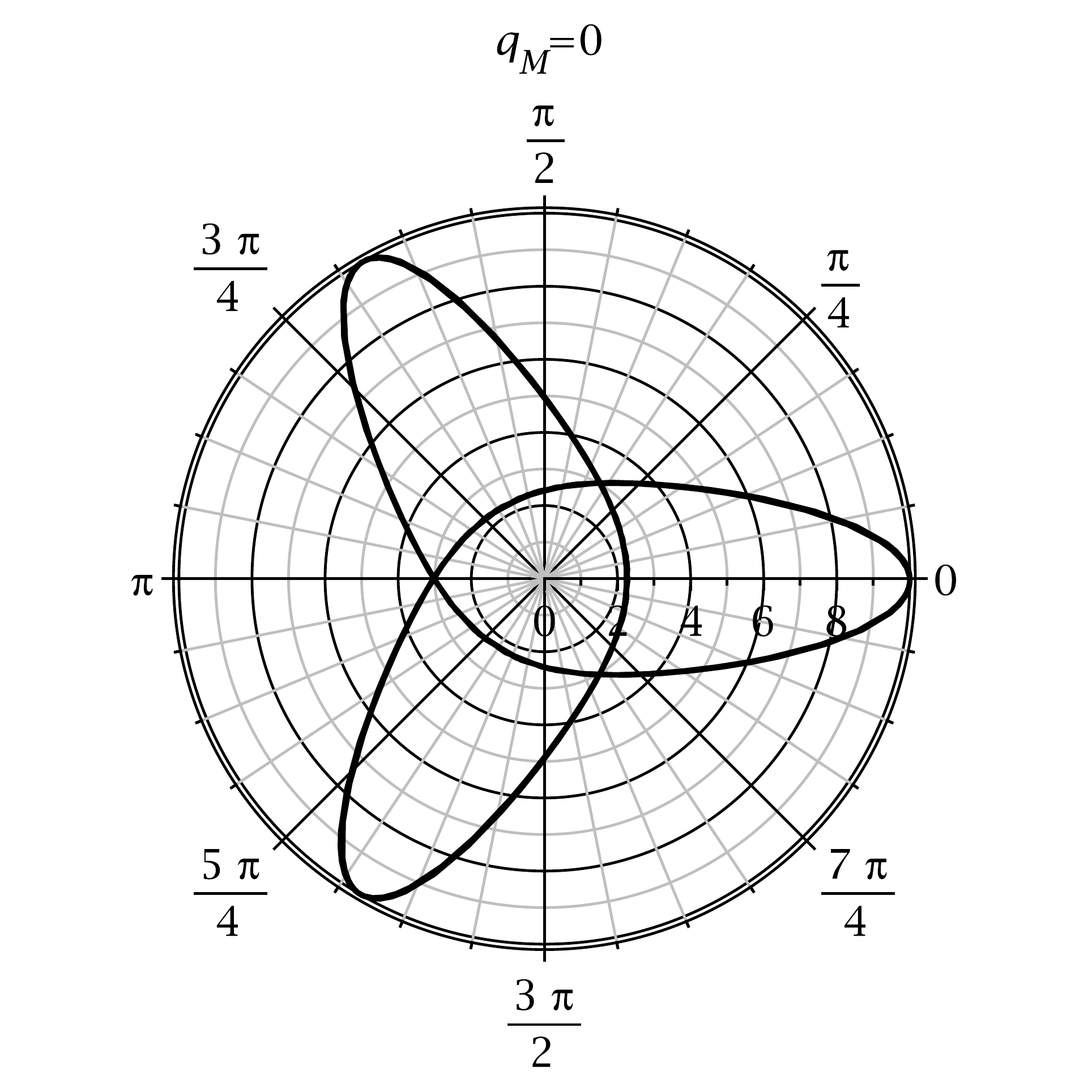}
        \includegraphics[width=0.27\textwidth]{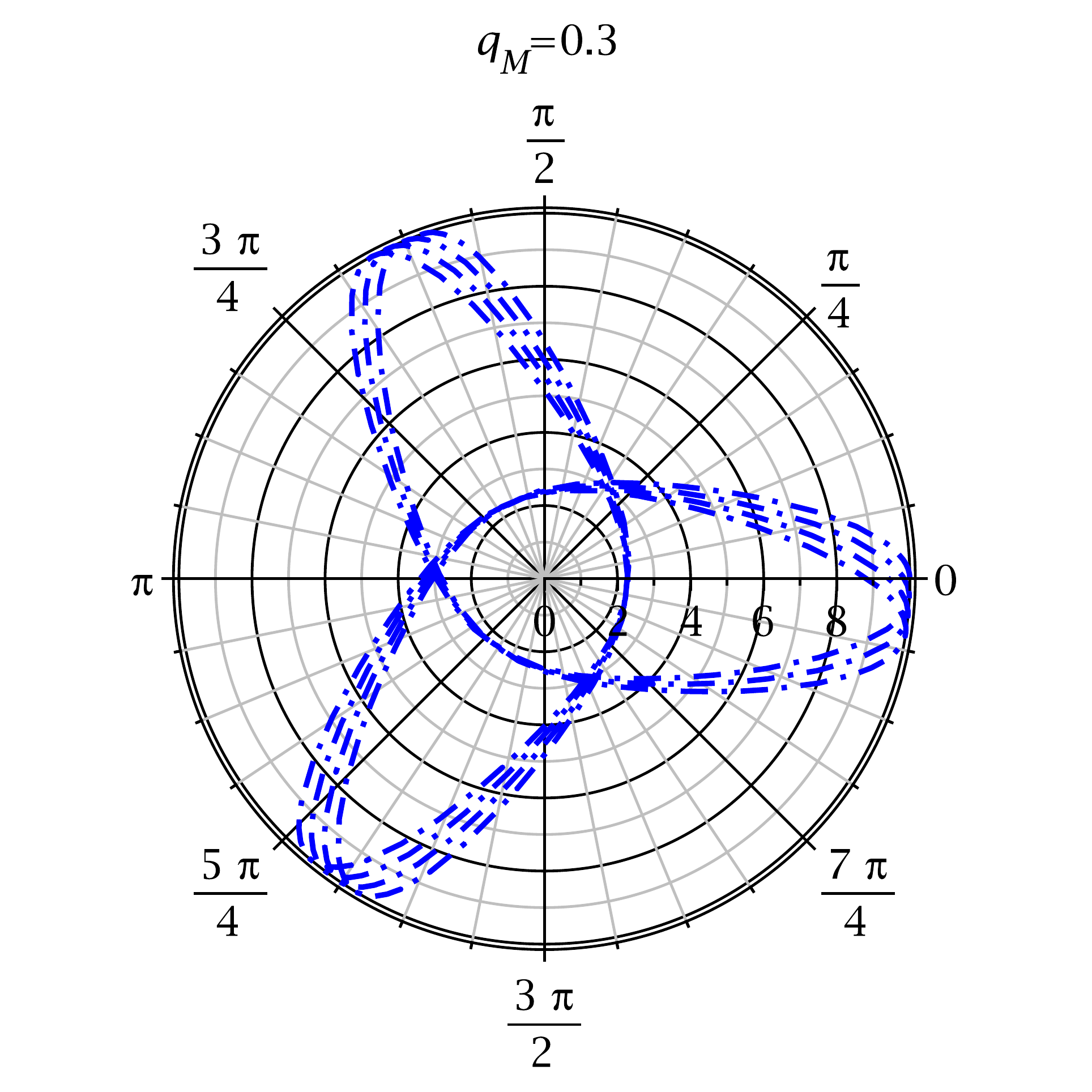}
        \includegraphics[width=0.27\textwidth]{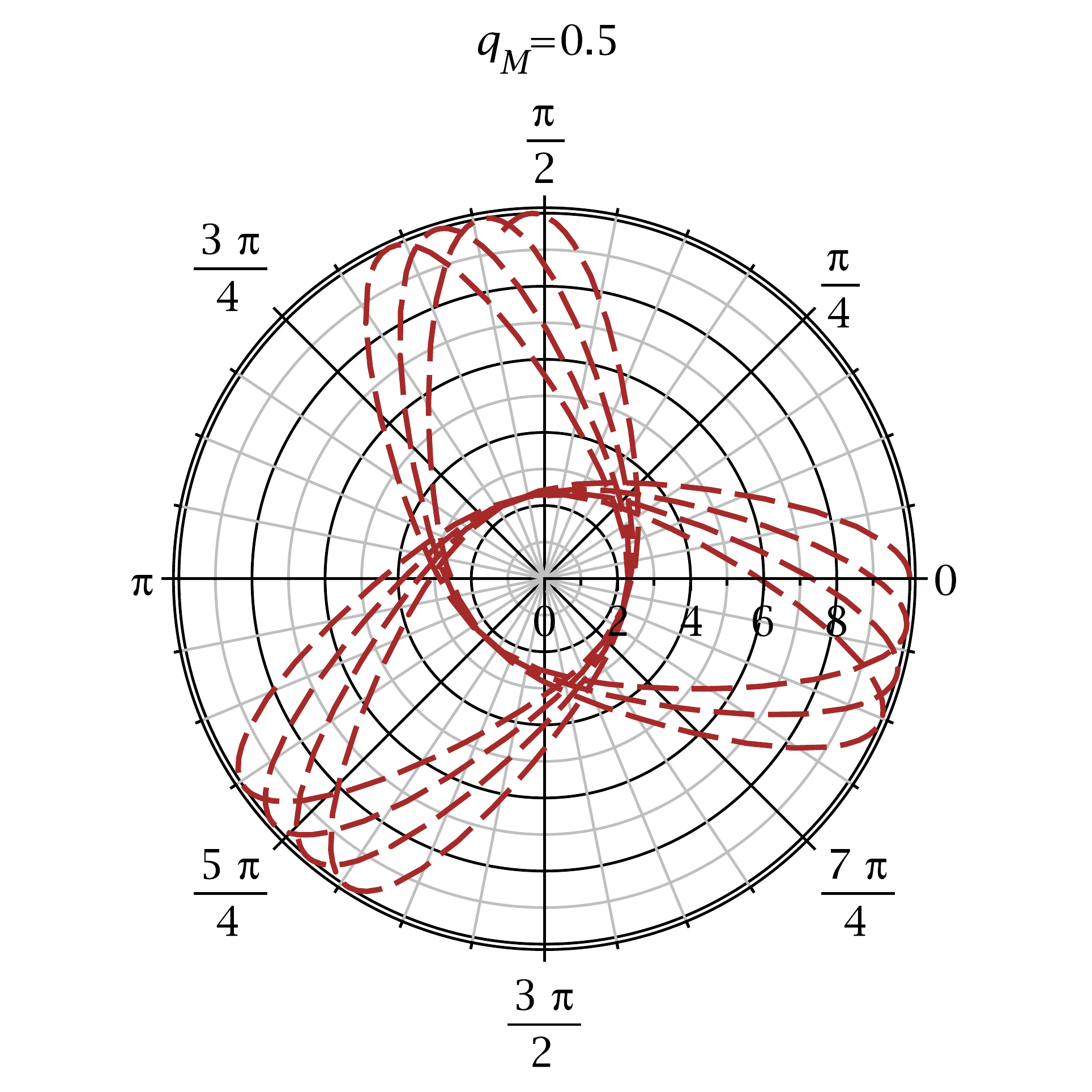}
        \includegraphics[width=0.27\textwidth]{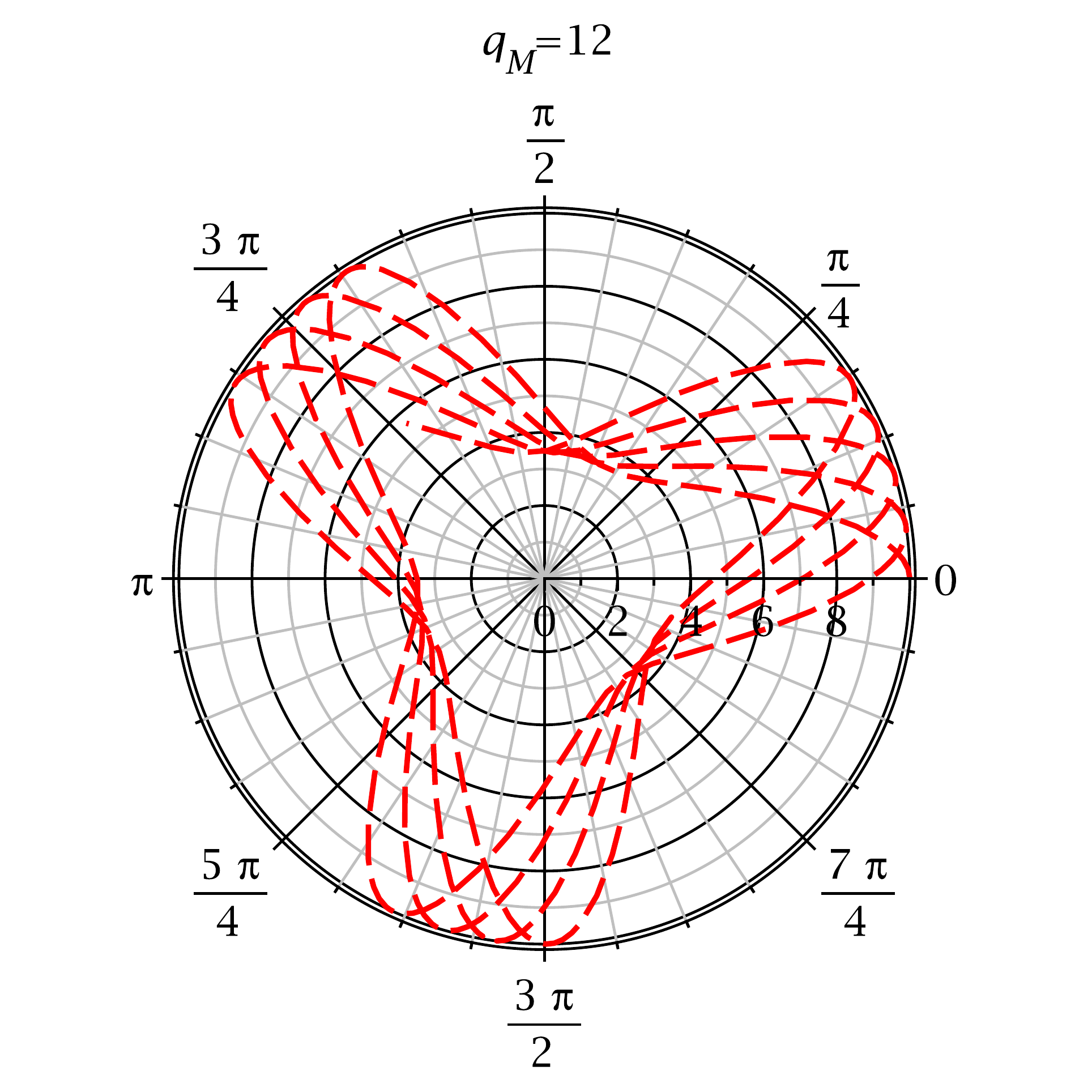}
        \label{2}} \\

    \subfloat[$\alpha=0.5$, $q_{M}=1$ and $L=5$.]{
        \includegraphics[width=0.27\textwidth]{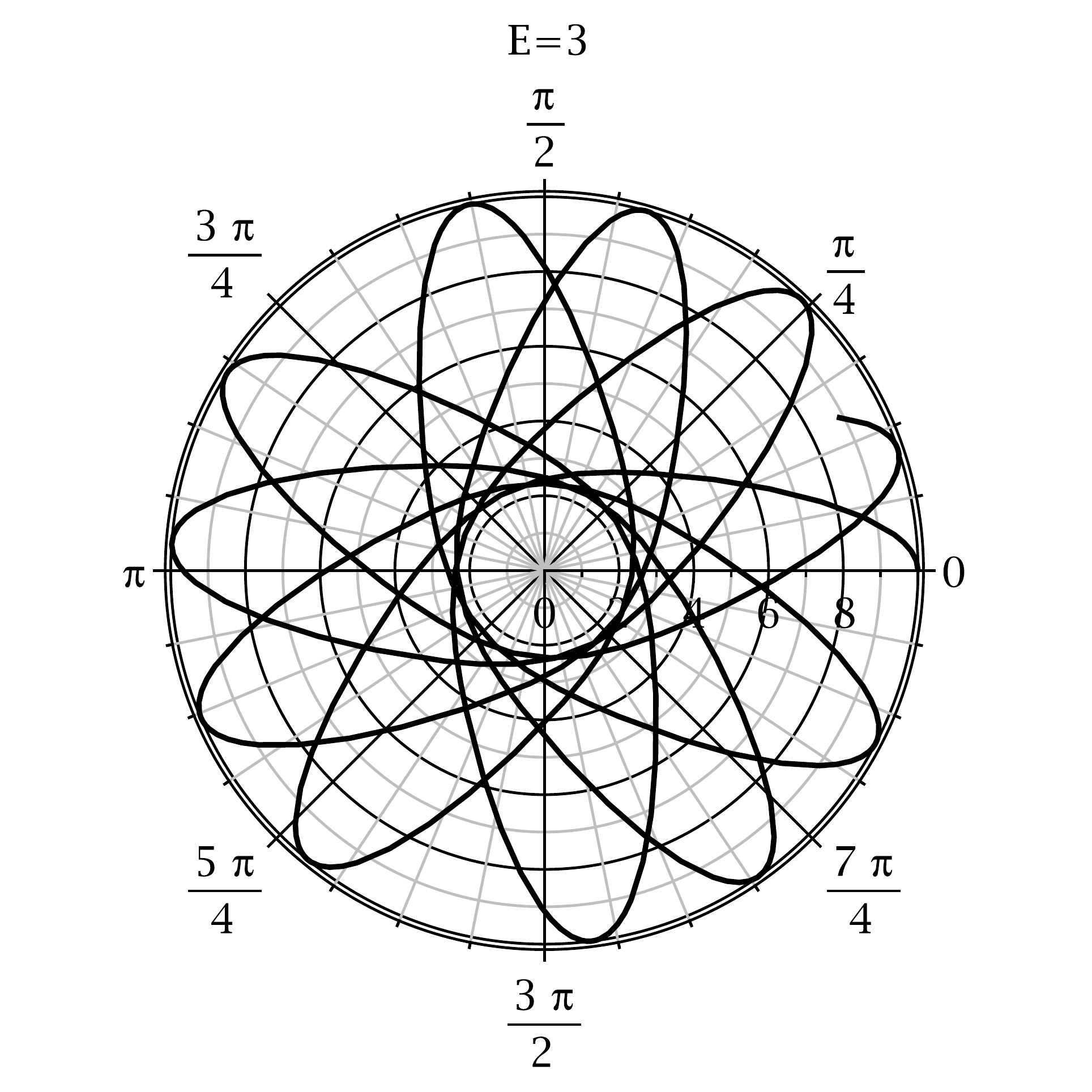}
        \includegraphics[width=0.27\textwidth]{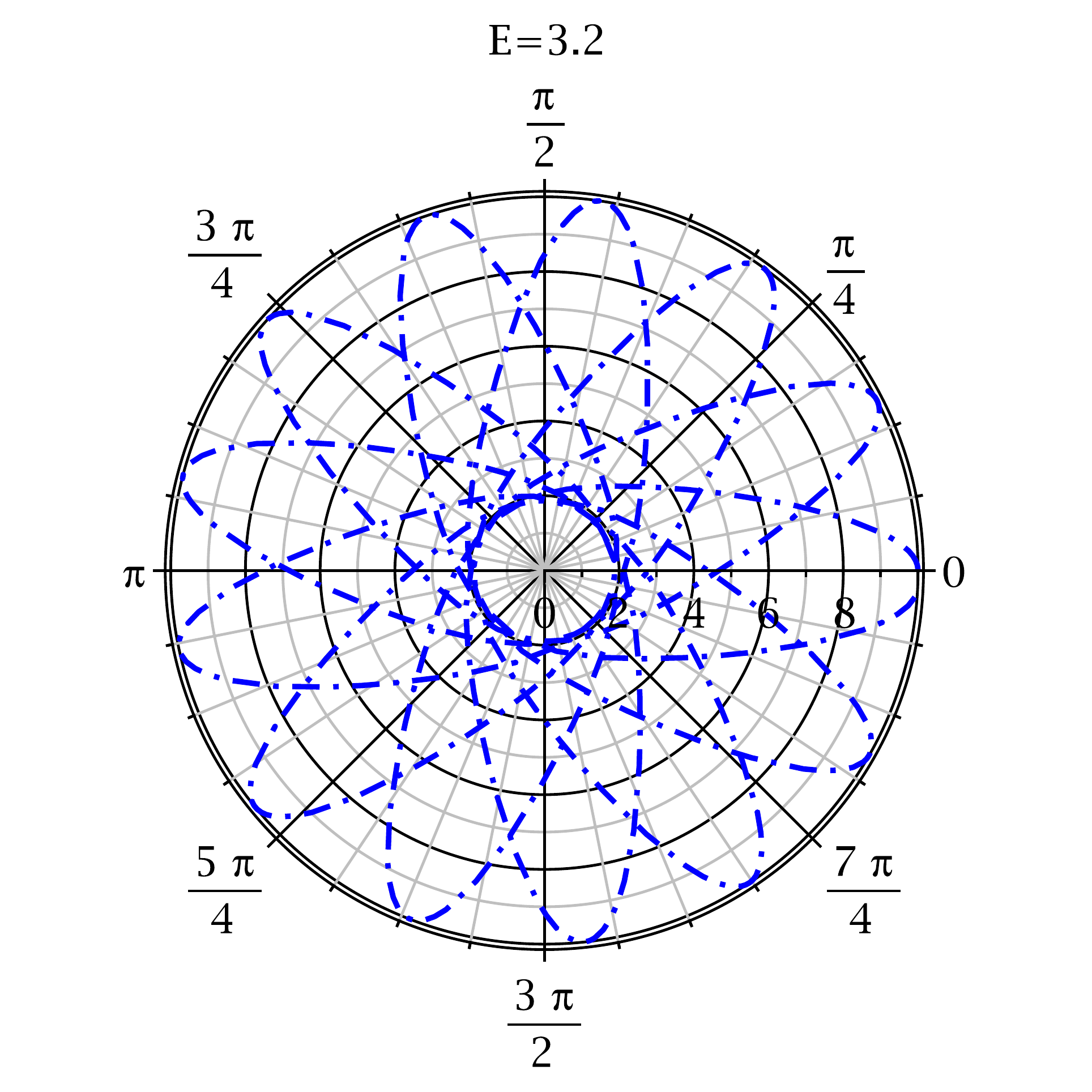}
        \includegraphics[width=0.27\textwidth]{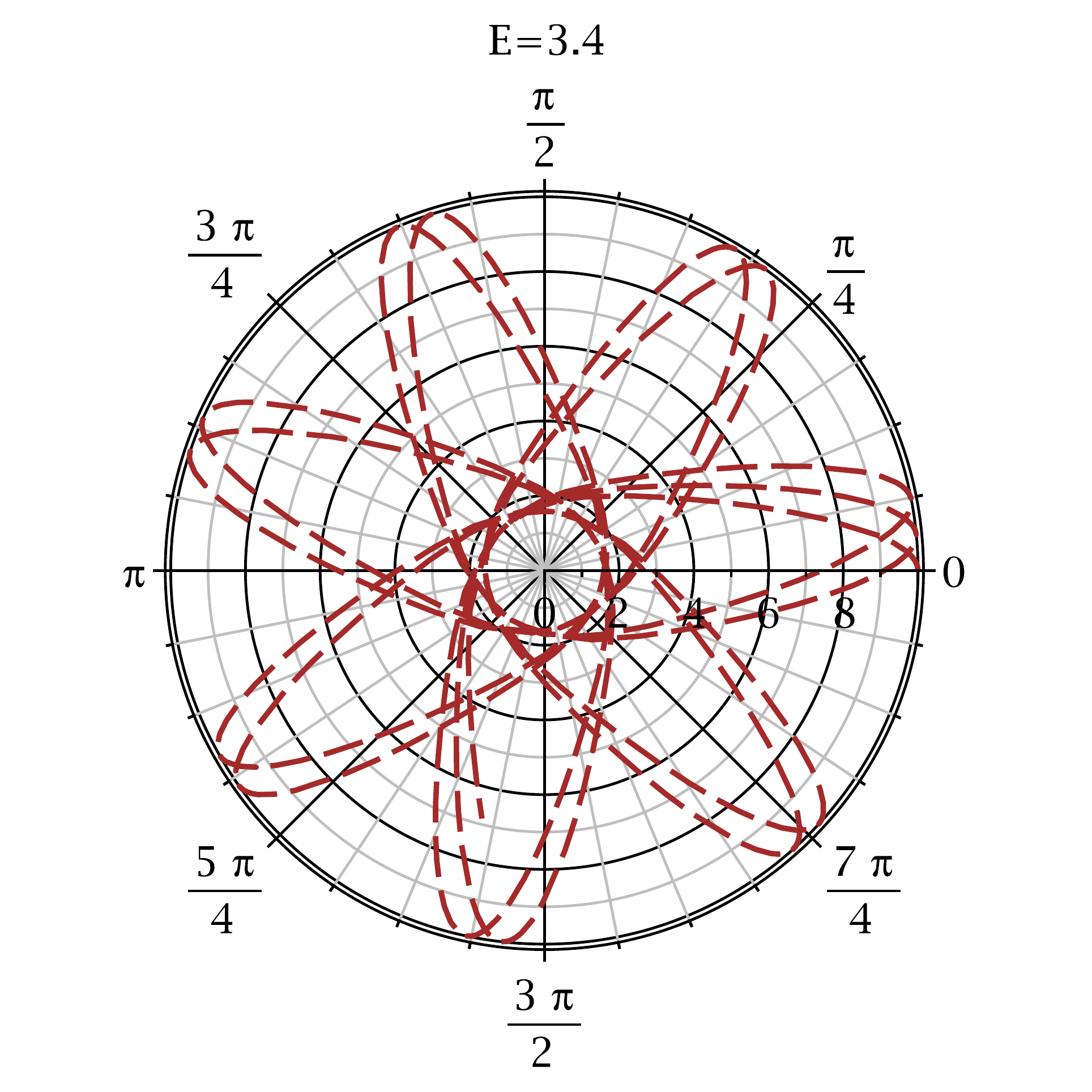}
        \includegraphics[width=0.27\textwidth]{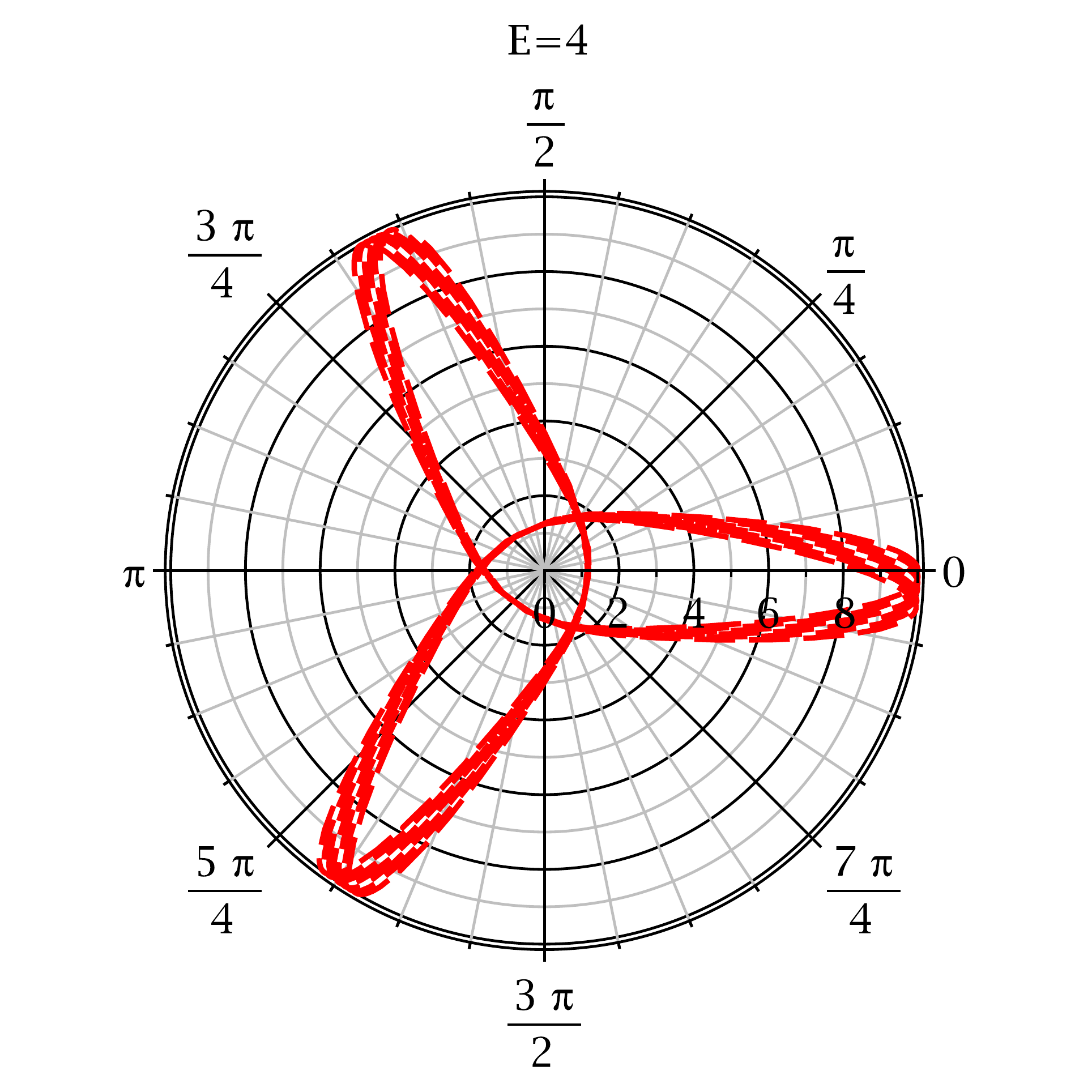}
        \label{3}} \\

    \subfloat[$\alpha=0.5$, $q_{M}=1$ and $E=3$.]{
        \includegraphics[width=0.27\textwidth]{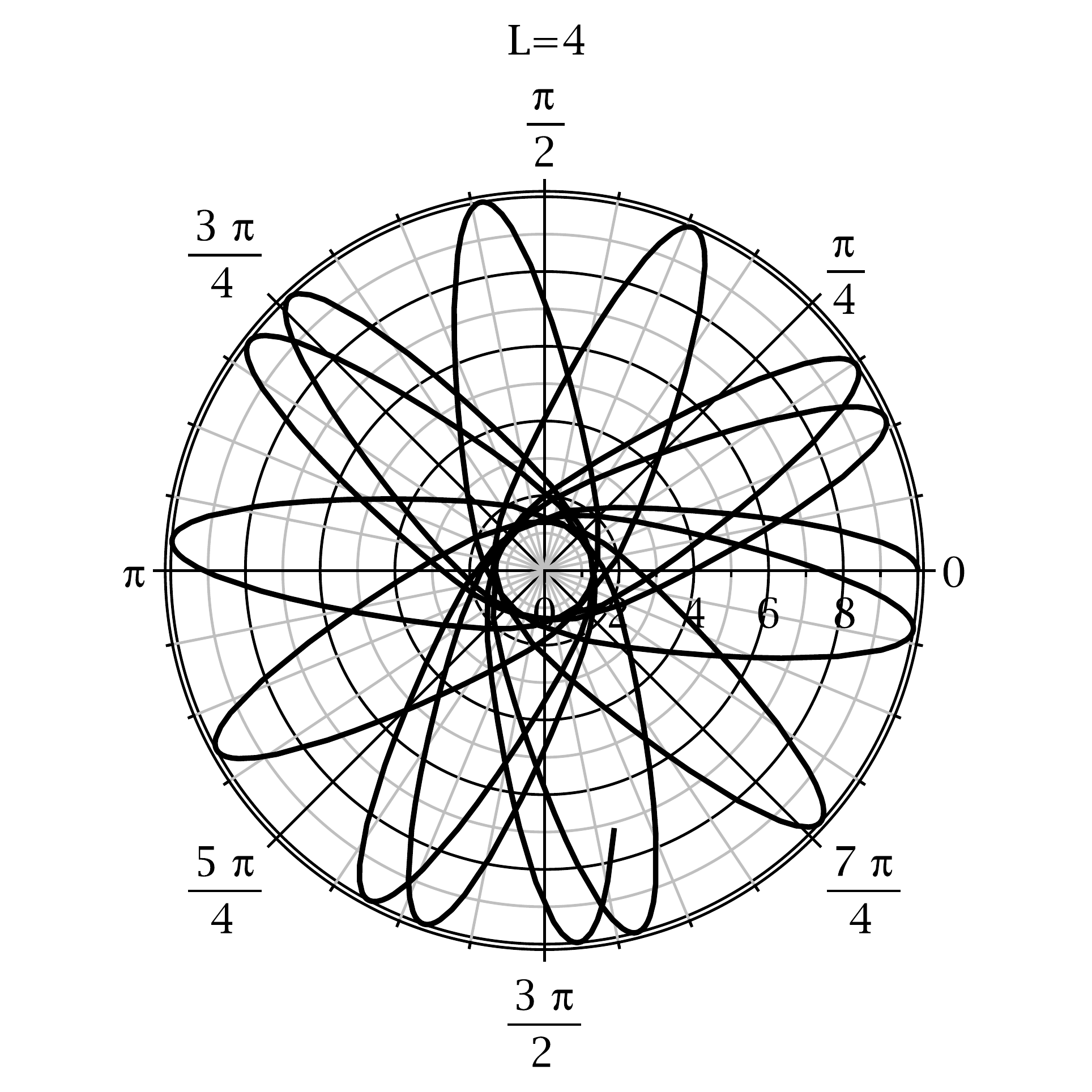}
        \includegraphics[width=0.27\textwidth]{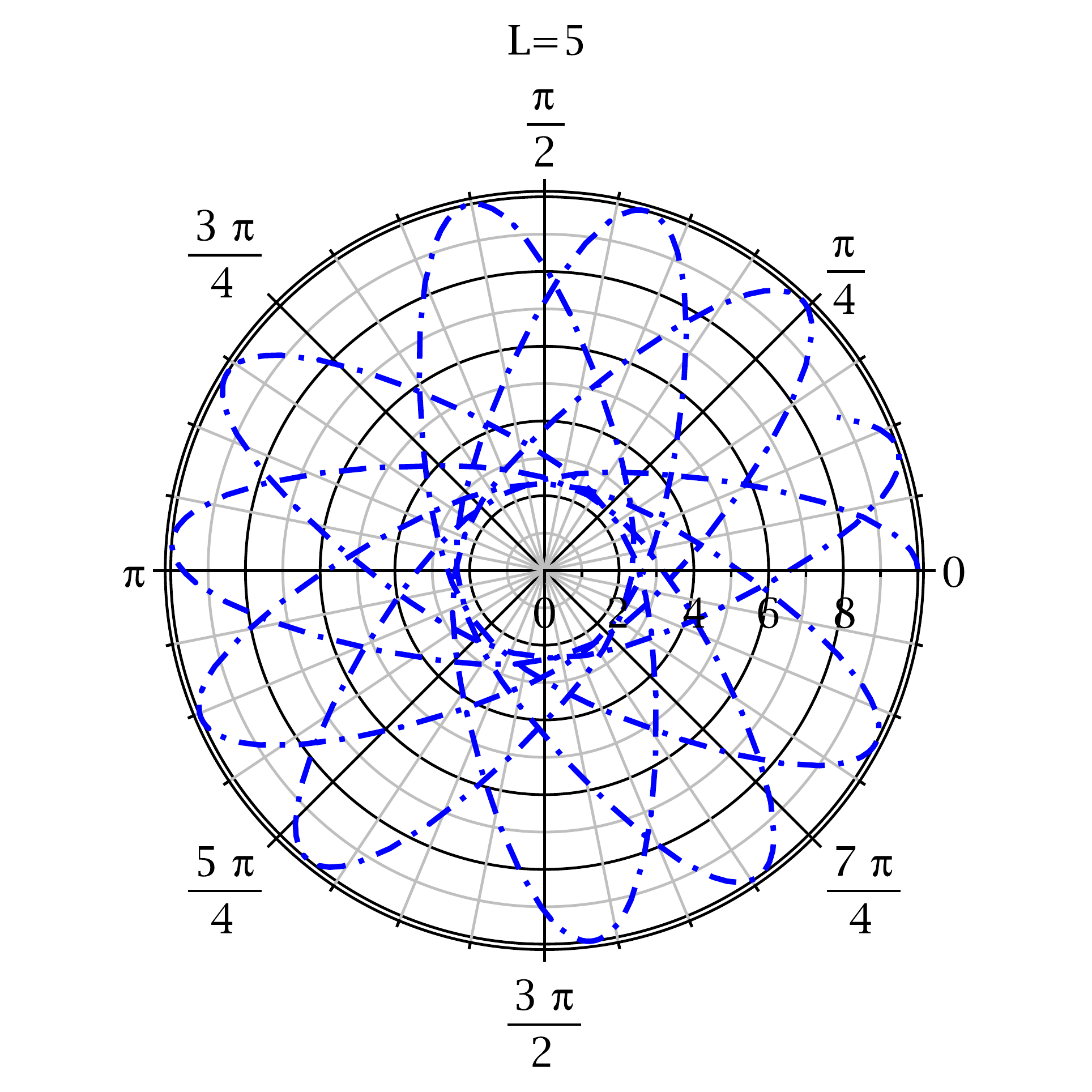}
        \includegraphics[width=0.27\textwidth]{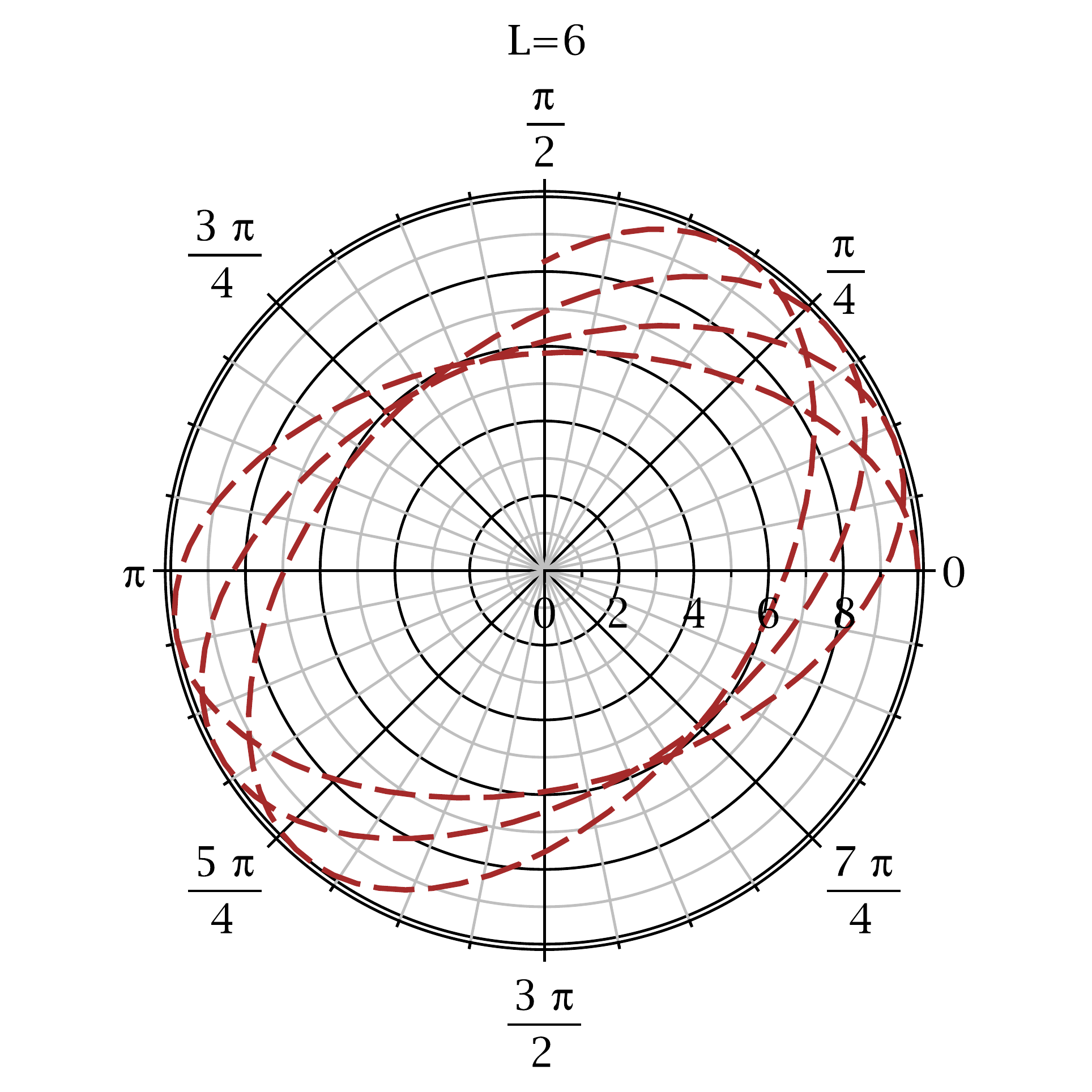}
        \includegraphics[width=0.27\textwidth]{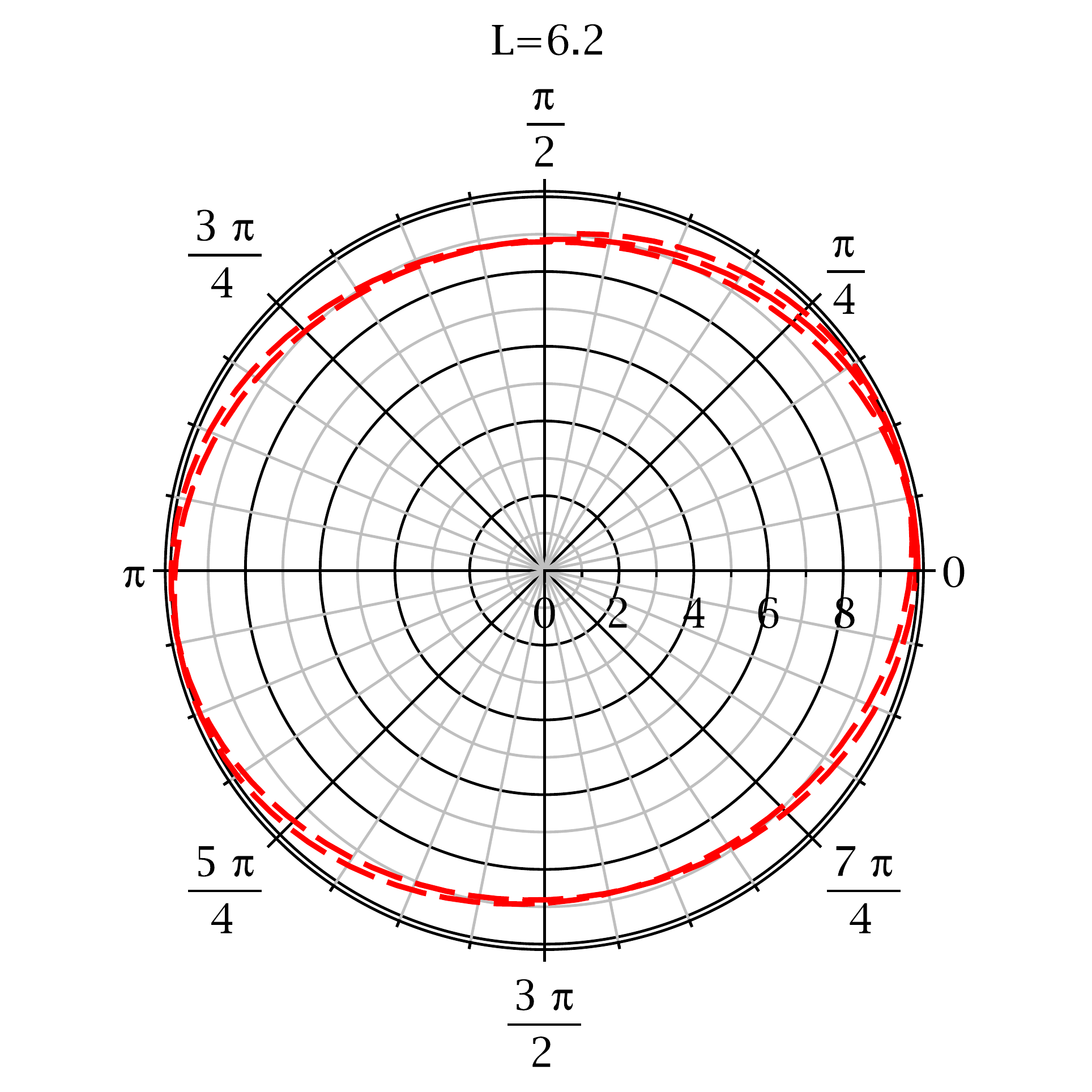}
        \label{4}} \\

    \caption{Trajectory of the particle around black holes for $m=k=q_{E}=1$,
    $\Lambda=-1$, $r(\tau=0)=10$ and $\tau$ running from $0$ to $200$.} \label{Fig5}
\end{figure}

\begin{figure}[!tbp]
    \centering
    \subfloat[$q_{M}=1$, $E=3$, $L=5$, $\alpha=0.7$]
    {\includegraphics[width=0.25\textwidth]{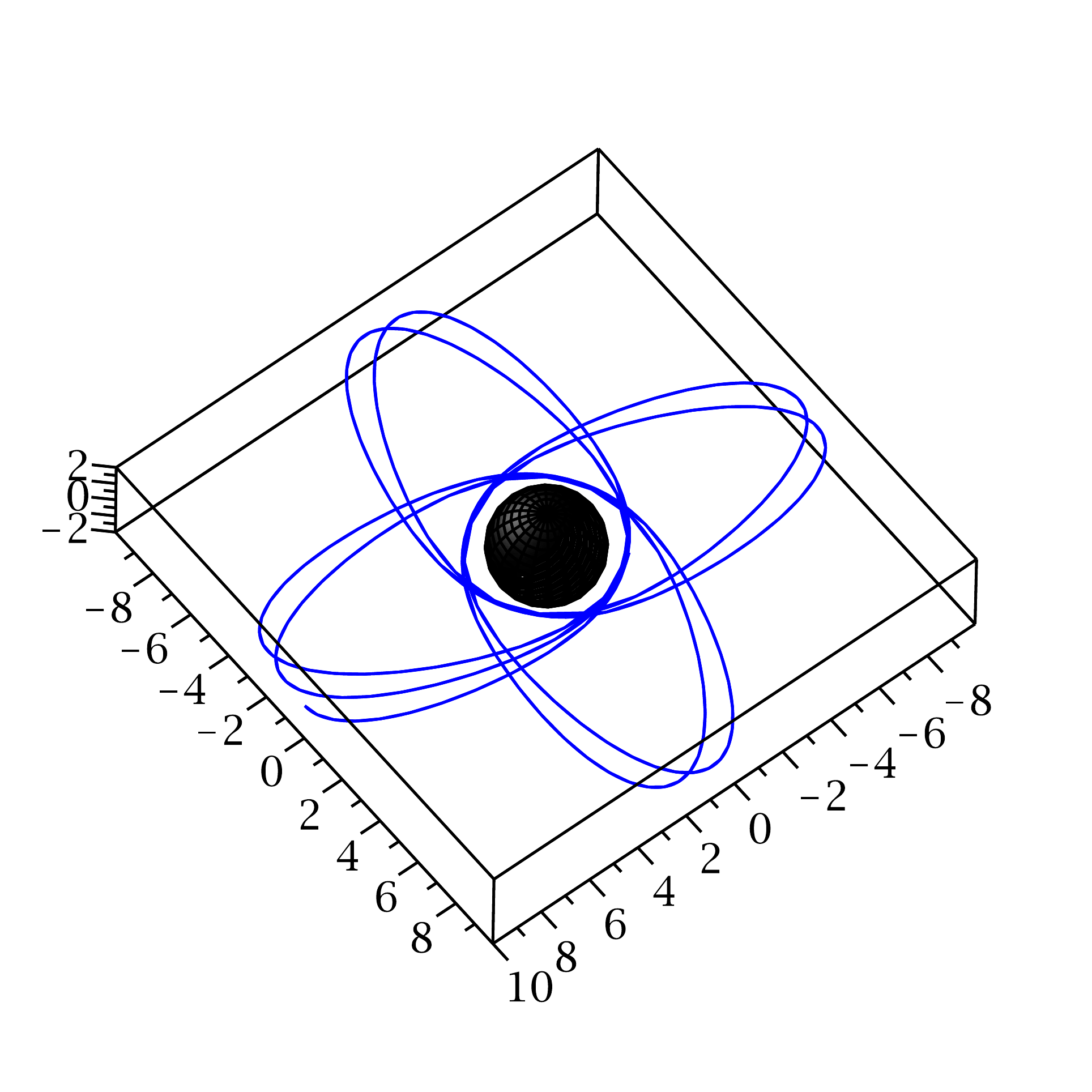}
        \includegraphics[width=0.25\textwidth]{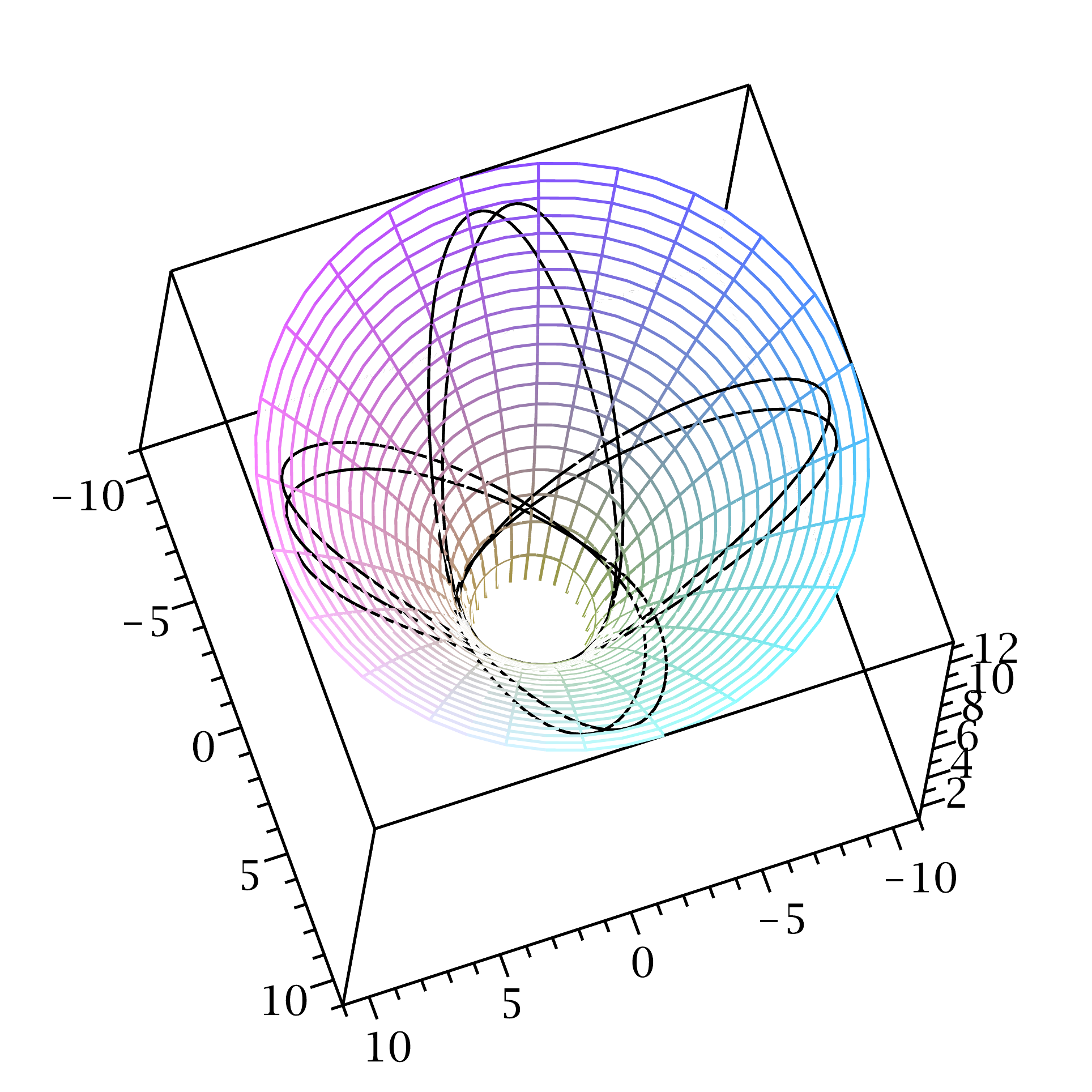}}
    \subfloat[$\alpha=0.5$, $E=3$, $L=5$ and $q_{M}=0.5$]
    {\includegraphics[width=0.25\textwidth]{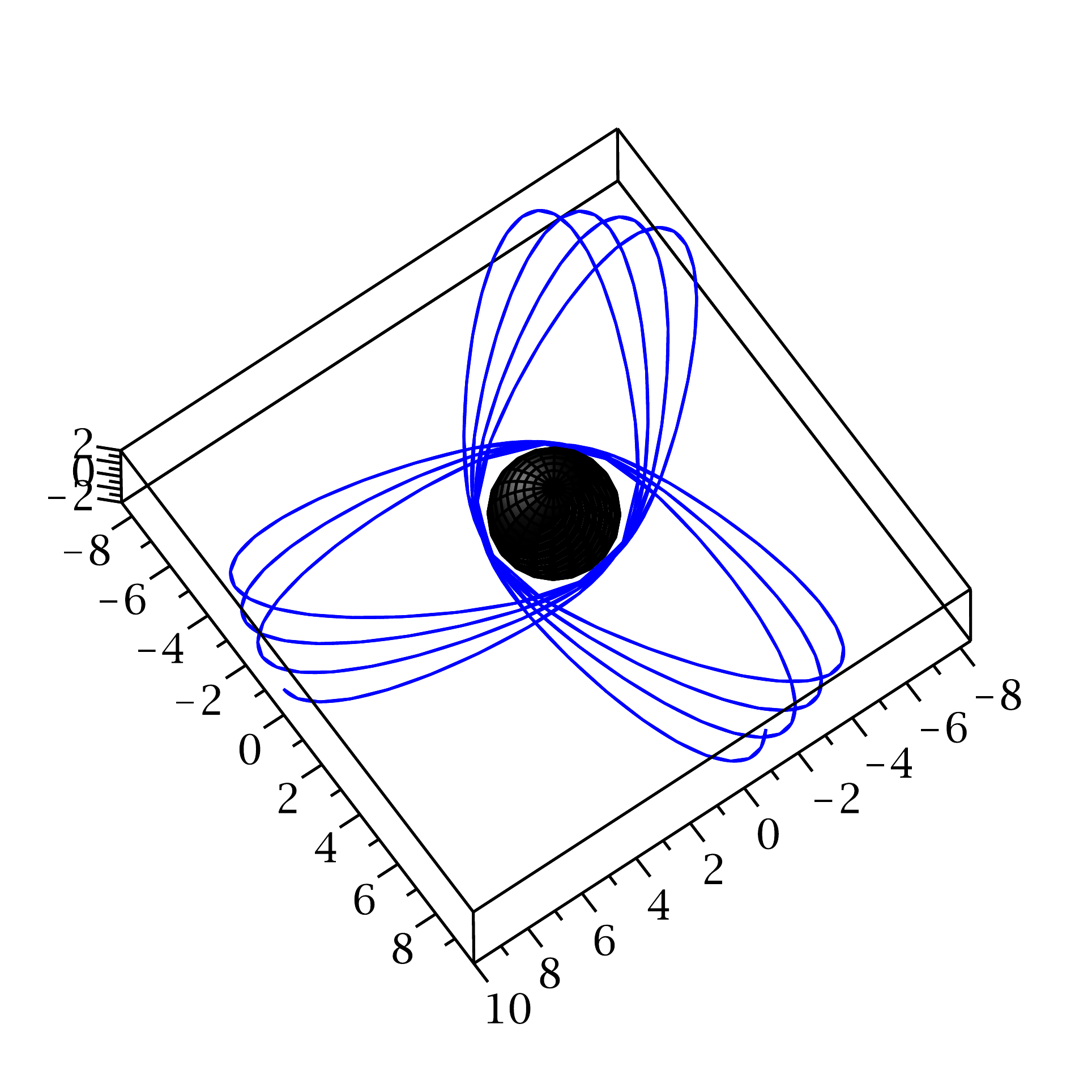}
        \includegraphics[width=0.25\textwidth]{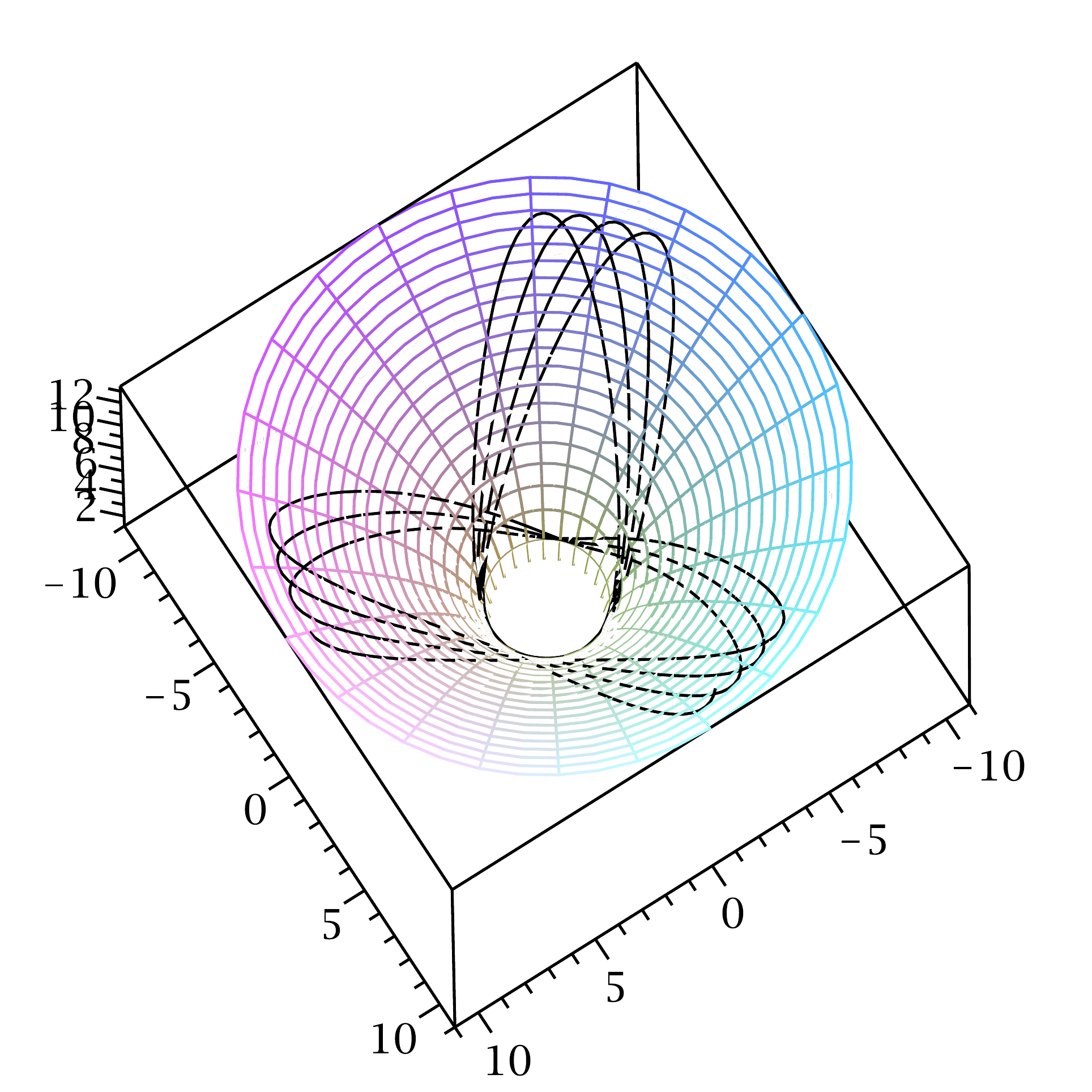}}     \\

    \subfloat[$q_{M}=1$, $L=5$, $\alpha=0.5$ and $E=3.2$]
    {\includegraphics[width=0.25\textwidth]{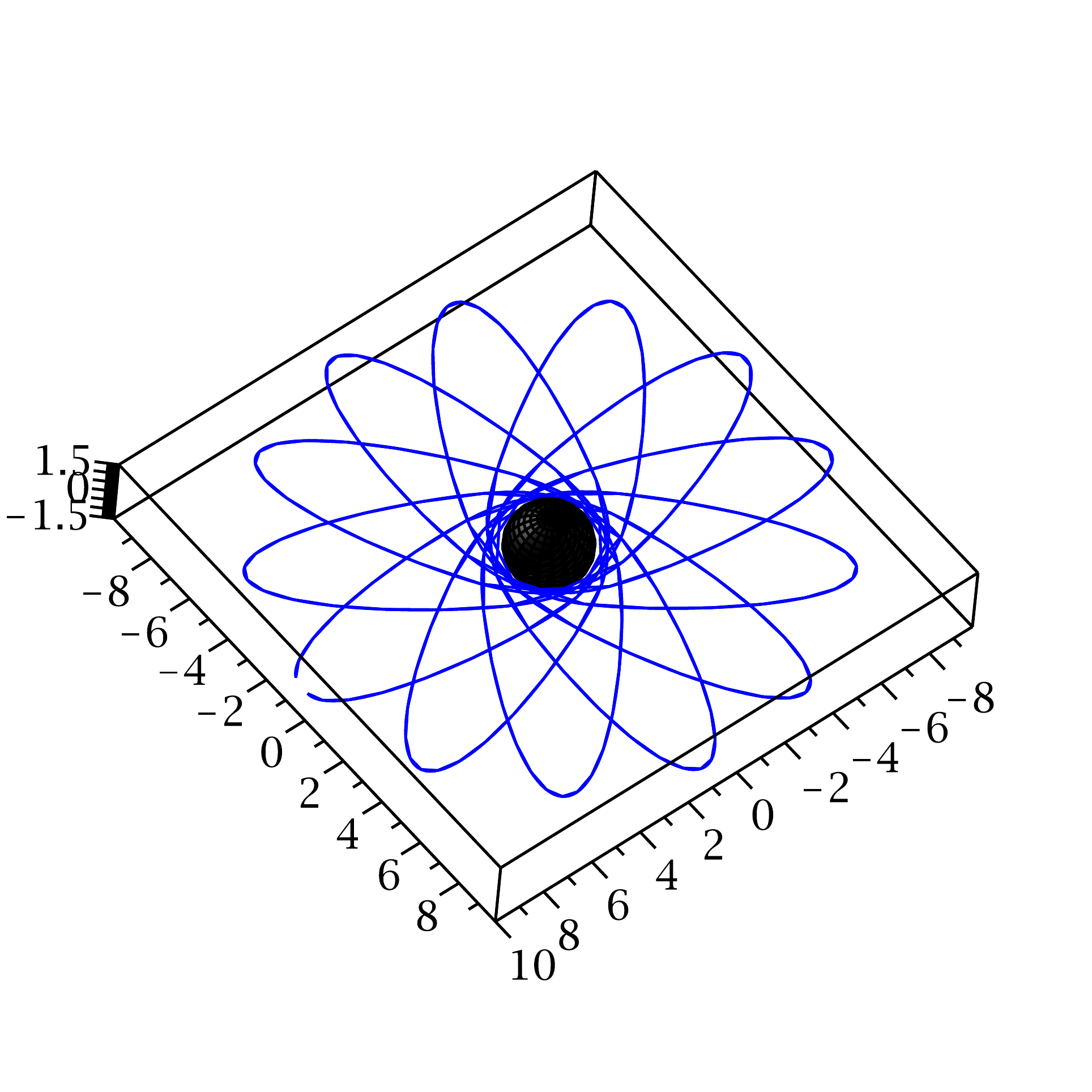}
        \includegraphics[width=0.25\textwidth]{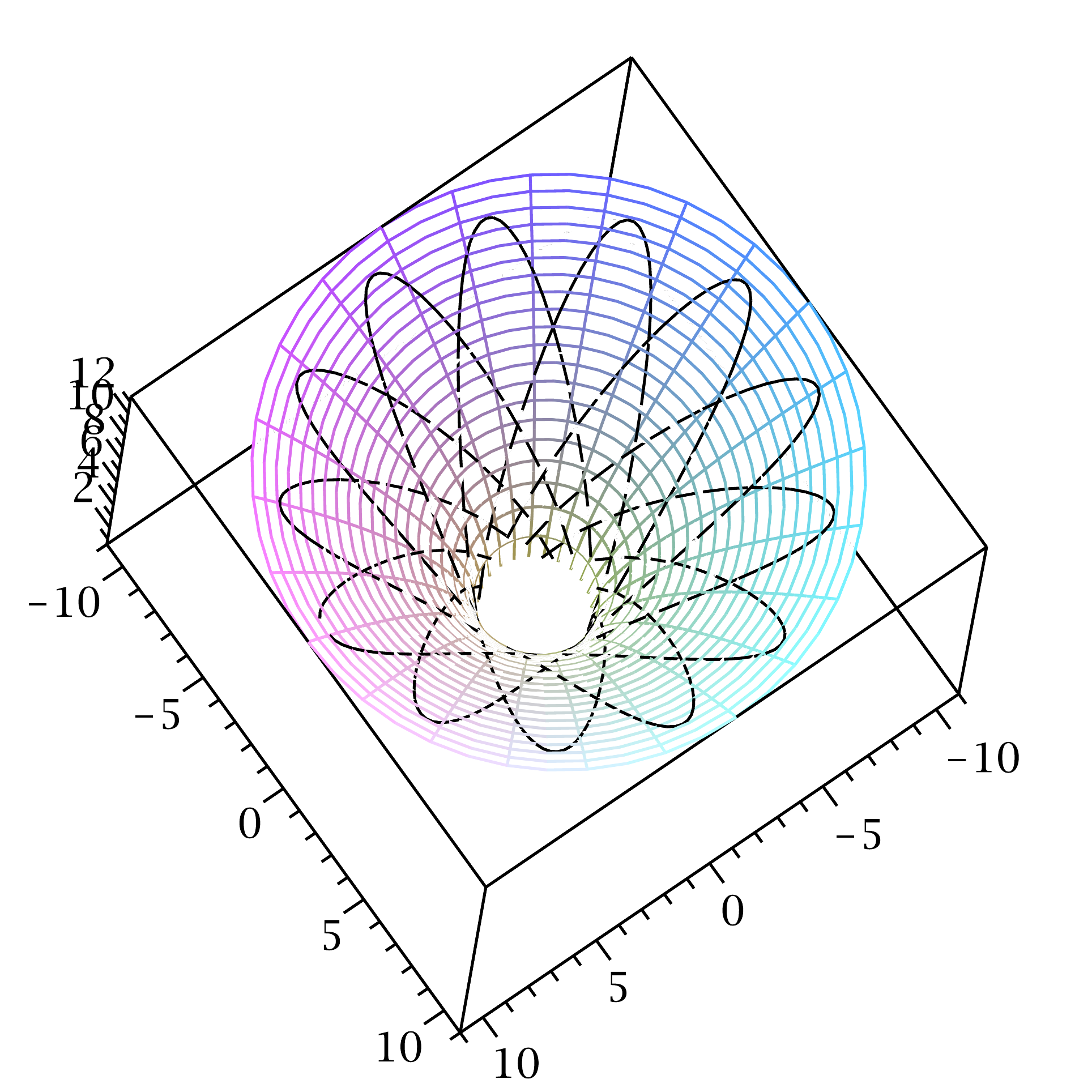}}
    \subfloat[$q_{M}=1$, $E=3$, $\alpha=0.5$ and  $L=6$]
    {\includegraphics[width=0.25\textwidth]{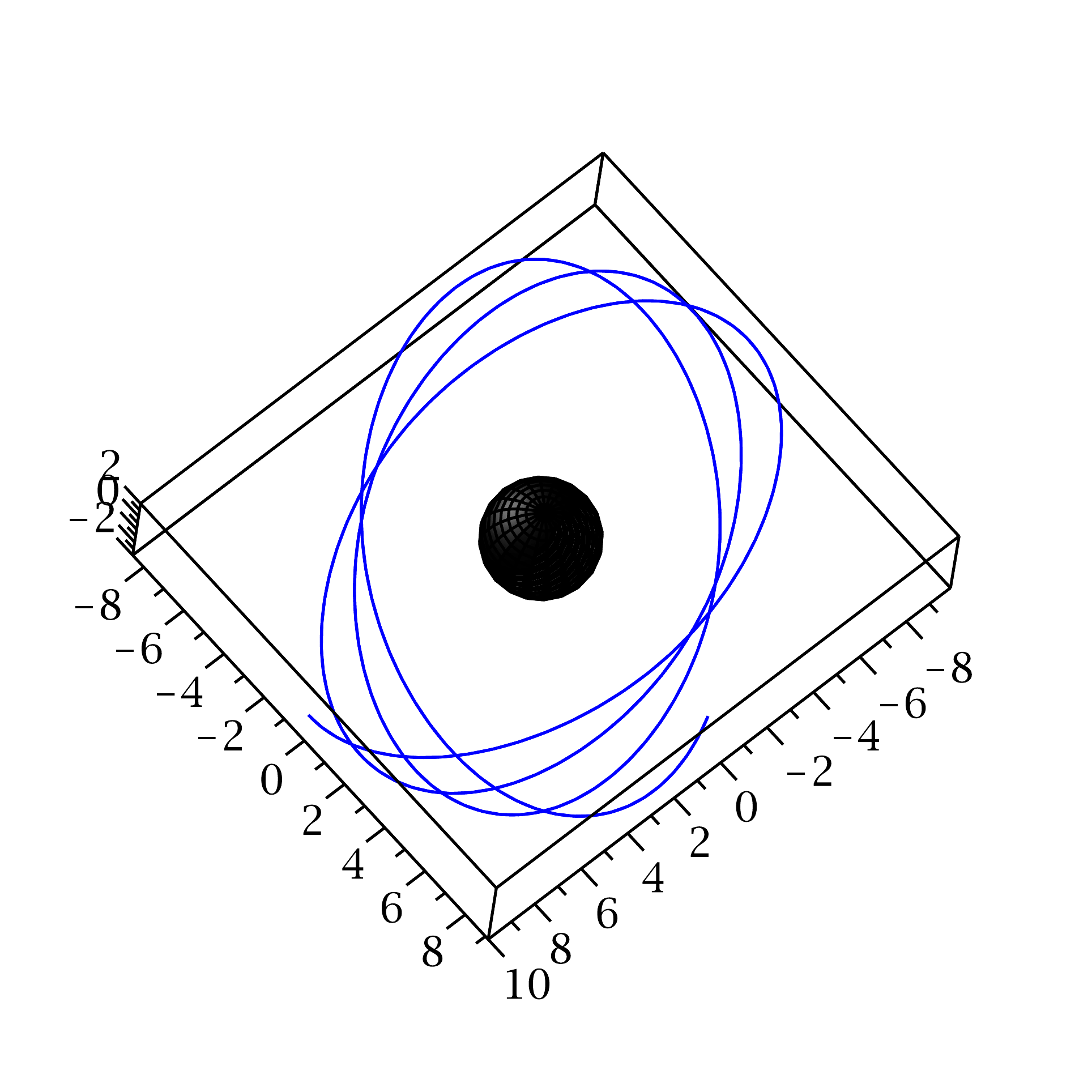}
        \includegraphics[width=0.25\textwidth]{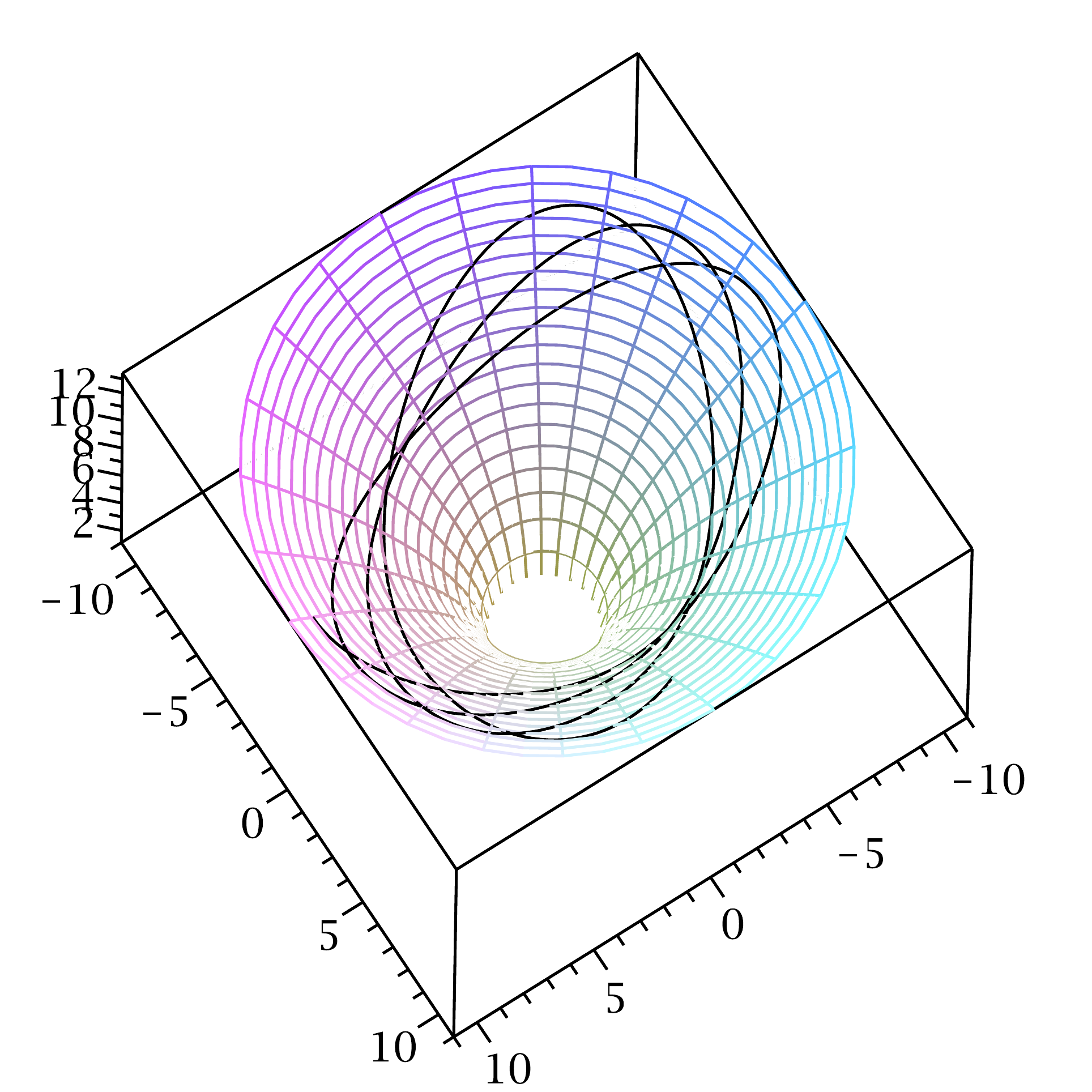}}     \\

    \caption{Trajectory of a particle around black holes for $m=k=q_{E}=1$,
    $\Lambda=-1$, $r(\tau=0)=10$ and $\tau$ running from $0$ to $200$.} \label{Fig6}
\end{figure}

Here, we focus only on the null orbits, requiring $\mathcal{L}=0$. For the
timelike orbits, one should consider $\mathcal{L}=1$. In order to obtain the
motion of the particle around black holes, one should replace Eqs. (\ref%
{Energy})-(\ref{Lag2}) in the Euler-Lagrange equation (\ref{EL}) and solve
it. Due to the complexity of the obtained relations, we employed numerical
computation to obtain expressions governing the motion of particle around
these black holes. We fixed the initial conditions and study the effects of
different parameters on the trajectory. The results are presented in Figs. %
\ref{Fig5} and \ref{Fig6}.

Evidently, depending on the choices of different parameters, the null orbits
of particle around the black holes would be different. The shape of orbital
movement of particle is highly sensitive to the variation of GB parameter,
energy and orbital angular momentum. Whereas, the effects of the variation
of magnetic charge on the shape is not as significant as other parameters.
The effects of this parameter could be better seen in the context of the
speed of particle. In order to have a better picture regarding the effects
of different parameters on the motion of particle, we calculate angular
frequency as measured at infinity. This is given by
\begin{equation}
\Omega_{c}= \frac{d\phi}{dt}=(\frac{d\phi}{d\tau})/(\frac{dt}{d\tau})=\frac{%
f(r)L}{r^2 E}.
\end{equation}

\begin{figure}[!tbp]
\centering
\subfloat[$q_{M}=1$, $E=3$ and
$L=5$]{\includegraphics[width=0.25\textwidth]{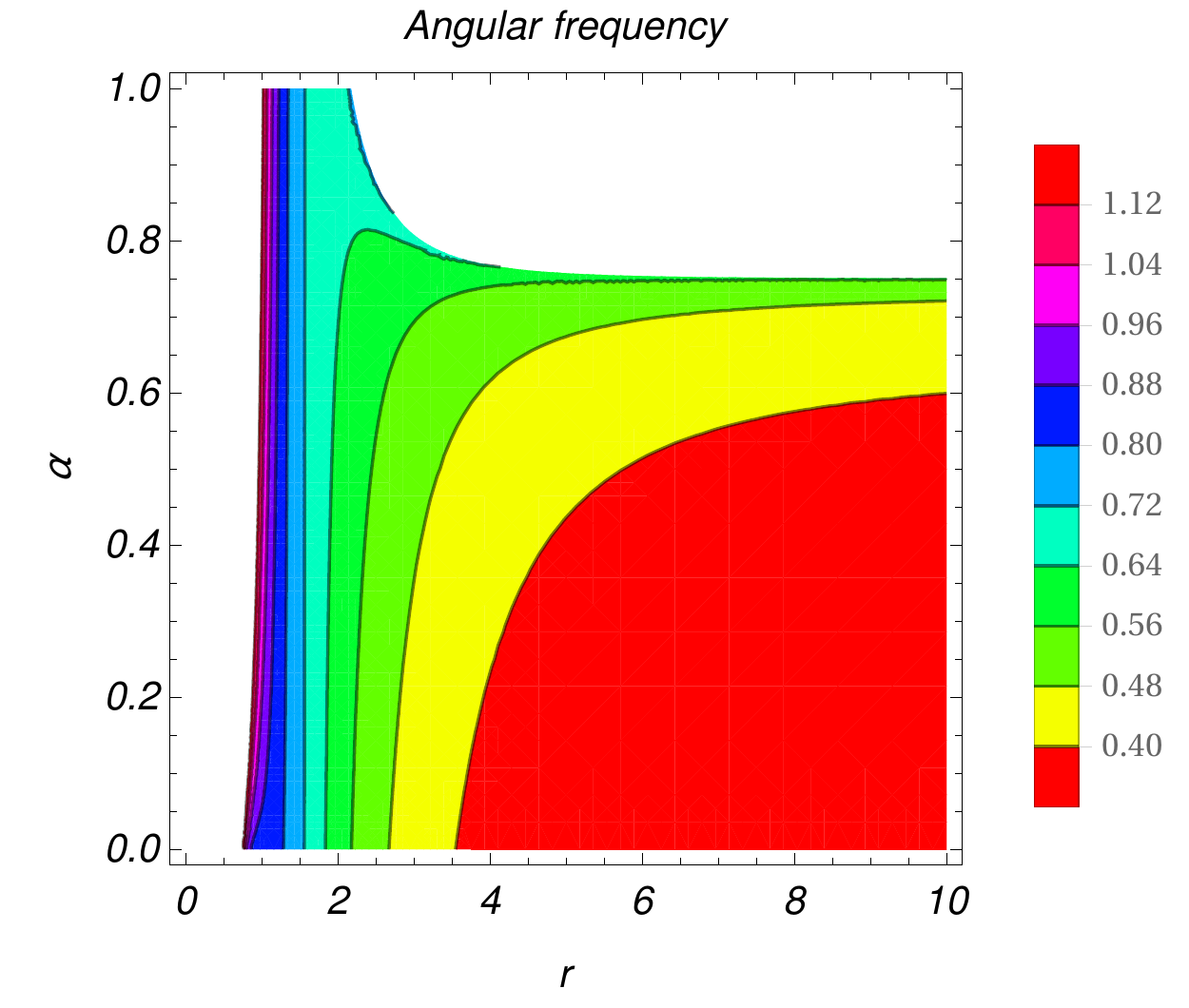}} \subfloat[$%
\alpha=0.5$, $E=3$ and $L=5$]{\includegraphics[width=0.25\textwidth]{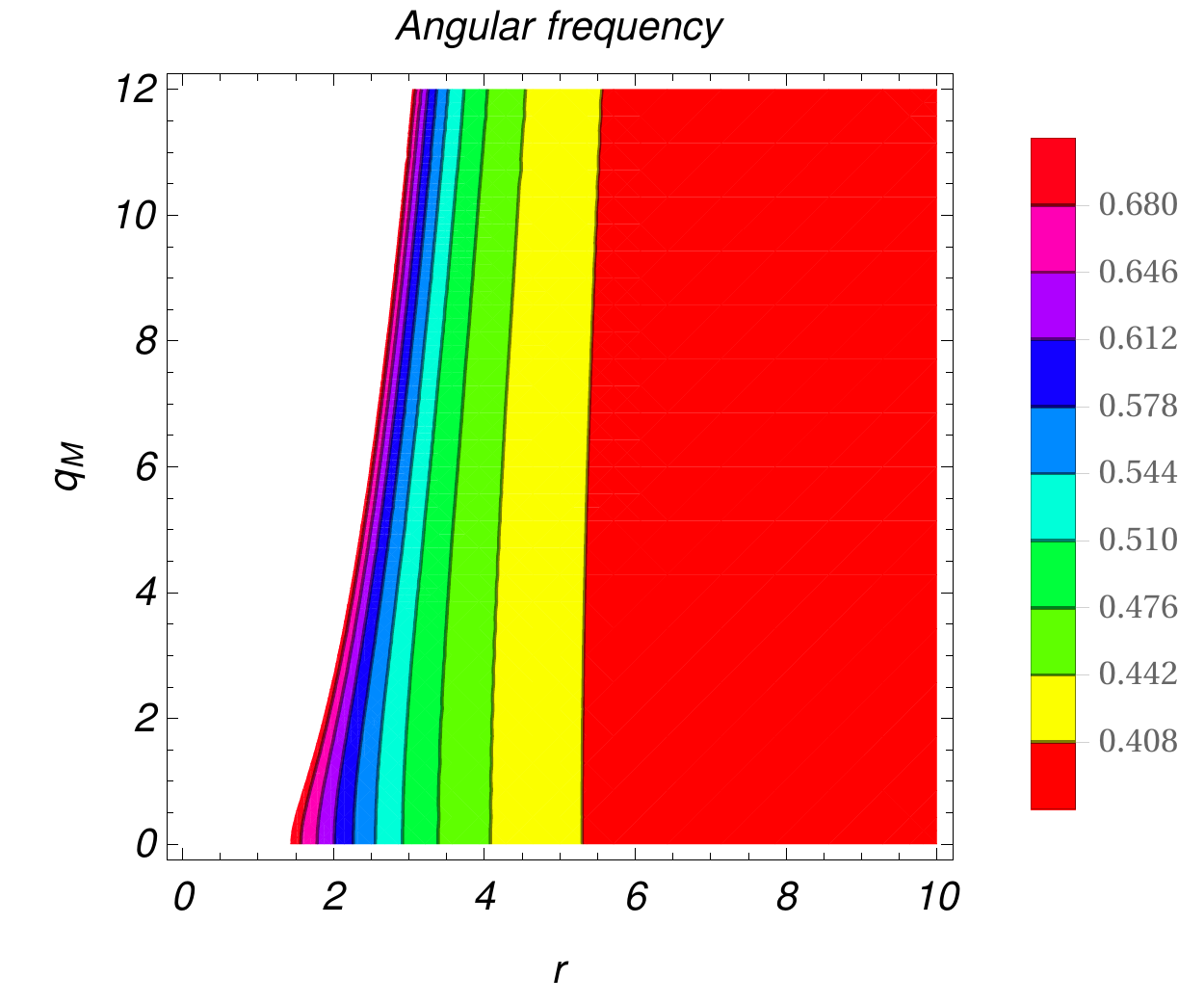}}
\subfloat[$q_{M}=1$, $\alpha=0.5$ and
$L=5$]{\includegraphics[width=0.25\textwidth]{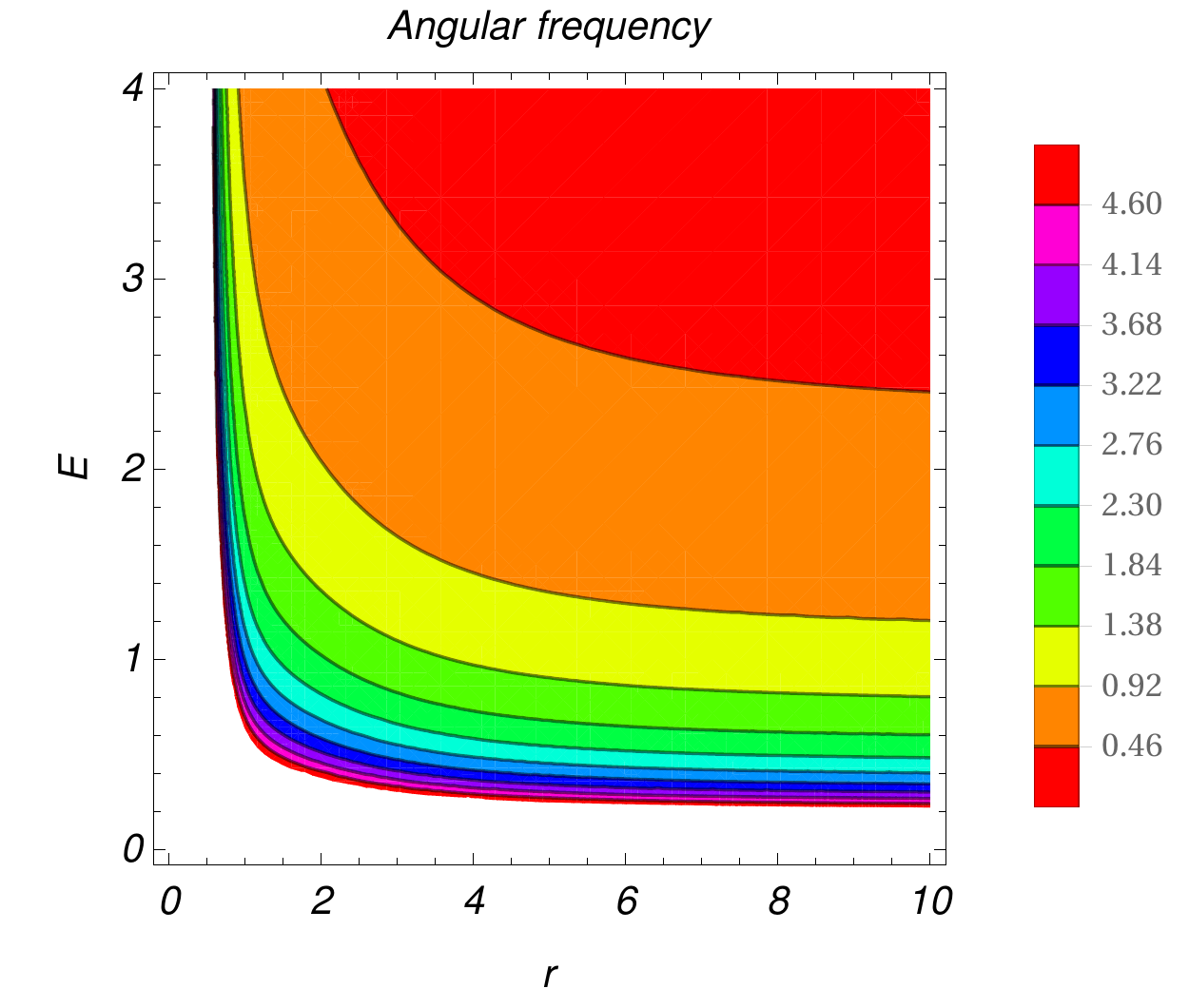}}
\subfloat[$q_{M}=1$,
$\alpha=0.5$ and $E=3$]{\includegraphics[width=0.25\textwidth]{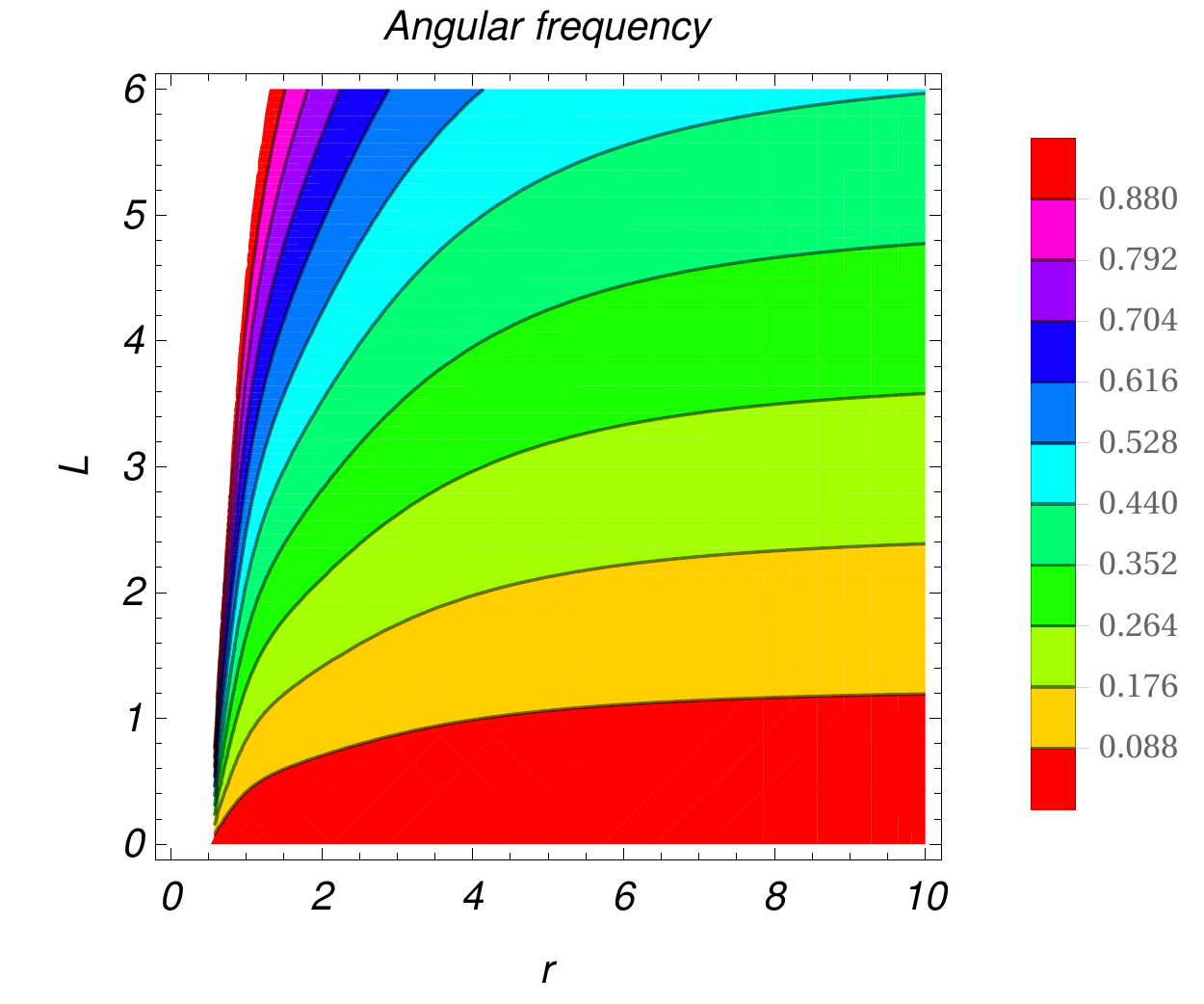}}
\newline
\caption{Angular frequency of the orbit for $m=k=q_{E}=1$ and $\Lambda=-1$.}
\label{Fig7}
\end{figure}

Using obtained metric function (\ref{metric function}) and considering
plotted diagrams in Fig. \ref{Fig5}, we plot the following diagrams for the
angular frequency (Fig. \ref{Fig7}). The blank space corresponds to the
non-real angular frequency. The diagrams confirm that angular frequency is a
decreasing function of magnetic charge, GB parameter, energy and angular
momentum. The plotted diagrams for angular frequency could also be used to
determine the regions where there is no physical geodesics for the particle
moving around the black holes. These regions are where no real valued
angular frequency exists (hence blank spaces in Fig. \ref{Fig7}).

\section{Conclusion}

In this paper, we have investigated dyonic black holes in the presence of GB
gravity. The paper was focused on three aspects of these black holes
including: solutions and geometrical properties, thermodynamic behavior and
possible phase transition, and finally the trajectory of a particle around
these black holes.

The metric function was obtained and it was shown that: I) These solutions
could enjoy from zero up to three roots in their structures. II) The
Kretschmann scalar admitted the existence of singularity and the fact that
this singularity is affected by both magnetic charge and GB gravity. III)
The asymptotic behavior of the Kretschmann admitted a modified version of
the AdS/dS behavior.

The investigations regarding the thermodynamics of these black holes
confirmed: I) Decoupling between electric and magnetic parts of the
solutions enabling possibility of studying purely magnetically charged
cases. II) Existence of the root for entropy and divergency for temperature
due to GB gravity contributions. III) Dependency of the high energy limit of
temperature and mass on both magnetic charge and GB gravity. IV) Dependency
of the high energy limit of the pressure on magnetic charge and, heat
capacity on GB gravity. V) Existence of van der Waals like phase transition
for these black holes restricted to satisfaction of certain conditions. VI)
Independency of heat capacity's root on GB gravity (except for hyperbolic
black holes) and its sensitivity on magnetic charge. VI) Dependency of the
heat capacity's divergencies on both GB gravity and magnetic charges which
results into dependency of the thermal stability/instability of the black
holes on these quantities.

Studying the trajectory of a particle around these black holes resulted
into: I) High sensitivity of the shape of orbit on the GB gravity, energy
and the orbital angular momentum whereas significantly less sensitivity to
variation of the magnetic charge. II) Angular frequency being a decreasing
function of the magnetic charge, GB gravity, energy and the orbital angular
momentum. III) The possibility of absence of angular frequency (hence
geodesics) for certain values of different parameters.

The obtained results of this paper could be employed in the context of
AdS/CFT. Especially, it would be interesting to study the effects of
magnetic charge and GB gravity on the magnetization and susceptibility of
the boundary theory, and the conditions for having diamagnetic and
paramagnetic behaviors. The dynamical stability, quasinormal modes and
gravitational waves of these black holes are other subjects of interest.

\section{Appendix}

\subsection{components of the field equations}

Considering the obtained field equation (\ref{fieldeq}) with the given
metric (\ref{Metric}), one can obtain the following non-zero components for
the field equation
\begin{equation}
e_{tt}=e_{rr}=12 r^{3}\alpha f \left( r \right) f^{\prime }(r) -2
q_{E}^{2}-2 q_{M}^{2} -6\,{r}^{4}f \left( r \right)-2\,{r}^{6}\Lambda-3
r^{5} f^{\prime 4} k (\frac{2 \alpha f^{\prime }(r)}{r}-1)=0,  \label{e1}
\end{equation}
\begin{equation}
e_{\theta \theta}= e_{\psi \psi}=e_{\phi \phi}=4r^{4}\alpha f^{\prime
2}-r^{6}f^{\prime \prime 4}\alpha f^{\prime \prime }(r) f \left( r \right)
+2 q_{E}^{2}+2 q_{M}^{2} -2 r^{6} \Lambda-4 r^{5} f^{\prime 4} f(r)-2 r^{4}
k (2 \alpha f^{\prime \prime }(r)-1) =0.  \label{e2}
\end{equation}

Solving these two set of field equations, one can obtain the metric function
that is given in Eq. (\ref{metric function}).

\subsection{roots of the metric function}

Using the obtained metric function (\ref{metric function}), it is a matter
of calculation to obtain the following roots
\begin{equation}
r_{f(r)=0}=\left\{
\begin{array}{cc}
\sqrt{\frac{A_{1}^{2/3}+2 \sqrt[3]{A_{1}} k+4 \alpha k^2 \Lambda + 4 k^2-2
\Lambda m} {\sqrt[3]{A_{1}} \Lambda }} &  \\
\sqrt{\frac{i \left(\sqrt{3}+i\right) A_{1}^{2/3}+4 \sqrt[3]{A_{1}} k+ 2
\left(1+i \sqrt{3}\right) \left(\Lambda m-2 k^2(\alpha \Lambda +1)\right)}{2%
\sqrt[3]{A_{1}} \Lambda }} &  \\
\sqrt{\frac{\left(-1-i \sqrt{3}\right) A_{1}^{2/3}+4 \sqrt[3]{A_{1}} k+2 i
\left(\sqrt{3}+i\right) \left(2 k^2 (\alpha \Lambda +1)-\Lambda m\right)}{2%
\sqrt[3]{A_{1}} \Lambda }} &
\end{array}%
\right. ,  \label{roots of metric}
\end{equation}
where
\begin{equation*}
A_{1}=8 k^3+6 k \Lambda \left(2 \alpha k^2-m\right)+\Lambda ^2
\left(q_{E}^2+q_{M}^2\right)+\sqrt{\left(4 k^3 (3 \alpha \Lambda +2)-6 k
\Lambda m+\Lambda ^2 \left(q_{E}^2+q_{M}^2\right)\right)^2+8 \left(\Lambda
m-2 k^2 (\alpha \Lambda +1)\right)^3}.
\end{equation*}

Evidently, under certain circumstances, it is possible to have from zero up
to three roots for the metric function. This issue is confirmed in Fig. \ref%
{Fig0}.

\subsection{Kretschmann}

The general formula for calculating the Kretschmann scalar of $5-$%
dimensional topological black holes is given in Eq. (\ref{Kretschmann}).
Using the obtained metric function (\ref{metric function}), it is a matter
of calculation to obtain Kretschmann scalar as
\begin{eqnarray}
K &=& \frac{1}{12 A_ {2}^3 r^12 \alpha^2}\left\{-2 \alpha (A_{2}-3) A_{3}^3
r^{12}+(A_{2}-3)^2 A_{2}^3 r^{12} +72 \alpha ^2 A_{2}^2 r^6 \left(7
\left(q_{E}^2+q_{M}^2\right)-10 m r^2\right) \right.  \notag \\
&&\left. -288 \alpha ^2 A_{2}^2 r^6 \left(-2 m r^2+q_{E}^2+q_{M}^2\right)+2
A_{2} \left((A_{2}-3) A_{2} r^6+36 \alpha \left(-2 m
r^2+q_{E}^2+q_{M}^2\right)\right)^2 \right.  \notag \\
&&\left. +5184 \alpha ^3 \left(-2 m r^2+q_{E}^2+q_{M}^2 \right)^2 \right\} ,
\label{Full Kretschmann}
\end{eqnarray}
in which
\begin{equation*}
A_{2}=\sqrt{12 \alpha \left(\Lambda + \frac{6 m}{r^4}-\frac{2
\left(q_{E}^2+q_{M}^2\right)}{r^6}\right)+9}.
\end{equation*}

\subsection{roots and critical values of thermodynamical quantities}

Using the obtained temperature (\ref{temp}), one is able to extract the its
roots in the following form
\begin{equation}
r_{T=0}=\left\{
\begin{array}{cc}
\sqrt{\frac{2^{2/3} A_{3}^2+6 A_{3} k+18 \sqrt[3]{2} k^2}{6 A_{3} \Lambda }}
&  \\
\sqrt{\frac{2^{2/3} \left(-1-i \sqrt{3}\right) A_{3}^2+12 A_{3} k+ 18 i \sqrt%
[3]{2} \left(\sqrt{3}+i\right) k^2}{12 A_{3} \Lambda }} &  \\
\sqrt{\frac{i 2^{2/3} \left(\sqrt{3}+i\right) A_{3}^2+12 A_{3} k- 18 i \sqrt[%
3]{2} \left(\sqrt{3}-i\right) k^2}{12 A_{3} \Lambda }} &
\end{array}%
\right. ,  \label{roots of temp}
\end{equation}
where
\begin{equation*}
A_{3}=\sqrt[3]{54 k^3-27 \Lambda ^2 (q_{E}^2+q_{M}^2)+ \sqrt{\left(54 k^3-27
\Lambda ^2 (q_{E}^2+q_{M}^2)\right)^2-2916 k^6}}.
\end{equation*}

In general, three roots exist for temperature that under satisfaction of
certain conditions, all three cases might yield positive real valued roots
for the temperature.

As for the total mass (\ref{mass}), it is a matter of calculation to obtain
the following roots
\begin{equation}
r_{M=0}=\left\{
\begin{array}{cc}
\sqrt{\frac{2^{2/3} A_{4}^2+12 A_{4} k+72 \sqrt[3]{2} \alpha k^2 \Lambda +72
\sqrt[3]{2} k^2}{6 A_{4} \Lambda }} &  \\
\sqrt{\frac{2^{2/3} \left(-1-i \sqrt{3}\right) A_{4}^2+24 A_{4} k+ 72 i \sqrt%
[3]{2} \left(\sqrt{3}+i\right) k^2 (\alpha \Lambda +1)} {12 A_{4}\Lambda }}
&  \\
\sqrt{\frac{i 2^{2/3} \left(\sqrt{3}+i\right) A_{4}^2+24 A_{4} k- 72 i \sqrt[%
3]{2} \left(\sqrt{3}-i\right) k^2 (\alpha \Lambda +1)}{12 A_{4} \Lambda }} &
\end{array}%
\right. ,  \label{roots of mass}
\end{equation}
in which
\begin{equation*}
A_{4}= \sqrt[3]{648 \alpha k^3 \Lambda +432 k^3+ 54 \Lambda ^2
(q_{E}^2+q_{M}^2)+\sqrt{\left(648 \alpha k^3 \Lambda +432 k^3+54 \Lambda
^2+54 \Lambda ^2 (q_{E}^2+q_{M}^2)\right)^2+4 \left(-36 \alpha k^2 \Lambda
-36 k^2\right)^3}}.
\end{equation*}

Evidently, the mass, similar to the temperature and metric function could
enjoy three roots in its structure.

Our next items of the interest are critical horizon radius, temperature and
pressure. By solving Eq. (\ref{critical expression}), one can obtain the
critical horizon radius as
\begin{equation}
r_{c}=\sqrt{\frac{\sqrt[3]{3} A_{5}^2+12 \alpha A_{5} k^2+48\ 3^{2/3} \alpha
^2 k^4+5\ 3^{2/3} k \left(q_{E}^2+q_{M}^2\right)}{3 A_{5} k}},
\label{critical horizon}
\end{equation}
in which
\begin{equation*}
A_{5}=\sqrt[3]{576 \alpha ^3 k^6+\sqrt{3} \sqrt{-k^3
\left(q_{E}^2+q_{M}^2\right) \left(-144 \alpha ^2 k^3+q_{E}^2+q_{M}^2\right)
\left(432 \alpha ^2 k^3+125 \left(q_{E}^2+q_{M}^2\right)\right)}+252 \alpha
k^3 \left(q_{E}^2+q_{M}^2\right)}.
\end{equation*}

Using obtained critical horizon, it is a matter of calculation to obtain
critical temperature and pressure in the following forms
\begin{equation}
T_{c}=-\frac{-k r_{c}^4+q_{E}^2+q_{M}^2}{12 \pi \alpha k r_{c}^3+\pi r_{c}^5}%
,  \label{critical temperature}
\end{equation}
\begin{equation}
P_{c}=\frac{-3 k r_{c}^4+24 \pi \alpha k r_{c}^3 T_{c}+q_{E}^2+q_{M}^2+6 \pi
r_{c}^5 T_{c}}{8 \pi r_{c}^6}.  \label{critical pressure}
\end{equation}

Finally, considering the calculated heat capacity (\ref{heat}), one can
extract its roots as
\begin{equation}
r_{C=0}=\left\{
\begin{array}{cc}
\sqrt{\frac{18 \sqrt[3]{2} k^2-6 k A_{6}+2^{2/3} A_{6}^2}{48 \pi A_{6} P}} &
\\
\sqrt{\frac{18 i \sqrt[3]{2} \left(\sqrt{3}+i\right) k^2-12 k A_{6}+2^{2/3}
\left(-1-i \sqrt{3}\right) A_{6}^2}{96 \pi A_{6} P}} &  \\
\sqrt{\frac{-18 i \sqrt[3]{2} \left(\sqrt{3}-i\right) k^2-12 k A_{6}+i
2^{2/3} \left(\sqrt{3}+i\right) A_{6}^2}{96 \pi A_{6} P}} &
\end{array}%
\right. ,  \label{roots of heat capacity}
\end{equation}
in which
\begin{equation*}
A_{6}= \sqrt[3]{-54 k^3+\sqrt{\left(-54 k^3+1728 \pi ^2 P^2
(q_{E}^2+q_{M}^2)\right)^2-2916 k^6}+1728 \pi ^2 P^2(q_{E}^2+q_{M}^2)}.
\end{equation*}

\begin{acknowledgements}
We thank both Shiraz University and Shahid Beheshti University
Research Councils. This work has been supported financially partly
by the Research Institute for Astronomy and Astrophysics of
Maragha, Iran.
\end{acknowledgements}

\end{document}